\documentclass[12pt]{article}

\usepackage[]{inputenc}
\usepackage[T1]{fontenc}
\usepackage{hyperref}
\usepackage{color}
\usepackage{lmodern}
\usepackage[toc,page]{appendix}
\usepackage{fullpage}
\usepackage{authblk}

\newtheorem{lemma}{Lemma}[section]
\newtheorem{definition}{Definition}[section]
\newtheorem{proposition}{Proposition}

\newcommand{\be}{\begin{equation}}
\newcommand{\ee}{\end{equation}}
\newcommand{\bea}{\begin{eqnarray}}
\newcommand{\eea}{\end{eqnarray}}

\newcommand{\bs}[1]{\boldsymbol{#1}}

\newcommand{\bS}{{\bs s}}

\newcommand{\bL}{{\bs L}}
\newcommand{\bA}{{\bs A}}
\newcommand{\bR}{{\bs R}}
\newcommand{\bLambda}{{\bs \Lambda}}
\newcommand{\bomega}{{\bs \omega}}
\newcommand{\btheta}{{\bs \theta}}
\newcommand{\bxi}{{\bs \xi}}
\newcommand{\bQ}{{\bs Q}}
\newcommand{\bE}{{\bs E}}
\newcommand{\bF}{{\bs F}}
\newcommand{\bC}{{\bs C}}
\newcommand{\bJ}{{\bs J}}
\newcommand{\bY}{{\bs Y}}
\newcommand{\bZ}{{\bs Z}}
\newcommand{\bI}{{\bs I}}
\newcommand{\bw}{{\bs w}}

\newcommand{\bc}{{\bs c}}

\newcommand{\bolde}{{\bs e}}
\newcommand{\bl}{{\bs l}}

\newcommand{\bepsilon}{{\bs \epsilon}}
\newcommand{\bbeta}{{\bs \beta}}

\newcommand{\boldeta}{{\bs \eta}}
\newcommand{\bgamma}{{\bs \gamma}}
\newcommand{\bmu}{{\bs \mu}}
\newcommand{\bb}{{\bs b}}
\newcommand{\bGamma}{{\bs \Gamma}}

\def\ud{{\rm d}}
\def\tr{{\rm tr}}

\newcommand{\cL}{\mathcal L}
\newcommand{\cN}{\mathcal N}
\newcommand{\cM}{\mathcal M}
\newcommand{\cS}{\mathcal S}

\newcommand{\cH}{\mathcal H}

\newcommand{\cD}{\mathcal D}
\newcommand{\cF}{\mathcal F}

\newcommand{\RR}{{\mathbb R}}

\usepackage{amsmath}
\allowdisplaybreaks
\usepackage{amssymb}

\title{The second law of black hole mechanics in effective field theory}

\author[1,2]{Stefan~Hollands,}
\author[3,4]{\'Aron~D.~Kov\'acs}
\author[5]{and~Harvey~S.~Reall}

\affil[1]{\small Institute for Theoretical Physics, Leipzig University, Br\"uderstrasse 16, Leipzig 04103, Germany}
\affil[2]{\small MPI-MiS, Inselstrasse 22, Leipzig 04103, Germany}
\affil[3]{\small SISSA and INFN Sezione di Trieste, Via Bonomea 265, Trieste 34136, Italy}
\affil[4]{\small IFPU- Institute for Fundamental Physics of the Universe, Via Beirut 2, Trieste 34014, Italy}
\affil[5]{\small Department of Applied Mathematics and Theoretical Physics, University of Cambridge, Wilberforce Road, Cambridge CB3 0WA, U.K.}
\affil[ ]{\small E-mail: \url{stefan.hollands@uni-leipzig.de}, \url{arkovac@sissa.it}, \url{hsr1000@cam.ac.uk}}

\begin{document}

\maketitle

\begin{abstract}
We investigate the second law of black hole mechanics in gravitational theories with higher derivative terms in the action. Wall has described a method for defining an entropy that satisfies the second law to linear order in perturbations around a stationary black hole. We show that this can be extended to define an entropy that satisfies the second law to quadratic order in perturbations, provided that one treats the higher derivative terms in the sense of effective field theory. We also address some outstanding issues with Wall's method, in particular its gauge invariance and its relation to the Iyer-Wald entropy. 

\end{abstract}

\section{Introduction and overview}

\subsection{Introduction}

The laws of black hole mechanics are a set of classical laws governing the behaviour of black holes. When combined with Hawking's discovery of black hole radiation, these laws provide compelling evidence for the interpretation of black holes as thermodynamic objects. These laws were first proved for Einstein gravity minimally coupled to matter. It is natural to ask whether they are also valid in extensions of Einstein gravity, for example in the presence of higher-derivative terms in the action that are expected from an effective field theory (EFT) perspective. 

The first law of black hole mechanics concerns linear perturbations of a stationary (i.e. time-independent) black hole. Wald has shown that a version of the first law holds for {\it any} diffeomorphism invariant theory of gravity coupled to matter \cite{Wald:1993nt}. In particular, this leads to a definition of the entropy -- the Wald entropy -- of a stationary black hole in such a theory. It provides a fully satisfactory definition of the entropy of an equilibrium black hole in this very large class of theories. 

The Wald entropy is defined unambiguously only for a stationary black hole. Iyer and Wald have made a proposal for the entropy of a dynamical (i.e. non-stationary) black hole solution of a general diffeomorphism-invariant theory \cite{Iyer:1994ys}. The procedure is based on classifying possible terms according to their ``boost weight'', which determines how they transform under a constant rescaling of affine parameter of the horizon generators. The Iyer-Wald entropy is built from quantities of zero boost weight. Iyer and Wald showed that their definition is independent of any choice of coordinates or basis. However, they left open the question of whether or not it satisfies a second law. 

Jacobson, Kang and Myers (JKM) investigated black hole entropy for the class of theories for which the gravitational Lagrangian is a function of the Ricci scalar, so-called $f(R)$ theories \cite{Jacobson:1995uq}. Such theories can be transformed into a conventional scalar-tensor theory using a field redefinition. Using this, JKM were able to define an entropy that satisfies a second law. Their entropy is proportional to the integral of $f'(R)$ over a horizon cross-section. For a stationary black hole, the JKM entropy coincides with the Wald entropy. However, for a dynamical black hole, the JKM entropy differs from the Iyer-Wald entropy.  

In general, one expects the Lagrangian of a gravitational EFT to include scalars built from contractions of the Riemann tensor and its derivatives. For such a theory, Wall has sketched a procedure for constructing an entropy that satisfies the second law to {\it linear order} in perturbations around a stationary black hole \cite{Wall:2015raa}. As we shall explain, the Wall procedure supplements the Iyer-Wald entropy with terms that are linear in quantities with positive boost weight. The simplest example of such a term is the integral over a horizon cross-section of $K \bar{K}$ where $K$ and $\bar{K}$ are the expansions of outgoing and ingoing null geodesic congruences orthogonal to the cross-section. We shall refer to the result of Wall's construction as the Iyer-Wald-Wall (IWW) entropy. For a stationary black hole it reduces to the Wald entropy.

To linear order, the second law does not imply an entropy increase, but only that the entropy does not change in time. To see why, assume that there exists a linear perturbation that leads to an entropy increase over time. Now multiply this perturbation by minus one. The result is a linear perturbation that decreases the entropy over time, in violation of the second law. Thus, to linear order, Wall's result implies that the IWW entropy can neither decrease nor increase but must remain constant in time.\footnote{Wall allowed for the presence of matter, which was described by an energy-momentum tensor satisfying the null energy condition. This acts as a source for the linearized gravitational field. In this case, the entropy does increase but the increase arises entirely from the matter, not from gravitational waves. We will not follow this approach for reasons explained in section \ref{sec:wall}.} 

For $f(R)$ theories, the JKM and IWW entropies agree if $f(R)$ is quadratic in $R$ but not if cubic or higher order terms are present. This suggests that, except for special theories (e.g. 2-derivative Einstein gravity), the IWW entropy is unlikely to satisfy a second law beyond the linearized approximation. For $f(R)$ theories, the JKM entropy involves adding terms to the IWW entropy that are quadratic, or higher, order in quantities with positive boost weight. So, except for special theories, it seems likely that such higher order terms will need to be included if the second law is to be extended beyond the linearized approximation.

In this paper we will establish a second law beyond the linearized approximation for a large class of theories. Our approach will be to treat these theories as EFTs, ordering terms in the equations of motion according to how many derivatives they contain. The lowest order terms are the 2-derivative terms familiar from conventional Einstein theory. These are then supplemented by terms with 4 derivatives, then 6 derivatives and so on. We assume that the coupling constants multiplying these terms are all powers of some fundamental (UV) length scale $\ell$. We restrict attention to solutions lying within the regime of validity of EFT. Roughly this means that if a solution varies over a length (or time) scale $L$ then $\ell/L \ll 1$. One then expects terms with many derivatives to be less important than terms with few derivatives. 

Consider truncating such a theory by retaining only terms with $N$, or fewer, derivatives. Rather than attempting to prove a second law that holds {\it exactly} in the truncated theory, our approach will be to prove a second law that holds {\it to the same expected accuracy as the theory itself}, i.e., up to neglect of terms with more than $N$ derivatives. We will show that, for any $N$, one can define an entropy that satisfies the second law to {\it quadratic order} in perturbations around a stationary black hole, in the sense that any violation of the second law will be of the same order (in $\ell/L$) as the terms with more than $N$ derivatives that have been neglected in truncating the theory. By increasing $N$ one improves the result, so the better one knows the EFT, the better the accuracy to which the second law is satisfied. 

Our entropy is defined by adding new terms to the IWW entropy. The new terms are of quadratic (or higher) order in quantities with positive boost weight. The simplest example is the integral over a horizon cross-section of $(K \bar{K})^2$. By counting derivatives, one can see that the new terms are required only once $N \ge 6$. In particular, if $N=4$ then our result implies that the IWW entropy satisfies the second law to quadratic order in perturbations, in the EFT sense described above.

An important question concerning the entropy is its gauge-invariance. The Iyer-Wald entropy of a horizon cross-section is independent of any choice of coordinates -- it depends only on the local geometry of the cross-section. Wall's procedure, which our approach extends, is based on a fixed Gaussian null coordinate system. Such a coordinate system has gauge freedom corresponding to a rescaling of the affine parameter along each horizon generator. The terms generated by Wall's procedure are not manifestly invariant if this rescaling differs from generator to generator. Nevertheless, we will prove below that the IWW entropy is invariant to linear order in perturbations around a stationary black hole. Thus the IWW entropy is gauge invariant to the same extent that it satisfies the second law, namely to linear order in perturbations. In fact, with quite a lot more work, we shall prove that the freedom to adjust higher order terms in the IWW entropy can be used to bring it to a form that is gauge invariant in a fully nonlinear sense. It is natural to guess that our extension of the IWW entropy is gauge invariant at least to quadratic order in perturbations, in the sense of EFT. We have not yet been able to prove this and we will discuss this issue further below. This is not an issue for $N=4$ since our entropy reduces to the IWW entropy in this case. 

In the rest of this section we shall explain our main results in more precise terms but omitting the proofs. Then section \ref{sec:preliminaries} will prove various technical results used in later sections. Section \ref{sec:entropy_current} will present proofs of our results concerning the IWW entropy. Section \ref{sec:improved_entropy} presents our construction of the improved entropy that satisfies the second law to quadratic order.

\subsection{Summary of conventions}

The metric signature is $(-++\dots+)$, and the conventions for the curvature tensors are as in Wald's text \cite{Wald:1984rg}. Bold face letters denote differential forms, with the conventions for $\wedge$ and $\ud$ as in \cite{Wald:1984rg}. The spacetime dimension is $n>2$. Lower case Greek letters $\mu,\nu,\dots$ are spacetime coordinate indices, upper case Roman letters $A,B, \dots$ are coordinate indices on an $(n-2)$ dimensional spacelike cut $C$ associated with the Gaussian null coordinates defined in section \ref{sec:GNCintro}. $(\mu_1 \dots \mu_s)$ resp. $[\mu_1 \dots \mu_a]$ denotes a total symmetrization resp. anti-symmetrization with combinatorial factors included to make these operations projections onto the space symmetric resp. anti-symmetric tensors. 

The notation $L[\Psi]$ (square brackets) indicates that $L$ is a local functional of some fields $\Psi$, i.e. at each point, $x$,
$L|_x$ depends on finitely many derivatives $\partial_{\mu_1} \dots \partial_{\mu_d} \Psi|_x$. We generally set $16\pi G = 1$ and $\ell$ denotes a UV length scale.

\subsection{Gaussian null  coordinates}
\label{sec:GNCintro}

Consider a spacetime of dimension $n$ containing a {\it smooth} null hypersurface $\cN$ that is ruled with affinely parameterized null geodesics with future-directed tangent vector $l^\mu$. We assume that the generators are future-complete, i.e., they extend to infinite affine parameter to the future. This hypersurface might be the event horizon of a black hole but some of our analysis applies more generally. Note that the smoothness assumption excludes important physical situations such as black hole mergers. However, it includes spacetimes describing a black hole ``settling down to equillibrium'', which is the physical situation we have in mind. 

Assume that every null geodesic generator of $\cN$ intersects a spacelike cross section $C$ precisely once. We can introduce Gaussian Null Coordinates (GNCs) in a neighbourhood of $\cN$ as follows. On $\cN$ we let $v$ be an affine parameter along each null generator such that $v=0$ on $C$ and such that $l^\mu \partial_\mu v = 1$. Then we transport $C$ by affine time $v$ into cross sections $C(v)$ thereby obtaining a null foliation of $\cN$. On each $C(v)$, a (past-directed) null vector field $n = n^\mu \partial_\mu$ is next defined uniquely by demanding that it is orthogonal to $C(v)$ and $g_{\mu\nu} l^\mu n^\nu = 1$. We consider the affinely parameterized null geodesics tangent to $n$ and call the affine parameter $r$ with $r=0$ on $\cN$. Finally, we choose on $C$ a coordinate chart $x^A, A=1, \dots, n-2$. Then we transport this first along the geodesics tangent to $l$ along $\cN$ and then along $n$ off $\cN$ at each fixed value of $v$. It can be shown that the metric and vector fields $n,l$ take 
the form
\be
\label{GNC1}
g = 2 \ud v(\ud r - \tfrac{1}{2} r^2 \alpha \ud v - r\beta_A \ud x^A) + \mu_{AB} \ud x^A \ud x^B, \quad
l = \partial_v, \quad n = \partial_r
\ee
in the GNCs $(v,r,x^A)$, where $\alpha$, $\beta_A$ and $\mu_{AB}$ are, at least for small $r$, smooth functions of the coordinates. $\cN$ is the surface $r=0$ and $C$ is the surface $v=r=0$. We denote the inverse of $\mu_{AB}$ as $\mu^{AB}$ and $A,B,\ldots$ indices will be raised and lowered with $\mu^{AB}$ and $\mu_{AB}$. The covariant derivative on $C(v)$ defined by $\mu_{AB}$ is denoted $D_A$. 

We also define
\be
\label{K:def}
K_{AB} \equiv \frac{1}{2} \partial_v \mu_{AB}, \quad 
\bar K_{AB} \equiv \frac{1}{2} \partial_r \mu_{AB} \qquad K \equiv K^A_A \qquad \bar{K} \equiv \bar{K}^A_A.
\ee
On $\cN$, $K$ is the expansions of the null geodesics tangent to $l$. Similarly $\bar{K}$ is the expansion of the (past-directed) null geodesics tangent to $n$. The tracefree parts of $K_{AB}$ and $\bar{K}_{AB}$ define the shear of these families of geodesics. 

The definition of GNCs is not unique. If one fixes the initial cut $C$ then one still has the freedom to rescale the affine parameter along each generator of $\cN$: if $a(x^A)>0$ then $v' \equiv v/a$ is also an affine parameter along the generators. The corresponding tangent vector to these generators is $l'=al$. This change of affine parameter leads to a new set of GNCs $(v',r',x^{A'})$. 

If $a$ is {\it constant} then we have simply 
\be
 v' = \frac{v}{a} \qquad r' = ar \qquad x^{A'} = x^A \qquad {\rm constant} \; a.
\ee
If a quantity $X$ transforms homogeneously under such a change of coordinates then we define its {\it boost weight} $b$ by
\be
\label{bw_def}
  X'(v',r',x^{A'}) = a^b X(v,r,x^A)
\ee
(see Section \ref{sec:GNC} for a precise definition); for example $\alpha,\beta_A$ and $\mu_{AB}$ have boost weight zero, $K_{AB}$ has boost weight $+1$ and $\bar{K}_{AB}$ has boost weight $-1$. A quantity $D_{A_1} \ldots D_{A_m} \partial_v^p \partial_r^q \psi$ with $\psi \in \{\alpha,\beta_A,\mu_{AB} \}$ has boost weight $p-q$. The boost weight of a tensor component $T^{\mu\nu\ldots}{}_{\rho \sigma \ldots}$ is given by the sum of $+1$ for each subscript $v$ or superscript $r$ index, and $-1$ for each superscript $v$ or subscript $r$ index e.g. $R_{vv}$ has boost weight $+2$. 

If $a$ is not constant then the transformation between the GNCs is highly non-trivial away from $\cN$. This implies that many quantities (e.g. $\alpha$, $\beta_A$, $\partial_v \partial_r \mu_{AB}$) transform inhomogeneously, even on $\cN$, with terms involving the derivative of $a$.

\subsection{Second law}

We now review a simple proof of the second law in conventional GR. The proof is simple because it makes various strong assumptions, specifically that the horizon is smooth, the horizon generators are future-complete, and that the black hole ``settles down to equilibrium'' at late time. One can of course prove the second law in conventional GR under much weaker assumptions than these \cite{Chrusciel:2000cu}. 

Let $\cH$ be the future event horizon of a black hole and take $\cN =\cH$ to define GNCs. The area of $C(v)$ is
\be
 A(v) = \int_{C(v)} \ud^{n-2} x \sqrt{\mu}
\ee
with rate of change
\be
 \dot{A}(v) = \int_{C(v)} \ud^{n-2} x \sqrt{\mu} K
\ee 
Recall Raychaudhuri's equation:
\be
 \partial_v K = -K^{AB} K_{AB} - R_{vv} \qquad {\rm on} \; \cN
\ee
We now {\it assume} that the expansion of the generators of $\cH$ vanishes at late time, i.e., $K \rightarrow 0$ as $v \rightarrow \infty$ on $\cH$. This would be the case if the black hole is ``settling down to equilibrium''. We can now write
\bea
\label{dotA}
 \dot{A}(v) &=& \int_{C(v)} \ud^{n-2} x \sqrt{\mu} \int_v^\infty dv' (-\partial_v K)(v',x) \nonumber \\ &=& \int_{C(v)} \ud^{n-2} x \sqrt{\mu} \int_v^\infty dv' \left(K^{AB} K_{AB} + R_{vv} \right)(v',x)
\eea
where we used Raychaudhuri's equation in the second step. If the spacetime satisfies the null convergence condition ($R_{\mu\nu} V^\mu V^\nu \ge 0$ for any null $V^\mu$) then the RHS is manifestly non-negative (as $R_{vv} = R_{\mu\nu} l^\mu l^\nu \ge 0$) and so we have $\dot{A}(v) \ge 0$, i.e, $A(v)$ is an increasing function.

If we consider a theory consisting of conventional 2-derivative GR coupled to matter satisfying the null energy condition then spacetimes satisfying the Einstein equation will obey the null convergence condition and so the second law holds in such a theory. However, we will be interested in more general theories in which higher derivatives are present in the Lagrangian. In this case there is no reason to expect the null convergence condition to be satisfied by solutions of the equations of motion and so the above argument no longer applies. 

\subsection{Stationary black holes}

Much of this paper will concern perturbations of stationary black holes. We will now describe the class of stationary black holes to be considered. In conventional GR coupled to various types of matter fields it is known that the event horizon of a stationary black hole must be a Killing horizon. This result has not been extended to theories of the type that we will be considering. We will simply {\it assume} that the theory admits a family $\cF$ of stationary black hole solutions for which the event horizon is a Killing horizon. Moreover, we will assume that this is a {\it bifurcate} Killing horizon. This implies that the zeroth law of black hole mechanics is satisfied,\footnote{
Very recently, Ref. \cite{Bhattacharyya:2022nqa} has proved a zeroth law for a large class of theories without assuming a bifurcate Killing horizon, under the assumption that higher derivative terms in the action are multiplied by powers of a dimensionful constant (analogous to our $\ell$ below) and that the metric depends analytically on this constant.} i.e., that the surface gravity is constant on the horizon (conversely, the zeroth law implies that the spacetime can be extended to contain a bifurcation surface \cite{Racz:1992bp}).  

These assumptions ensure that if $\cH$ is the future event horizon of a black hole in $\cF$ then all positive boost-weight quantities vanish on $\cH$. Furthermore, all non-zero boost-weight quantities vanish on the bifurcation surface.

\subsection{Wall's procedure}

\label{sec:wall}

Wall \cite{Wall:2015raa} has sketched an approach that, for any diffeomorphism invariant theory of gravity, produces an entropy $S$ which satisfies the second law to linear order, i.e., $\delta \dot{S}=0$ for perturbations of a black hole in the family $\cF$.\footnote{
See \cite{Chatterjee:2011wj,Bhattacharjee:2015yaa} for earlier work establishing a linearized second law in particular theories.
} Wall's approach has been discussed in more detail in \cite{Bhattacharya:2019qal,Bhattacharyya:2021jhr}, where it has been reformulated in terms of an {\it entropy current}. To explain this, let $\sqrt{-g} E^{\mu\nu} = \delta I/\delta g_{\mu\nu}$, where $I$ is the action, so the Einstein equation of our theory is $E_{\mu\nu}=0$. Then, for a general null hypersurface $\cN$ the entropy current is a vector field $s^v \partial_v + s^A \partial_A$ tangent to $\cN$, where $s^v$ and $s^A$ are functions of the GNC components of the metric, and their derivatives, with boost-weights $0$ and $1$ respectively. It satisfies the identity
\be
\label{s_def}
 E_{vv}= \partial_v \left[ \frac{1}{\sqrt{\mu}} \partial_v \left( \sqrt{\mu} s^v \right) + D_A s^A \right]+ \ldots
\ee
where the ellipsis denotes terms that are of quadratic or higher order in quantities of positive boost weight. This is an off-shell identity, i.e., it holds independently of the equations of motion. To obtain the linearized second law, we consider this equation linearized around a member of $\cF$, taking $\cN$ to be the event horizon $\cH$. Since positive boost weight quantities vanish on the event horizon of the unperturbed spacetime we obtain
\be
\label{deltas_eq}
\delta E_{vv}= \partial_v \delta \left[ \frac{1}{\sqrt{\mu}} \partial_v \left( \sqrt{\mu} s^v \right) + D_A s^A \right]
\ee
On-shell this becomes
\be
\label{deltas_eq_on_shell}
  \partial_v \delta \left[ \frac{1}{\sqrt{\mu}} \partial_v \left( \sqrt{\mu} s^v \right) + D_A s^A \right]=0
\ee
The quantity inside square brackets vanishes for the background solution and can be expected to decay as $v \rightarrow \infty$ if the perturbed spacetime settles down to a stationary black hole belonging to $\cF$. Hence integrating the above equation w.r.t. $v$ gives
\be
\label{divs_zero}
\frac{1}{\sqrt{\mu}} \delta \partial_v \left( \sqrt{\mu} s^v \right) + D_A \delta s^A = 0
\ee
Equation \eqref{deltas_eq_on_shell} shows that if this condition holds on $C$ then it holds on all of $\cH$. This equation can be regarded as a gauge condition on the metric perturbation. It arises because we have chosen our coordinates such that the perturbation does not change the location of $\cH$, i.e., the horizon of the perturbed black hole remains at $r=0$. For conventional GR (where \eqref{divs_zero} is simply $\delta K=0$) this was explained in \cite{Hollands:2012sf}. 

Integrating \eqref{divs_zero} over $C(v)$ (with measure $\sqrt{\mu}$) the divergence drops out and we obtain $\delta \dot{S}_{\rm IWW} = 0$ where the Iyer-Wald-Wall entropy is\footnote{Recall our choice of units $16\pi G=1$ so $1/(4G)=4\pi$. This factor is included for convenience, so that we have $s^v=1$ for standard GR.}
\be
\label{Ssv}
 S_{\rm IWW}(v) = 4\pi \int_{C(v)} \ud^{n-2} x \sqrt{\mu} s^v
\ee
Thus this definition of entropy satisfies the second law to linear order in perturbations. 

Wall considered coupling the original theory to a matter source so that the Einstein equation becomes $E_{\mu\nu} = (1/2) T_{\mu\nu}$ (units: $16\pi G = 1$) where $T_{\mu\nu}$ is the energy-momentum tensor of the matter. If one treats $T_{\mu\nu}$ as a term of linear order then equation \eqref{deltas_eq_on_shell} has $(1/2)T_{\mu\nu}$ on the RHS and if the matter obeys the null energy condition $T_{vv} \ge 0$ then $\delta \dot{S}_{\rm IWW} \ge 0$ so one can have a genuine increase in entropy driven by the matter source. However, we shall not include such a matter source below for several reasons: (i) if the matter fields are treated using the linearized approximation, as with the gravitational field, then $T_{\mu\nu}$ is of quadratic, not linear, order; (ii) in EFT one does not expect a clear division of the Lagrangian into a ``gravitational part'' and a ``matter part''; instead there will be higher derivative terms mixing matter and gravitational fields; (iii) higher derivative matter terms will not satisfy the null energy condition. We take the view that one should treat the gravitational and matter fields on an equal footing.\footnote{Wall's analysis does allow for a scalar field which is treated on an equal footing with the metric, i.e., this field is not included in $T_{\mu\nu}$.}

Section \ref{sec:entropy_current} of this paper will address some outstanding issues concerning Wall's approach. 

First, one needs to make sure that this definition of entropy also satisfies the {\it first} law of black hole mechanics, which relates $\delta S$ evaluated on the bifurcation surface of the black hole to the perturbations in mass and angular momentum. Wall argues that this must be true as follows. His result for $s^v$ can be divided into terms built entirely from quantities with vanishing boost weight and a part that is at least quadratic in quantities with non-vanishing boost weight. Wall claims that the integral of the former part is the same as the entropy $S_{\rm IW}$ defined by Iyer and Wald. From this claim it follows that $\delta S_{\rm IWW} = \delta S_{\rm IW}$ on the bifurcation surface of a member of $\cF$ and since $S_{\rm IW}$ is known to satisfy the first law, so must $S_{\rm IWW}$. However, a proof of this claim is lacking. We shall present one in section \ref{sec:IWW}. 

Second, Wall's procedure is carried out using a particular set of GNCs and produces an expression for $s^v$ that depends on the metric components, and their derivatives, w.r.t. those GNCs. These quantities transform in a complicated way under transformations between GNCs with non-constant $a(x^A)$. Therefore it is very unclear whether the result of Wall's procedure must be gauge-invariant. (This has also been noted in \cite{Bhattacharyya:2022njk}.) 

The issue of gauge invariance is important if we wish to compare the entropy of two cuts $C,C'$ of $\cH$ with $C'$ strictly to the future of $\cH$. The linearized second law described above lets us do this for the special case where $C'=C(v)$. So to compare the IWW entropy of $C'$ and $C$ we must choose our GNCs such that $C'$ is a constant $v$ cut of $\cH$. This can always be achieved by rescaling the affine parameters of the generators of $\cH$, i.e., by a choice of the function $a(x^A)$ defined in section \ref{sec:GNCintro}. But we do not want the definition of the entropy of $C$ to depend on our choice of $C'$ (or vice versa). Hence we must demonstrate that this definition is gauge-invariant under changes of GNCs involving non-constant $a(x^A)$. 

In section \ref{sec:gauge_inv}, we shall give a short proof that the IWW entropy is gauge-invariant at the linearized level. However, in the class of examples studied by Wall \cite{Wall:2015raa} (a Lagrangian that is a function of the Riemann tensor) one can see that the result is actually gauge-invariant at the fully nonlinear level and one can ask whether this is true generally. In Wall's analysis, he implicitly makes use of the fact that his procedure only determines terms in $S_{\rm IWW}$ that are of up to linear order in quantities with positive boost weight since terms of quadratic (e.g. $K^2 \bar{K}^2$) or higher order vanish when linearized around a member of $\cF$. Wall makes a specific choice of these higher order terms in his explicit examples. The question is whether it is always possible to choose these higher order terms to render $S_{\rm IWW}$ fully gauge-invariant. We shall show in section \ref{sec:gauge_inv} that the answer is yes.

\subsection{Second law in EFT}

\label{sec:EFT_2nd_law}

We can now explain our main result, derived in section \ref{sec:vacuum_gravity}. For simplicity, we shall consider pure gravity, without matter fields. Consider an EFT for a diffeomorphism invariant theory of gravity. We assume that the Lagrangian is a formal sum of terms with increasing numbers of derivatives of the fields, multiplied by suitable powers of some UV length scale $\ell$. A term with $k+2$ derivatives will be multiplied by $\ell^k$. For pure gravity (assuming parity symmetry if $n$ is odd) only terms with even numbers of derivatives can occur. The terms with 2 or fewer derivatives are assumed to take the standard Einstein-Hilbert form, with a possible cosmological constant. 

Validity of EFT requires that terms with increasing numbers of derivatives are increasingly less important. This is the case if we restrict ourselves to spacetimes varying over some length/time scale $L$ satisfying $\ell/L \ll 1$. This $L$ should be a lower bound for the size of the final black hole equilibrium state and any length/time scales associated with the perturbation away from equilibrium. More precisely, we assume that there exists $L>0$ with $\ell/L \ll 1$ such that, in a neighbourhood of the event horizon, w.r.t. some set of GNCs, any quantity $X_k$ involving $k$ derivatives of the metric obeys a uniform bound of the form $|X_k| \le C_k/L^k$ for some dimensionless constants $C_k$ depending on the initial data. We assume that the cosmological constant satisfies $|\Lambda|L^2 \le 1$.

In practice, an EFT Lagrangian will not be known to all orders. We assume that only the terms with $N$ or fewer derivatives are known explicitly. The EFT equation of motion is then
\be
\label{Einstein_EFT}
 E_{\mu\nu} = {\cal O}(\ell^{N})
\ee 
where $E_{\mu\nu} = -\Lambda g_{\mu\nu} - G_{\mu\nu} + \ldots$ denotes the terms with up to $N$ derivatives and the RHS represents the effects of the terms with $N+2$ or more derivatives. We should really write this as ${\cal O}(\ell^{N}/L^{N+2})$ but we shall mostly suppress the $L$-dependence below. 

Consider first the case $N=2$, i.e., conventional GR viewed as an EFT. Without matter we have $R_{vv} = -E_{vv} = {\cal O}(\ell^2)$. Hence equation \eqref{dotA} is
\be
\dot{A}(v) = \int_{C(v)}  \ud^{n-2} x  \sqrt{\mu} \int_v^\infty \ud v' \left(K^{AB} K_{AB} (v',x)+ {\cal O}(\ell^2) \right)
\ee
The RHS may become negative but only by a small ${\cal O}(\ell^2)$ amount. From an EFT perspective the second law no longer holds exactly, but only to the same ${\cal O}(\ell^2)$ accuracy as the theory itself. If $\dot{A}(v_0)<0$ then the above equation implies (reinstating $L$) $||K_{AB}||_{L^2(v_0)} = {\cal O}(\ell/L^2)$ where $|| \cdot ||_{L^2(v_0)}$ denotes the norm defined by the above integral on the portion $v \ge v_0$ of $\cH$. Hence if $\dot{A}(v_0)<0$ then the expansion and shear of the generators of $\cH$ must be small for all times $v \ge v_0$. This suggests that the black hole is close to equilibrium. Thus the process of relaxation to equilibrium seems the most likely situation in which the higher derivative EFT terms could cause a (small) decrease in horizon area.

Our main result is to show that, for general $N$, the following EFT generalization of \eqref{s_def} (evaluated on-shell) holds on a null hypersurface $\cN$:\footnote{
A similar equation was derived for Lovelock theories in \cite{Bhattacharyya:2016xfs}.}
\be
\label{gen_Ray}
 \partial_v  \left[ \frac{1}{\sqrt{\mu}} \partial_v \left( \sqrt{\mu} S^v \right) + D_A S^A \right] = - \left(K^{AB} + X^{AB} \right)\left( K_{AB} + X_{AB} \right) - D_A Y^A +{\cal O}(\ell^N)
\ee 
Here $S^A=s^A$, and $S^v$ is defined by adding new terms to the expression for $s^v$ arising from $E_{\mu\nu}$. These new terms are quadratic (or higher order) in terms with positive boost weight so they do not affect any of the results described in the previous section, i.e., such terms are not fixed by Wall's procedure. On the RHS, $X_{AB}$ (which is symmetric) and $Y^A$ are both ${\cal O}(\ell^2)$ and have boost weight $1,2$ respectively, and $Y^A$ is a sum of terms that each contain at least two factors of positive boost weight. In contrast with \eqref{s_def}, the above equation holds on-shell, i.e., its derivation makes use of most of the components of the Einstein equation \eqref{Einstein_EFT}. The ${\cal O}(\ell^N)$ term arises not only from the ${\cal O}(\ell^N)$ corrections on the RHS of  \eqref{Einstein_EFT} but also from various steps in the derivation. In other words, even if we worked with the exact equation $E_{\mu\nu}=0$ we would still generate a ${\cal O}(\ell^N)$ term. (In this regard the $N=2$ case is exceptional.)

To obtain a second law we take $\cN$ to be the horizon $\cH$ of a black hole and use $S^v$ to define an entropy $S$ as in \eqref{Ssv}:
\be
\label{SSv2}
 S(v) = 4\pi \int_{C(v)} \ud^{n-2} x \sqrt{\mu} S^v
\ee
Equation \eqref{gen_Ray} implies
\be
\label{int_S}
 \dot{S}(v) = \frac{1}{4} \int_{C(v)}  \ud^{n-2} x  \sqrt{\mu} \int_v^\infty \ud v' \left[\left(K^{AB} + X^{AB} \right)\left( K_{AB} + X_{AB} \right) + D_A Y^A + {\cal O}(\ell^N)\right](v',x)
\ee
where again we assume that our black hole settles down to a member of $\cF$ at late time, which implies that all quantities with positive boost weight vanish at late time, hence the (boost-weight $1$) quantity in square brackets on the LHS of \eqref{gen_Ray} vanishes as $v \rightarrow \infty$. In constrast with the $N=2$ case, we do not know anything about the sign of $D_A Y^A$ on the RHS of \eqref{int_S} in a fully nonlinear situation. However, we can make progress by resorting to second order perturbation theory about a member of $\cF$. 

Positive boost weight quantities vanish on the horizon of a member of $\cF$, which implies that $S^v$ and $s^v$ agree at zeroth and first order in perturbation theory. Since $Y^A$ is at least quadratic in positive boost weight quantities we have 
\be
 \delta^2 \int_{C(v)}  \ud^{n-2} x  \sqrt{\mu} \int_v^\infty \ud v' (D_A Y^A)(v',x)=\int_{C(v)} \ud^{n-2} x \sqrt{\mu} \int_v^\infty \ud v' (D_A \delta^2 Y^A)(v',x)
\ee
We are working to quadratic order so on the RHS $D_A$ and $\sqrt{\mu}$ are evaluated in the background spacetime, where they are independent of $v$. Hence we can interchange the order of integration and use the divergence theorem to see that the RHS vanishes. This leaves
\be
\label{2ndlaw_quad}
 \delta^2 \dot{S}(v) = 
 \frac{1}{4} \int_{C(v)}  \ud^{n-2} x  \sqrt{\mu} \int_v^\infty \ud v' \left[\mu^{AC} \mu^{BD} \delta \left(K_{AB} + X_{AB} \right)\delta \left( K_{CD} + X_{CD} \right) +{\cal O}(\ell^N)\right](v',x)
\ee
where we used the fact that $K_{AB}$ and $X_{AB}$ have boost weight one and hence vanish on $\cN$ in the background spacetime. Since the first term on the RHS is positive definite, this shows that, to quadratic order in perturbations, $\dot{S}(v)$ can become negative only by an amount ${\cal O}(\ell^N)$, i.e., of the same size as the unknown effects caused by our ignorance of higher order EFT terms. In particular, the better we know the EFT, the larger $N$ is, and the smaller the amount by which $\dot{S}(v)$ can become negative.\footnote{
There is another way of looking at the second law to quadratic order. If $S$ satisfies a second law then any horizon cross-section satisfies $\dot{S} \ge 0$. A horizon cross-section of a stationary black hole satisfies $\dot{S}=0$ and therefore minimizes $\dot{S}$ within the ``space of cross-sections of black hole horizons''. Hence if the second law holds then variations around a stationary black hole must satisfy the conditions for a local minimum, i.e., $\delta \dot{S}=0$ and $\delta^2 \dot{S} \ge 0$.
}

Since our equations of motion involve $N$ or fewer derivatives, it follows that $S^v$ and $S^A$ involve $N-2$ or fewer derivatives. These quantities are sums of monomials where each monomial is at most quadratic in terms with positive boost weight. It is natural to group such monomials into those with zero, one or two (or more) factors with positive boost weight. The former terms generate the Iyer-Wald entropy, including the second set of terms gives the Iyer-Wald-Wall entropy and including the final set of terms gives our generalization of this. A term with non-zero boost weight $b$ costs at least $b$ derivatives so a term quadratic in positive boost weight quantities, with overall boost weight zero, must contain at least $4$ derivatives and therefore occurs only for $N \ge 6$. Hence for $N=4$ we have $S^v = s^v$ and $S^A=s^A$ so in this case our result implies that, without further modification, the Iyer-Wald-Wall entropy satisfies the second law to quadratic order in perturbations around a member of $\cF$. 

Equation \eqref{gen_Ray} holds for a purely gravitational EFT. We shall make a few remarks about the generalization to include matter fields. With matter fields there is a greater variety of higher derivative terms that can occur on the RHS of \eqref{gen_Ray}. However, there are also further terms on the RHS of \eqref{gen_Ray} arising from the energy-momentum tensor of the $2$-derivative part of the matter Lagrangian. If this $2$-derivative Lagrangian satisfies the null energy condition then these terms have a ``good sign'' (i.e. they are negative definite). The hope is that this can be used to help control the behaviour of higher derivative terms involving the matter fields, by completing the square in the same way that we did with the $K_{AB}K^{AB}$ term in vacuum gravity. For example, a minimally coupled $2$-derivative scalar field contributes $-(1/2)(\partial_v \Phi)^2$ to the RHS of \eqref{gen_Ray}. So one can try to use this to control higher derivative terms involving $\partial_v \Phi$ by completing the square. This will lead to an extra term of the form $-(1/2)(\partial_v \Phi+P)^2$ on the RHS of \eqref{gen_Ray}. (We will study an example of this in section \ref{sec:EdGB}.) This will give an extra term $[\delta (\partial_v \Phi +P )]^2$ inside the integral of \eqref{2ndlaw_quad}. Since this term has a good sign, the above argument that the second law holds to quadratic order, in the sense of EFT, is still valid. In section \ref{sec:scalar_tensor_all_orders} we shall sketch a proof that this can be done for any scalar-tensor EFT. 

\subsection{Example: vacuum gravity}

Field redefinitions can be used to simplify the Lagrangian of an EFT. For example, in vacuum gravity, a field redefinition can be used to bring the terms with up to $4$ derivatives to the ``Einstein-Gauss-Bonnet'' form:\footnote{
See \cite{Bhattacharjee:2015qaa} for previous work on the second law in this theory under the assumption of spherical symmetry.
}
\be
\label{EGB_lag}
 L = -2\Lambda + R +\frac14 k \ell^2  L_{\rm GB}
\ee
where $k$ is a dimensionless constant and $L_{\rm GB}$ is the Euler density associated with the Gauss-Bonnet invariant:
\be
L_{\rm GB} = \frac{1}{4} \delta^{\mu_1 \mu_2 \mu_3 \mu_4}_{\nu_1 \nu_2 \nu_3 \nu_4} R_{\mu_1 \mu_2}{}^{\nu_1\nu_2} R_{\mu_3 \mu_4}{}^{\nu_3 \nu_4} 
\ee
This term is topological in $n=4$ dimensions but non-trivial in higher dimensions. This theory has second order equations of motion and admits a well-posed initial value problem \cite{Kovacs:2020pns,Kovacs:2020ywu} as long as it remains within the regime of validity of EFT.

Since we have $N=4$ here, we have $S^v = s^v$ and $S^A=s^A$ as explained above. Using previous results for $s^v$ and $s^A$ \cite{Wall:2015raa,Bhattacharya:2019qal,Bhattacharyya:2021jhr}  gives 
\be
 S^v = 1 + \frac{k}{2} \ell^2 R[\mu] \qquad S^A = -k \ell^2  \left( D_B K^{AB} -D^A K \right)
\ee
where $R[\mu]$ is the Ricci scalar of $\mu_{AB}$. In this case $S^v$ involves only quantities of zero boost weight, so our entropy is simply the Iyer-Wald entropy $S_{\rm IW}$. Hence, for Einstein-Gauss-Bonnet theory, our result implies that the Iyer-Wald entropy satisfies the second law to quadratic order, in the EFT sense explained above, i.e., $\delta^2 \dot{S}_{\rm IW}$ is positive up to ${\cal O}(\ell^4)$ terms.

We can also demonstrate how to obtain \eqref{gen_Ray} for this theory. A calculation gives the off-shell identity
\bea
\label{Evv:EGB}
-E_{vv}
&=&-\partial_v \left[ \frac{1}{\sqrt{\mu}} \partial_v \sqrt{\mu} \right] - K^{AB} K_{AB}  \\&+&
\partial_v\left[\frac{k \ell^2}{\sqrt{\mu}}\partial_v \left(-\frac12\sqrt{\mu}  R[\mu]\right)+k \ell^2D^A\left(D^BK_{AB}-D_A K\right)\right]  - K_{AB} W^{AB}- {\cal D}_AY^A  \nonumber
\eea
where the terms on the first line are the usual terms arising from the Einstein-Hilbert Lagrangian, and the terms on the second line arise from $L_{GB}$. These involve
\bea
W^{AB}&=&-k \ell^2 \left[ \frac12 \mu^{AB} \partial_v R[\mu]-\partial_v R[\mu]^{AB}+R_{vv}\left({\bar K}^{AB}-{\bar K}\mu^{AB}\right) \right. \nonumber \\
& &+2R_v{}^{(A}{}_{vC}{\bar K}^{B)C}-R_{vCvD}{\bar K}^{CD}\mu^{AB}-R_v{}^A{}_{v}{}^B{\bar K} \nonumber \\
& &+\left. {\cal D}^{(A}R_{vC}{}^{B)C}-{\cal D}_C R_{vD}{}^{CD}\mu^{AB}+{\cal D}_C R_v{}^{(A|C|B)} \right]
\eea
and
\be
 Y^A = -k \ell^2\left[K R_{vC}{}^{AC}-K^{AC}R_{vBC}{}^B-K_{BC}R_v{}^{CAB} \right].
\ee
where $R[\mu]_{AB}$ is the Ricci tensor of $\mu_{AB}$. These expressions involve a new derivative operator $\cD_A$. Acting on a quantity of boost weight $b$ this is defined as $\cD_A = D_A - b\beta_A/2$ ($b=1$ in the above expressions). As we shall explain below, this a connection on the normal bundle of the horizon cross-section. Note that each term in $Y^A$ contains a factor of $K_{BC}$ so we can write $K^{AB} K_{AB}+ K_{AB}W^{AB} + \cD_A Y^A = (K_{AB} + X_{AB} )(K^{AB} + X^{AB}) + D_A Y^A + {\cal O}(\ell^4)$ for some ${\cal O}(\ell^2)$ quantity $X_{AB}$, with the ${\cal O}(\ell^4)$ ``error'' term arising from completing the square. So in this case, we have the off-shell result that $-E_{vv}$ can be written as the difference between the LHS and RHS of \eqref{gen_Ray} up to ${\cal O}(\ell^4)$ terms. On-shell,  the equation of motion gives $E_{vv} = {\cal O}(\ell^4)$ so \eqref{gen_Ray} holds as claimed.\footnote{
In this example (and the one of the next subsection) we could also include extra matter fields satisfying the null energy condition, as discussed in section \ref{sec:wall}, which would contribute negatively to the RHS of \eqref{gen_Ray} and so the second law would still hold to quadratic order in the gravitational perturbation.
} We have used only the $vv$-component of the Einstein equation. This is atypical: in general, the derivation of \eqref{gen_Ray} makes use of multiple components of the Einstein equation. 
 
In this example, the entropy is simply the IW entropy, which is manifestly gauge-invariant, in agreement with our general result. However note that $S^A$ is not gauge-invariant. This is because it is written in terms of $D_A$ rather than $\cD_A$. So the entropy current is not gauge-invariant in this example (see also \cite{Bhattacharyya:2022njk} for discussion of this point).

 \subsection{Example: scalar-tensor EFT}
 
 \label{sec:EdGB}
 
As a second example, we will consider the EFT of gravity coupled to a scalar field in $n=4$ dimensions. As above, field redefinitions can be used to simplify the Lagrangian \cite{Weinberg:2008hq}. In particular, assuming a parity symmetry, terms with up to $4$ derivatives can be written
\be
\label{4dST}
L = -V(\Phi)+ R + X + \frac{1}{2}\ell^2 \alpha(\Phi) X^2
 + \frac{1}{4} \ell^2 \beta(\Phi) L_{\rm GB} 
 \ee 
where $X \equiv - \frac{1}{2} g^{\mu\nu} \partial_\mu \Phi \partial_\nu \Phi$, and $V,\alpha,\beta$ are arbitrary functions. (The coupling functions $\alpha,\beta$ should not be confused with the GNC metric components $\alpha,\beta_A$.) The coupling functions $\alpha,\beta$ are dimensionless. This theory has second order equations of motion and admits a well-posed initial value problem \cite{Kovacs:2020pns,Kovacs:2020ywu} as long as it remains within the regime of validity of EFT.

The Einstein equation is $E_{\mu\nu}={\cal O}(\ell^4)$ where the RHS arises from the unkown EFT terms with $6$ or more derivatives and the LHS is
\bea
\label{eq:EGB}
 -E_{\mu\nu} &=&  G_{\mu\nu} -\left( \frac{1}{2} + \alpha \ell^2 X \right) \partial_\mu \Phi \partial_\nu \Phi - \frac{1}{2} g_{\mu\nu} \left( X- V + \frac{1}{2} \ell^2 \alpha X^2 \right) \nonumber \\
 &-& \frac{1}{4} \ell^2 \epsilon_\mu{}^{\alpha_1 \alpha_2 \alpha_3} \epsilon_\nu{}^{\beta_1 \beta_2\beta_3} R_{\alpha_1 \alpha_2 \beta_1 \beta_2} \nabla_{\alpha_3}  \nabla_{\beta_3} \beta 
\eea
A calculation gives the off-shell result 
\bea
\label{Evv:EdGB}
-E_{vv}
&=&-\partial_v \left[ \frac{1}{\sqrt{\mu}} \partial_v \left( \sqrt{\mu} S^v \right) + D_A S^A \right] - K^{AB} K_{AB}  -\left( \frac{1}{2} + \alpha \ell^2 X \right) (\partial_v \Phi)^2  \nonumber \\
& &-D_A Y^A -K_{AB} W^{AB}- \beta^\prime\partial_v \Phi W
\eea
where
\bea
S^v&=&1+\frac12 \ell^2 \beta(\Phi) R[\mu] \nonumber \\
S^A &=&\ell^2 \left(\beta(D_BK^{AB}-D^B K)-D_B\beta(K^{AB}-K \mu^{AB})\right).
\eea
and $Y^A,W^{AB},W$ are ${\cal O}(\ell^2)$ quantities that we shall not write out explicitly. We can complete the square on $K_{AB}$ as in our previous example, generating an ${\cal O}(\ell^4)$ error term. We can also complete the square on $\partial_v \Phi$, again generating an ${\cal O}(\ell^4)$ error term. So on-shell we obtain an equation of the form \eqref{gen_Ray} with an extra term of the form $-(1/2) (\partial_v \Phi +P)^2$ on the RHS for some ${\cal O}(\ell^2)$ quantity $P$. This term has a good sign and so the second law holds to quadratic order in perturbations, in the sense of EFT, as explained above. 

As in our previous example, only the $vv$ component of the Einstein equation is used above, and $S^v$ involves only quantities of zero boost weight, so for this theory our entropy is simply the Iyer-Wald entropy $S_{\rm IW}$, which is manifestly gauge-invariant. 
 
 \subsection{Example: Ricci squared gravity}

The above examples were atypical because the IWW entropy coincides with the IW entropy, and because equations \eqref{Evv:EGB} and \eqref{Evv:EdGB} hold off-shell. As a more typical example consider a theory of vacuum gravity with
\be
\label{ricci2}
 L=-2\Lambda+R + k_1 \ell^2 R_{\mu\nu}R^{\mu\nu}+ k_2 \ell^2 R^2
\ee
For $n=4$ dimensions this includes the most general (non-topological) terms with $N=4$. These terms could be eliminated via a field redefinition but we choose not to do so here. Since we have $N=4$ we must have $S^v=s^v$ and $S^A=s^A$ as in our previous examples. Using previous results for $s^v,s^A$ \cite{Wall:2015raa,Bhattacharya:2019qal,Bhattacharyya:2021jhr} gives
\bea
\label{ricci2_entropy}
S^v &=&1+k_1 \ell^2(2R_{rv}-K{\bar K}) + 2 k_2 \ell^2 R \nonumber \\ 
S^A &=& k_1 \ell^2 \left(2D_B K^{AB}-D^A K+\mu^{AB}\partial_v \beta_B\right)
\eea
The $k_2$ term in $S^v$ agrees with the JKM entropy \cite{Jacobson:1995uq}. In this example, the IWW entropy differs from the IW entropy because of the $K \bar{K}$ term and similar terms contained in $R_{rv}$ and $R$. On $C$, $R_{rv} \equiv R_{\mu\nu} \ell^\mu n^\nu$ and $K \bar{K}$ are both invariant under a change of GNCs. (This is explained in more detail in section \ref{sec:GNC}.) Hence the entropy is gauge invariant, in agreement with our general result. Once again, the entropy current is not gauge-invariant if $k_1 \ne 0$.

We shall not write out equation \eqref{gen_Ray} explicitly. However, we emphasise that, for this example, it is necessary to use several components of the Einstein equation to obtain \eqref{gen_Ray}. In particular the $vA$ component is used to eliminate a term on the RHS quadratic in $\partial_v \beta_A$. 

\subsection{Discussion}

\label{sec:discuss}

We shall conclude this overview with a discussion of some important open issues, namely the gauge invariance and uniqueness of the entropy. 

Section \ref{sec:entropy_current} of this paper establishes that the IWW entropy can be defined in a gauge-invariant manner. In more detail, what this means is that if we pick a cross-section $C$ of the horizon and define GNCs based on $C$ then the resulting definition of the IWW entropy of $C$ is the same for {\it any} choice of these GNCs. Clearly it is important to determine whether our improved entropy (based on $S^v)$ is also gauge-invariant in this sense. In the examples discussed above, i.e. EFTs with up to $4$ derivatives, our improved entropy is the same as the IWW entropy and therefore the entropy is gauge-invariant in these examples. But it is unclear whether this remains true if we include terms with $6$ or more derivatives. 

To understand why this is important, let $C,C'$ be two cross-sections of $\cH$ with $C'$ lying entirely to the future of $C$. We can choose GNCs based on $C$ and normalize the affine parameter along the horizon generators such that $C'$ is given by $v=v_0$ for some $v_0>0$. The entropy defined using our approach will then satisfy a second law: $S[C'] \ge S[C]$ (to quadratic order, in the sense of EFT). However, if the entropy is not gauge-invariant then the definition of $S[C]$ depends on the choice of GNCs, and hence on $C'$, which is clearly unsatisfactory.  

Our entropy satisfies the second law only to quadratic order in perturbations, in the sense of EFT (i.e. modulo higher derivative terms), so maybe a proof of gauge invariance would also only hold to quadratic order, in the sense of EFT. We leave the construction of such a proof to future work.   

Now we turn to uniqueness of the entropy current. The definition of entropy in non-equilibrium thermodynamics is known to suffer from ambiguities. An example is a relativistic viscous fluid. The fluid equations of motion can be viewed as an expansion in increasing numbers of derivatives of the fields and one can define an entropy current also in terms of an expansion in increasing numbers of derivatives of the fields. The aim is to find a definition of the entropy current that satisfies a second law on-shell. It has been shown that such an entropy current is not unique: there are multi-parameter families of entropy currents that satisfy the second law \cite{Bhattacharyya:2008xc,Romatschke:2009kr,Bhattacharyya:2013lha}. 

Something similar can be seen for black hole entropy in the perturbative context we are considering. For example, consider conventional vacuum GR. The IWW entropy is simply given by the horizon area: $s^v=1$. Now consider $\tilde{s}^v = 1 + c \ell^2 K \bar{K}$. Recall that $K=0$ on $\cH$ in the background, and that, on-shell, linear perturbations satisfy \eqref{divs_zero}, which here reduces to $\delta K=0$. Hence we have $\delta(\sqrt{\mu} \tilde{s}^v ) = \delta (\sqrt{\mu} s^v)$ on-shell. So, to linear order, $s^v$ and $\tilde{s}^v$ define the same entropy on shell. However, off-shell they differ at linear order. If we restrict to linear perturbations then the only sense in which $s^v$ is preferred over $\tilde{s}^v$ appears to be that the former satisfies the off-shell equation \eqref{s_def}. 

Going beyond linear order we must adopt the EFT perspective. For $n=4$ vacuum gravity, field redefinitions can be used to eliminate $4$-derivative terms from the equations of motion, i.e., the equation of motion is $G_{\mu\nu} + \Lambda g_{\mu\nu} = {\cal O}(\ell^4)$. Our ``improved'' entropy $S^v$ differs from $s^v$ by terms with at least $4$ derivatives, which appear at higher order in EFT than the order to which we are working. Hence we have $S^v=s^v=1$ here. Now we could use $\tilde{s}^v$ as the starting point in our algorithm for improving the entropy and we would then obtain $\tilde{S}^v = \tilde{s}^v= 1 + c \ell^2 K \bar{K}$. On-shell $S^v$ and $\tilde{S}^v$ agree at linear order but they differ at quadratic order. Since $c$ is arbitrary, we therefore have a non-unique entropy current. 

Generalising this example, for a general theory we denote the term in square brackets on the RHS of \eqref{s_def} as $\partial \cdot s$. Equation \eqref{divs_zero} says that on-shell we have $\delta(\partial \cdot s)=0$. Now consider
\be
 \tilde{s}^v = s^v + {\cal L}(\partial \cdot s) \qquad \tilde{s}^A = s^A + {\cal L}^A (\partial \cdot s)
\ee
where ${\cal L},{\cal L}^A$ are arbitrary linear operators built from $\partial_v$, $D_A$ and the metric components. On-shell we have $\delta(\sqrt{\mu} \tilde{s}^v ) = \delta (\sqrt{\mu} s^v)$ as above, so the entropies agree to linear order. But, as we saw above, $\tilde{S}^v$ and $S^v$ will differ at quadratic order. Clearly there is a lot of freedom in the choice of ${\cal L}$ and ${\cal L}^A$ so there seems to be a lot of freedom in defining an entropy that satisfies a second law in the perturbative sense we have discussed.\footnote{We should also note that we can freely adjust terms in $S^v,S^A$ that are of cubic or higher order in positive boost weight quantities whilst continuing to satisfy the second law to quadratic order. This is because such terms vanish to quadratic order.}

This non-uniqueness in the entropy can arise from field redefinitions.\footnote{
In a ``note added'' to the published version of \cite{Iyer:1994ys}, Iyer and Wald criticized their own definition of black hole entropy because it is not invariant under (non-derivative) field redefinitions in an arbitrary theory of gravity and matter.}  In general, a redefinition of the metric would change the location of $\cH$ but we can restrict to field redefinitions that are trivial on-shell to avoid having to deal with this. An example is $n=4$ vacuum gravity viewed as an EFT: as mentioned above we can eliminate $4$-derivative terms from the equations of motion to obtain the equations arising from the Lagrangian $L = R + {\cal O}(\ell^4)$ (ignoring $\Lambda$ for simplicity). Starting from this Lagrangian we could perform a field redefinition of the form $g_{\mu\nu} \rightarrow g_{\mu\nu} + a_1 \ell^2 R_{\mu\nu} + a_2 \ell^2 R g_{\mu\nu}$ to obtain a new Lagrangian $L'$ of the form \eqref{ricci2}. This field redefinition is trivial on-shell (as the original equation of motion is $R_{\mu\nu}={\cal O}(\ell^4)$). The IWW entropy resulting from $L'$ is given by \eqref{ricci2_entropy}, which differs off-shell from that of $L$, although they agree on-shell to linear order. Going beyond linear order, the expressions for $S^v$ for the two Lagrangians differ on-shell by a multiple of $K \bar{K}$, so at quadratic order the entropy is different before and after the field redefinition.\footnote{
Note that the procedure just described differs from substituting the field redefinition into the original expression for the entropy \cite{Jacobson:1993vj}, which does not change the on-shell entropy because the field redefinition is trivial on-shell.} 

\section{Preliminaries}

\label{sec:preliminaries}

\subsection{More on Gaussian Null Coordinates (GNCs)}
\label{sec:GNC}

In a spacetime $(\cM,g)$, consider a smooth co-dimension 1 null surface $\cN$ that is ruled by 
affinely parameterized null geodesics with tangent $l = l^\mu \partial_\mu$ such 
that every null geodesic intersects a spacelike cross section $C$ precisely once. Associated with this 
structure one can construct GNCs \eqref{GNC1} as described in Section \ref{sec:GNCintro}.

The tensors $\beta_A \ud x^A$ and $\mu_{AB} \ud x^A \ud x^B$ appearing in the Gaussian null form of $g$
\eqref{GNC1} have an invariant geometric meaning on the cross section $C$ (or more generally on each fixed leaf $C(v)$ of the foliation) of $\cN$. The meaning of $\mu_{AB}$ is obvious: it is the induced metric on $C$, with Levi-Civita connection $D$. To understand the role of $\beta_A$, note that the foliation defines a split
\be
T_C \cM = T_C C \oplus (T_C C)^\perp,
\ee
where the normal bundle $(T_C C)^\perp$ is spanned by the null vectors $n=\partial_r,l=\partial_v$. Now $\bbeta = \beta_A \ud x^A$ can be regarded as a connection 1-form on the 2-dimensional Lorentzian vector bundle $(T_C C)^\perp$. Said differently, on $C$ we can reduce the $SO(n-1,1)$ principal fibre bundle $F^g \cM|_C$ of orthonormal frames defined from the metric $g$ to the product of the $SO(n-2)$ principal 
fibre bundle $F^\mu C$ associated with orthonormal frames of $\mu_{AB} \ud x^A \ud x^B$ and the (trivial) $SO(1,1)$ principal fibre $P$ bundle of pairs of null directions (rather than vectors) 
in $(T_C C)^\perp$. 
The group $SO(1,1)$ may be identfied with $\RR_+$ and it acts in this parameterization on a pair of null vectors $n,l$ by the local rescaling $al, a^{-1}n$, where 
$a>0$ is a smooth function on $C$ identified with a local gauge transformation.  The representation of $\RR_+$ on a line $\RR$ given by $\pi_b: a \mapsto a^b$ for a given 
boost weight $b$ gives rise to a line bundle $P \ltimes_{\pi_b} \RR$ over $C$ via the associated vector bundle construction, 
with corresponding covariant derivative operator $\cD = D - \tfrac{1}{2} b \bbeta$. If a given tensor field on $\cM$ is decomposed into its tangential components along $C$, its components 
along $l$ and along $n$ (corresponding to the coordinate indices $A,v,r$ in GNCs), then $b$ corresponds exactly to the boost weight of such a component as described briefly in Section \ref{sec:GNCintro} and more 
precisely in Section \ref{lctGNC} below. 

We shall consider theories including matter fields, assumed to be either scalar fields or abelian $p$-form fields. In the latter case, we shall fix the gauge as follows. A $p$-form field $\bA = A_{\mu_1 \dots \mu_p} \ud x^{\mu_1} \wedge \dots \wedge \ud x^{\mu_p}$ transforms under a gauge transformation by a $(p-1)$-form $\bLambda$ as in $\bA \to \bA + \ud \bLambda$. By a suitable choice of $\bLambda$, we can always achieve that $n \cdot \bA = 0$ in some neighborhood of $\cN$ and $l \cdot \bA = 0$ on $\cN$. For a 1-form field this means that in such a gauge, 
\be
\label{eq:bw}
\bA = \phi r \ud v + A_A \ud x^A, 
\ee
so the differential $\ud r$ does not appear and the other differentials occur in a boost invariant combination.

As mentioned in section \ref{sec:GNCintro}, there is freedom in the choice of GNCs on $\cN$. Understanding this freedom will be important in our discussions of gauge-invariance below.\footnote{
Some of this discussion overlaps with the recent paper \cite{Bhattacharyya:2022njk} which appeared while the current work was in preparation.} The freedom present in the choice of GNCs is:
(1) we may start from a different cross section, $C'$; (2) we may choose a different set of coordinates $x^{\prime A}$ on $C'$; (3) we may choose a different affine parameterization. Making such a change implies that the relationship between the GNCs is 
\be
\label{affine0}
x^A = x^A(x^{\prime C}), \quad v = a(x^{\prime C})v' + b(x^{\prime C}) \quad \text{on $\mathcal N$,}
\ee
where $a$ is positive and gives the change of affine parameter and where $b$ corresponds to a change of the reference cross section $C'$
defined by $v'=0$. 

The relationship between $(x^\mu)=(v,r,x^C)$ and $(x^{\prime \mu})=(v',r',x^{\prime C})$ away from $\mathcal N$ is in general 
complicated. Let us, for later purposes, consider only a change of the affine parameter leaving $C$ as it is and keep the coordinates on $C$ as they are 
(invariance under a change of coordinates on $C$ will be manifest below). 
Thus, we take $b=0$ and $x^{\prime C} = x^C$ on the cut $C$ for simplicity, 
so that on $\cN$, the change of GNCs is 
\be
\label{affine1}
x^C = x^{\prime C}, \quad v = a(x^{\prime C})v' \quad \text{on $\mathcal N$.}
\ee
We will determine the relationship between $(x^\mu)=(v,r,x^C)$ and $(x^{\prime \mu})=(v',r',x^{\prime C})$ away from $\mathcal N$
order by order in $r'$ in the following way. First note that 
\be
l'=\frac{\partial}{\partial v'} = a \frac{\partial}{\partial v} , \quad 
\frac{\partial}{\partial x^{\prime B}} = \frac{\partial}{\partial x^B} + v' \frac{\partial a}{\partial x^{\prime B}} \frac{\partial}{\partial v}  \quad \text{on $\mathcal N$,}
\ee
and then imposing the defining relations for $n'$ on $\cN$, $g(n',n')=g(n',\partial/\partial x^{\prime C})=0, g(l', n')=1$ gives that
\be
n'=a^{-1} \frac{\partial}{\partial r} - v' \mu^{AB} \frac{\partial \log a}{\partial x^{\prime B}} \frac{\partial}{\partial x^{A}} 
- \frac{a v'^{2}}{2} \mu^{AB} \frac{\partial \log a}{\partial x^{\prime A}} \frac{\partial \log a}{\partial x^{\prime B}} \frac{\partial}{\partial v}
\quad \text{on $\mathcal N$.}
\ee 
In other words, since $n' = \partial/\partial r'$ in the primed GNCs,
\be
\label{in1}
\left( \frac{\partial r}{\partial r'} \right)_\cN = \, a^{-1}, \qquad
\left( \frac{\partial v}{\partial r'} \right)_\cN = \, - \frac{a v'^{2}}{2} \mu^{AB} \frac{\partial \log a}{\partial x^{\prime A}} \frac{\partial \log a}{\partial x^{\prime B}}, \qquad
\left( \frac{\partial x^A}{\partial r'} \right)_\cN = \, - v' \mu^{AB} \frac{\partial \log a}{\partial x^{\prime B}}
\ee
Using \eqref{affine1}, we then have, in a neighborhood of $\cN$,
\bea
\label{in2}
r&=&\frac{r'}{a(x^{\prime C})} + O({r'}^2) , \quad v = a v' - \frac{a r' v'^{2}}{2} \mu^{AB} \frac{\partial \log a}{\partial x^{\prime A}} \frac{\partial \log a}{\partial x^{\prime B}} + O({r'}^2) \nonumber \\
 \quad x^C &=& x^{\prime C}- r' v' \mu^{AB} \frac{\partial \log a}{\partial x^{\prime B}} + O({r'}^2)
\eea
Eqs. \eqref{in1}, \eqref{in2} are the initial conditions for the geodesic equation for $n'$, which since 
$n^{\prime \mu} = \partial x^\mu/\partial r'$, is
\be
\label{geo_eq}
0 = \frac{\partial^2 x^\mu}{\partial r^{\prime 2}} + \Gamma^\mu_{\sigma\rho}(x^\alpha) \frac{\partial x^\sigma}{\partial r^{\prime}} \frac{\partial x^\rho}{\partial r^{\prime}}.
\ee
Here, $\Gamma^\mu_{\sigma\rho}$ is the Christoffel symbol of the metric in the coordinates $(x^{\alpha})=(r,v,x^C)$ (see Appendix \ref{app:GNC}). By integrating the 
geodesic equations we can determine $x^\mu = x^\mu(r',v',x^{\prime C})$ order by order in $r'$. 
Note that the integration is trivial for $v=v'=0$ and yields
\be
r=a(x^{\prime C})^{-1}r', \quad v = 0, \quad x^C = x^{\prime C} \quad \text{for $v'=0$.}
\ee
In general the GNC components of the metric in $(r',v',x^{\prime C})$ coordinates are rather 
complicated functions of the original metric components. However, some expressions simplify on $\cN$ or $C$. For example using \eqref{in2} one obtains, 
on $\cN$ (i.e. for $r=r'=0$)
\be
\label{eq:homog}
\begin{split}
\mu'_{AB} =& \, \mu_{AB}^{}\\
\partial_{v'}^N K'_{AB} =& \, a^{N+1} \partial_{v}^N K_{AB}^{} 
\quad
\text{on $\cN$,}
\end{split}
\ee
subject to the identification $(v,x^A) = (a(x^{\prime C}) v', x^{\prime A})$ where $K_{AB}, \bar K_{AB}$ are the extrinsic curvatures \eqref{K:def}. 
The transformation law of $\beta_A$, 
though not of its $r$-derivatives, is also relatively simple on $\cN$,
\be
\label{beta_trans}
\beta_A' = \, \beta_A + 2D_{A} \log a - 2vK_A{}^B D_B \log a \quad \text{on $\mathcal N$,}
\ee
subject to the same identification. The formulae for the transformation of $\bar K_{AB}$ and its $r$-derivatives, or the Riemann tensor $R_{ABCD}[\mu]$ of $\mu_{AB}$, are complicated on $\cN$. However, they simplify on $C$, i.e., for $v=v'=r=r'=0$:
\be
\begin{split}
D'_{(B_1} \cdots D'_{B_r)} R_{ABCD}[\mu'] =& \, D_{(B_1} \cdots D_{B_r)} R_{ABCD}[\mu] \\
\partial_{r'}^N \bar K'_{AB} =& \, a^{-N-1} \partial_{r}^N \bar K_{AB}^{} \quad
\text{on $C$}
\end{split}
\ee
A quick way to determine the transformation law of a given tensor in GNCs of the form 
$D_{(B_1} \cdots D_{B_r)} \partial^p_v \partial_r^q \{ \mu_{AB}, \beta_A, \alpha\}$ is to transform to one of the covariant 
bases of monomials described in lemmas \ref{lem:0a}, \ref{lem:0b} below. 

We can think of the above change of GNCs as a diffeomorphism $f$ of some open neighborhood of $\cN$ by 
assigning to a point $p$ with a given set $(x^{\mu})$ of GNCs a new point $p'=f(p)$ with the same values 
$(x^{\prime \mu})$ in the unique GNC system related by $(v,x^A) = (a(x^{\prime C}) v', x^{\prime A})$ on $\cN$. 
Then we can say that $g' = f^* g$, where $g'$ is the metric \eqref{GNC1} defined by the transformed $ \mu_{AB}', \beta_A', \alpha'$. 
It is important to note that, although the restriction of $f$ to $\cN$ only refers to the function $a>0$ on $C$
but not to a particular metric, the diffeomorphism $f$ depends on both $a$ and the metric $g$ off of $\cN$, 
because its construction involves solving for the transverse geodesics relative to the metric $g$. Thus, 
we should write $f[a,g]$ to indicate properly the dependencies on $a$ and $g$. Then we have the cocycle condition
\be
\label{cocycle}
f[g,a] \circ f[g',a'] = f[g,aa'], 
\ee
where $\circ$ indicates the composition of diffeomorphisms and $g'=f[g,a]^* g$. This relationship can be proven by noting that 
both sides trivially have the same action on $\cN$ which together with the preservation of the Gaussian null form 
uniquely determines the diffeomorphisms off of $\cN$ on both sides. 

\subsection{Local covariant tensors and GNCs}
\label{lctGNC}

Consider a tensor field such as $t^{\alpha_1 \dots \alpha_r}{}_{\beta_1 \dots \beta_s}[g]$ that is constructed locally and covariantly out of the 
metric. By the Thomas replacement lemma \cite{Iyer:1994ys} it is built from contractions of $g_{\mu\nu}, g^{\mu\nu},  \nabla_{(\alpha_1} \cdots \nabla_{\alpha_k)} R_{\mu\nu\alpha\beta}$ (and possibly the volume form). 
We may evaluate such a tensor in GNCs, laboriously computing the curvature tensors and covariant 
derivatives in these coordinates. This will lead to, in general complicated, expressions involving $\mu_{AB}, \beta_A, \alpha,r$ and their derivatives, see Appendix \ref{app:GNC} for the Riemann and Ricci components, for example. Here we would like to make certain general statements 
about the resulting expressions. These general features arise from the functional property $t^{\alpha_1 \dots \alpha_r}{}_{\beta_1 \dots \beta_s}[f^*g]
= f^* t^{\alpha_1 \dots \alpha_r}{}_{\beta_1 \dots \beta_s}[g]$, taking $f$ to be a diffeomorphism preseriving the Gaussian null form, as described in the previous section. 

Recall that a general change of coordinates preserving the Gaussian null form is given on $\cN$ by \eqref{affine0}. We first specialize this equation to 
$a=1,b=0$. Then it is easy to see that, also away from $\cN$, the diffeomorphism $f$ is $v=v',r=r',x^A=f^A(x^{\prime C})$. If we now take a GNC component
of $t^{\alpha_1 \dots \alpha_r}{}_{\beta_1 \dots \beta_s}$, and view this as a tensor relative to the indices chosen as $\alpha_i=A_i, \beta_j=B_j$ 
(i.e. those not chosen as $r,v$), then the resulting tensor is on $C$ a local covariant functional of $\mu_{AB}, \beta_A, \alpha$ and $r$. 
In other words, it is a functional which is built out of contractions (with the inverse metric $\mu^{AB}$) of 
$D_{(A_1} \cdots D_{A_k)}  \partial_v^{p} \partial_r^{q} \{\mu_{AB}, \alpha, \beta_A\}$ 
or $D_{(A_1} \cdots D_{A_k)}  R_{ABCD}[\mu]$, multiplied by non-negative powers of $r$. The latter will be absent on $\cN$ (as $r=0$). It is convenient to introduce the following terminology (here $\epsilon[\mu]$ denotes the volume form induced by $\mu_{AB}$ on $C(v)$):

\begin{definition}
\label{def:primitive}
A ``primitive factor'' is one of the following: $\mu^{AB}$,  $\epsilon[\mu]_{A_1\dots A_{n-2}}$, $D_{(A_1} \cdots D_{A_k)}  \partial_v^{p} \partial_r^{q} \psi$ with $\psi \in \{\mu_{AB}, \alpha, \beta_A\}$, or $D_{(A_1} \cdots D_{A_k)}  R_{ABCD}[\mu]$. A ``primitive monomial'' is a (possibly contracted) product of primitive factors. 
\end{definition}

\noindent On $\cN$, any component of the Riemann tensor (or its derivatives) is a sum of primitive monomials. 

If we specialize the change of GNCs defined by \eqref{affine1} to the case when $x^C = x^{\prime C}$, $b=0$ and $a>0$ is {\em constant} then globally we have $v=av',r=a^{-1}r',x^C = x^{\prime C}$. Recall that we say that a quantity $X$ has {\it boost weight} $b$ if it transforms as in \eqref{bw_def} under such a change of GNCs. Recall also the simple rule described in section \ref{sec:GNCintro} for computing the boost weight of a tensor component:
\be
\label{bw}
\begin{split}
&\text{boost weight of GNC component of $t^{\alpha_1 \dots \alpha_r}{}_{\beta_1 \dots \beta_s}$} \\
=&\text{$+\#$(up $r$) $-\#$(down $r$) $-\#$(up $v$) $+\#$(down $v$),}
\end{split}
\ee
A given GNC component of $t^{\alpha_1 \dots \alpha_r}{}_{\beta_1 \dots \beta_s}$ is a sum of monomials built from $\mu^{AB}$-contractions of $D_{(A_1} \cdots D_{A_k)}  \partial_v^{p} \partial_r^{q} \{\mu_{AB}, \alpha, \beta_A\}$ 
or $D_{(A_1} \cdots D_{A_k)}  R_{ABCD}[\mu]$, multiplied by non-negative powers of $r$. The boost weight of the various terms appearing in such expressions is given as follows:
\begin{definition}
The boost weight of $r$ is $+1$, the boost weight of $v$ is $-1$; The boost weight of $D_{(A_1} \cdots D_{A_k)}  \partial_v^{p} \partial_r^{q} \psi$ with 
$\psi \in \{\mu_{AB}, \alpha, \beta_A\}$  is $p-q$, the boost weight of $\mu^{AB}$, $\epsilon[\mu]_{A_1\dots A_{n-2}}$, $D_{(A_1} \cdots D_{A_k)}  R_{ABCD}[\mu]$
is zero. The boost weight is additive under products.
\end{definition}
The boost weight of each monomial in the expression for a given tensor component is equal to the boost weight of that tensor component. For example, the boost weight of each monomial in the Riemann component $R_{rAvB}$ is zero,  as can be seen explicitly by the expressions in Appendix \ref{app:GNC}. Likewise, the boost weight of each monomial in the Ricci component $R_{rr}$ is $-2$, etc.

When matter fields with a tensorial character such as $1$-forms $\bA_J$ are present, then we should first decompose them into GNC components and pick a suitable gauge as e.g. in (\ref{eq:bw}) for a 1-form field. In such a gauge, $A_{JA}, \phi_J$ count as having boost weight zero and are treated on the same footing as the $\mu_{AB}, \beta_A, \alpha$ or scalar fields $\Phi_I$, so in this case tensor components are expressed in terms of derivatives of $\psi  \in \{\Phi_I, A_{JA}, \phi_J, \mu_{AB}, \alpha, \beta_A\}$ and non-negative powers of $r$, and similarly for higher form fields. $D_A$ should be replaced by an appropriate charged covariant derivative using $A_{BJ} \ud x^B$ on any charged
scalar fields. Quantities descending from a diffeomorphism and gauge covariant quantity remain gauge covariant under the restricted gauge transformations preserving the gauge (\ref{eq:bw}), that is $\Lambda_J$ that do not depend on $r,v$. We will usually suppress discussion of matter fields below and only comment on then where they lead to important differences. 

The notion of boost weight depends on the chosen cut $C$ and it refers to a particular choice of GNCs. If we perform a non-trivial change of GNCs, corresponding to non-constant $a$, then the definition of boost weight w.r.t. the new GNCs will not agree with the definition w.r.t. the old GNCs. However, (for fixed $C$) the definitions will agree on $\cN$. More precisely, let $\psi'(x^{\prime \mu}) \in \{\mu_{A'B'}'(x^{\prime \mu}), \beta_{A'}'(x^{\prime \mu}), \alpha'(x^{\prime \mu})\}$ be the quantities defined by the components of the metric $g$ w.r.t. a second set of GNCs $x^{\prime \mu}$. Then we have the following lemma.

\begin{lemma}
\label{lem:bw_def}
On $\cN$, the definition of boost weight is independent of the choice of GNCs. In other words, the expressions for 
$D'_{(A_1} \cdots D'_{A_{j})} \partial_{v'}^{p'} \partial_{r'}^{q'} \psi'(x^{\prime \mu})$
in terms of $D_{(A_1} \cdots D_{A_{k})} a(x^C)$, $D_{(A_1} \cdots D_{A_{j})} \partial_{v}^p \partial_{r}^q \psi(x^{\mu})$ and powers of $v'$ will only contain terms of boost weight $p'-q'$, with 
$v'$ counting as boost weight $-1$ and $a$ counting as boost weight 0.
\end{lemma}

\noindent
{\it Proof.}
On $\cN$, the quantities $\alpha,\beta_A,\mu_{AB}$ and their derivatives can all be written as $\partial_{\mu_1} \ldots \partial_{\mu_N} g_{\nu \rho}$ for some choice of indices (e.g. $\beta_A = -\partial_r g_{vA}$ on $\cN$). If we consider the corresponding quantity $\partial_{\mu'_1} \ldots \partial_{\mu'_N} g_{\nu'\rho'}$ in another set of GNCs then this will be given by some contraction of $\partial_{\mu_1} \ldots \partial_{\mu_N} g_{\sigma \tau}$ with expressions of the form $J^\mu_{\nu'_1\ldots \nu'_p} \equiv \partial^p x^\mu/\partial {x'}^{\nu_1} \ldots \partial {x'}^{\nu_p}$ for $p = 1, \ldots, N$. We claim that, on $\cN$, each component of such an expression has a definite boost weight determined by the same rule described above for tensor components, i.e., by counting the number of up and down indices of each type. (For example $J^r_{v'A'}|_\cN$ will have boost weight $1+1+0=2$.) Hence, when contracted with $\partial_{\mu_1} \ldots \partial_{\mu_N} g_{\nu \rho}$, the additivity of boost weight under products will ensure that the result will have a definite boost weight. An example of this can be seen in \eqref{beta_trans} where both the LHS and RHS have boost weight zero (since we assign $a$ boost weight zero).

To see why each component of $J^\mu_{\nu'_1\ldots \nu'_p}|_\cN$ has definite boost weight, we proceed inductively. Consider first the case where $\nu'_i \ne r'$ for all $i$. The result is then immediate from \eqref{affine1} (and $r'=0$ on $\cN$). For example $(\partial^2 v/\partial v' \partial {x'}^A)_\cN=\partial_A' a$, which has boost weight zero, in agreement with the above rules applied to the LHS. Next assume exactly one of the $\nu'_i$ is $r'$. For $p=1$ the result is immediate from \eqref{in1}, where each term on the RHS has the correct boost weight to match the above rules applied to the LHS. Taking derivatives w.r.t. $v'$ and $x^{A'}$ respects these rules and so this result extends to any $p$. If exactly two of the $\nu'_i$ are $r'$ then, for $p=2$, the result follows by evaluating \eqref{geo_eq} on $\cN$ and using the fact that the components of the Christoffel symbols have definite boost weight. The result for $p>2$ follows by taking $v'$ and $x^{A'}$ derivatives of \eqref{geo_eq}. If exactly three of the $\nu'_i$ are $r'$ then one takes another $r'$ derivative of \eqref{geo_eq} and evaluates on $\cN$ and so on. Q.E.D.

\medskip

Since we will be dealing with equations involving higher derivatives, we will sometimes need to keep track of the number of derivatives associated with a given quantity. We are mostly interested in doing this on $\cN$. In the above proof we saw that, on $\cN$, the quantities of interest can all be expressed in terms of partial derivatives of the metric tensor: $\alpha$ involves $2$ $r$-derivatives and $\beta_A$ involves $1$ $r$-derivative. We make the following definition to count derivatives:

\begin{definition}
\label{def:dim}
The ``dimension'' of $\alpha$, $\beta_A$ and $\mu_{AB}$ are $2,1,0$ respectively. Taking a derivative w.r.t. $v$, $r$ or $x^A$ increases the dimension by $1$. Dimension is additive under products. 
\end{definition}
For example $D_{(A_1} \cdots D_{A_k)}  \partial_v^{p} \partial_r^{q}\alpha$ has dimension $k+p+q+2$. A quantity with boost weight $b$ involves at least $|b|$ derivatives and so its dimension is bounded below by $|b|$. The dimension of a component of the $k$th covariant derivative of the Riemann tensor is $k+2$. 

The primitive factors of definition \ref{def:primitive} are employed in Wall's method. However, in various places we will find it more convenient to express quantities on $\cN$ in terms of a different set of quantities. These are described in lemmas \ref{lem:0a} and \ref{lem:0b}.

\begin{lemma}
\label{lem:0a}
On $\cN$, any of the expressions $D_{(A_1} \cdots D_{A_k)}  \partial_v^{p} \partial_r^{q} \psi$ with 
$\psi \in \{\mu_{AB}, \alpha, \beta_A\}$ can be expressed uniquely as a sum of monomials where each monomial is a product of factors of the following form (and possible factors of $\mu^{AB}$)
\begin{itemize}
\item $D_{(A_1} \cdots D_{A_j)} \partial_{v}^N K_{AB}^{}$, $D_{(A_1} \cdots D_{A_k)} \partial_{r}^{\bar N} \bar K_{AB}^{}$,
\item $D_{(A_1} \cdots D_{A_j)} \beta_B$, 
\item $D_{(A_1} \cdots D_{A_j)}  R[\mu]_{BCDE}$. Note that, by the Bianchi identities, these components are not all independent. When $j \ge 1$, an 
independent set is obtained e.g. by choosing $D_{(A_1} \cdots D_{A_j}  R[\mu]^B{}_{A_{j+1} A_{j+2})C}$. \footnote{In fact, \cite{Muller} shows that in Riemann normal coordinates any number of partial derivatives of the metric can be expressed using this set of quantities.}
\item A GNC component of $\nabla_{(\alpha_1} \cdots \nabla_{\alpha_j)} R_{\mu\nu}$, except for:
$(\mu\nu) = (rr)$ and $\alpha_j \in \{r,A\}$ or $(\mu\nu) = (vv)$ and $\alpha_j \in \{v,A\}$, and 
except for GNC components that are linearly dependent on a suitably chosen minimal set via 
the Bianchi identities.
\end{itemize}
\end{lemma}

\noindent
{\em Proof:} First consider the case of $\psi=\alpha$. To express $D_{(A_1} \cdots D_{A_k)}  \partial_v^{p} \partial_r^{q} \alpha$ in the desired form, we start by writing the Ricci component $R_{vr}$ in terms of GNC components (see Appendix \ref{app:GNC})
\be\label{eq:Rvr}
R_{vr}= -\frac{1}{2} \partial_r^2 \left( r^2 \alpha\right) - \frac{1}{2} \bar{K} \partial_r \left( r^2 \alpha \right) + \ldots
\ee
where the ellipsis denotes terms that do not depend on $\alpha$. Evaluating this equation at $r=0$ uniquely determines $\alpha|_{r=0}$ in terms of $\beta_A$, $D_A\beta_B$ and quantities of the form $\partial_v^{p'}\partial_r^{q'}\mu_{AB}$ with $p',q'\leq 1$. Next, we act on the above equation with $\partial_r$ and use\footnote{To clarify, in this equation $\nabla_r R_{rv}$ means the $rrv$ component of the tensor obtained by taking the covariant derivative of the Ricci tensor. Of course, the component $\nabla_r R_{rv}$ depends on multiple Ricci components $R_{\mu\nu}$, not just on $R_{rv}$. We continue to use similar conventions hereafter. Note, however, that covariant derivatives mix up components only at the subprincipal level.}
\be
\partial_r R_{rv}=\nabla_r R_{rv}+\Gamma^A_{rv}R_{r A}+\ldots
\ee
where the ellipsis denotes terms that vanish at $r=0$. By virtue of the identities of Appendix \ref{app:GNC}, this can be used to determine $\partial_r \alpha|_{r=0}$ in terms of quantities of the following form: $D_A\partial_r^{q'}\beta_B$ with $q'\leq 2$; $\partial_v^{p''}\partial_r^{q''}\mu_{AB}$ with $p''\leq 1$, $q''\leq 2$; and the GNC component $\nabla_r R_{rv}$. Proceeding inductively, we assume that we can express $\partial_r^{q-1} \alpha|_{r=0}$ in terms of the following quantities:
\begin{itemize}
    \item $D_A \partial_r^{q'}\beta_B$ with $q'\leq q-1$, and
    \item $\partial_v^{p''} \partial_r^{q''}\mu_{AB}$ with $p''\leq 1$ $q''\leq q$, and
    \item a GNC component $\nabla_{\alpha_1} \cdots \nabla_{\alpha_j} R_{\mu\nu}$ with $j\leq q-1$.
\end{itemize} 
We have already shown that this is true for $q=1$ and $q=2$. Now we act on \eqref{eq:Rvr} with $\partial_r^q$ and evaluate at $r=0$. Then we determine $\partial_r^q \alpha|_{r=0}$ in terms of primitive factors of the form $D_A \partial_r^{q'}\beta_B$ with $q'\leq q$, $\partial_v^{p''} \partial_r^{q''}\mu_{AB}$ with $p''\leq 1$, $q''\leq q+1$, and $\partial_r^q R_{rv}$. One can then rewrite $\partial_r^q R_{rv}$ in terms of GNC components $\nabla_{\alpha_1} \cdots \nabla_{\alpha_j} R_{\mu\nu}$ with $j\leq q$ and terms of the form $\partial_r^{\bar q}\Gamma^\mu_{r\nu}$ with ${\bar q}<q$. It follows from the formulae of Appendix \ref{app:GNC} that the terms involving Christoffel symbols can be written in terms of primitive factors of the form $\partial_r^{q'} \beta_A$ with $q'\leq q$, $ \partial_r^{q''} \mu_{AB}$ with $q''\leq q+1$ and $\partial_r^{q'''}\alpha$ with $q'''\leq q-1$. However, the terms of the form $\partial_r^{q'''}\alpha$ with $q'''\leq q-1$ can be eliminated in favour of derivatives of $\beta_A$, $\mu_{AB}$ and covariant derivatives of $R_{\mu\nu}$, as specified above. This shows that if the induction hypothesis holds then $\partial_r^{q} \alpha|_{r=0}$ can be expressed with the quantities
\begin{itemize}
    \item $D_A \partial_r^{q'}\beta_B$ with $q'\leq q$, and
    \item $\partial_v^{p''} \partial_r^{q''}\mu_{AB}$ with $p''\leq 1$ $q''\leq q+1$, and
    \item a GNC component $\nabla_{\alpha_1} \cdots \nabla_{\alpha_j} R_{\mu\nu}$ with $j\leq q$.
\end{itemize}
Since this statement is true for $q=1$, it follows that it is true for general $q$.

By taking $v$ or $x^A$ derivatives of this result, we determine $D_{A_1} D_{A_2} \ldots D_{A_m} \partial_v^p \partial_r^q \alpha|_{r=0}$ as a sum of products of factors depending only on
\begin{itemize}
    \item $D_{A_1} D_{A_2} \ldots D_{A_{m'}} \partial_v^{p'} \partial_r^{q'}\beta_A$ with $q'\leq q$, and
    \item $D_{A_1} D_{A_2} \ldots D_{A_{m''}} \partial_v^{p''} \partial_r^{q''}\mu_{AB}$ with $q''\leq q+1$, and
    \item a GNC component $\nabla_{\alpha_1} \cdots \nabla_{\alpha_j} R_{\mu\nu}$ with $j\leq m+p+q$.
\end{itemize} 
Each monomial in this sum must have boost weight $p-q$. In the rest of this argument it is understood that whenever $D_{A_1} D_{A_2} \ldots D_{A_m} \partial_v^p \partial_r^q \alpha|_{r=0}$ appears, we will rewrite it in terms of the above quantities. 

The next step is to show that we can use similar methods to eliminate certain derivatives of $\beta_A$. First we consider the following expression for $R_{rA}$ (see Appendix \ref{app:GNC})
\be
\label{rA_eq}
R_{rA}= -\frac{1}{2}\partial_r^2 \left( r \beta_A \right) - \frac{1}{2} \bar{K} \partial_r \left( r \beta_A \right)+ \bar{K}_A{}^B \partial_r \left( r \beta_B \right) + r \beta^B \left( \partial_r \bar{K}_{AB} + \ldots \right) + \ldots 
\ee
where the first ellipsis on the RHS denotes terms quadratic in $\bar{K}_{AB}$ and the second ellipsis denotes terms linear in $D_A \bar{K}_{BC}$ (contracted in some way with $\mu^{AB}$ in both cases). 
Evaluating this equation at $r=0$ determines $\partial_r \beta_A|_{r=0}$ in terms of $\beta_A$, $\mu_{AB}$, $\partial_r \mu_{AB}$, $D_C\partial_r \mu_{AB}$ and $R_{rA}$. Acting with $\partial_r^{q-1}$ for $q \ge 1$ and evaluating at $r=0$ lets us determine, inductively (similarly to the $\psi=\alpha$ case above), $\partial_r^q \beta_A|_{r=0}$ in terms of $\beta_A$, $D_C\partial_r^{q'}\mu_{AB}$ with $q'\leq q$ and GNC components $\nabla_{\alpha_1}\ldots \nabla_{\alpha_j} R_{\mu\nu}$ with $j\leq q$. 


Next we consider $R_{vA}$ evaluated at $r=0$:
\be
R_{vA}|_{r=0}= \frac{1}{2} \partial_v  \beta_A  + \frac{1}{2} K  \beta_A -D_A K + D_B K_A{}^B 
\ee
This determines $\partial_v \beta_A|_{r=0}$ in terms of $\beta_A$, $\mu_{AB}$, $\partial_v\mu_{AB}$, $D_C\partial_v \mu_{AB}$ and $R_{vA}$. Acting with $\partial_v^{p-1}$ for $p \ge 1$ lets us determine, inductively, $\partial_v^p \beta_A|_{r=0}$ in terms of $\beta_A$, terms of the form $D_C \partial_v^{p'}\mu_{AB}$ with $p'\leq p$ and $\partial_v^{p-1}R_{vA}$. The last one of these can then be written in terms of GNC components $\nabla_{\alpha_1}\ldots \nabla_{\alpha_j} R_{\mu\nu}$ with $j\leq p-1$ and $\partial_v^{\bar p}\Gamma^\mu_{v \nu} \bigr|_{r=0}$ with $\bar{p}<p$. Employing the expressions of Appendix \ref{app:GNC} relating Christoffel symbols to primitive factors then gives a formula for $\partial_v^p \beta_A|_{r=0}$ in terms of $\beta_A$, terms of the form $D_C \partial_v^{p'}\mu_{AB}$ with $p'\leq p$ and GNC components $\nabla_{\alpha_1}\ldots \nabla_{\alpha_j} R_{\mu\nu}$ with $j\leq p-1$.

Now we return to \eqref{rA_eq}: taking a $v$-derivative of this equation and using our result for $\partial_v \beta_A$, we can write $\partial_v \partial_r \beta_A|_{r=0}$ in terms of $R_{vA}$, $\nabla_v R_{rA}$, $\beta_A$, $\mu_{AB}$, and terms of the form $\partial_v^{p'}\partial_r^{q'}\mu_{AB}$ and $D_C\partial_v^{p'}\partial_r^{q'}\mu_{AB}$ with $p'\leq 1$ and $q'\leq 1$. Similarly, taking multiple $v$-derivatives of \eqref{rA_eq} and employing an induction on the number of $v$-derivatives gives us an expression for $\partial_v^p \partial_r \beta_A|_{r=0}$ containing the following quantities: $\beta_A$, primitive factors with $\psi=\mu_{AB}$ and components of the form $\nabla_{\alpha_1}\ldots \nabla_{\alpha_j} R_{\mu\nu}$ with $j\leq p$. Furthermore, taking $v$-derivatives of our previous expressions for $\partial_r^q \beta |_{r=0}$ and proceeding inductively as before we obtain an expression for $\partial_v^p \partial_r^q \beta_A|_{r=0}$ ($p,q \ge 1$) in terms of $\beta_A$, $\mu_{AB}$, primitive factors of the form $D_{A_1}\ldots D_{A_m} \partial_v^{p'}\partial_r^{q'}\mu_{AB}$ with $p'\leq p$, $q'\leq q$, and components $\nabla_{\alpha_1}\ldots \nabla_{\alpha_j} R_{\mu\nu}$. Taking derivatives w.r.t. $x^A$ of this result lets us write $D_{A_1} \ldots D_{A_m} \partial_v^p \partial_r^q \beta_B|_{r=0}$ in terms of
\begin{itemize}
    \item $D_{A_1} \ldots D_{A_n} \beta_B$ with $0 \le n \le m$,
    \item $D_{A_1} \ldots D_{A_{n'}} \partial_v^{p'} \partial_r^{q'}\mu_{AB}$ with $p'\leq p$, $q'\leq q$
    \item and covariant components $\nabla_{\alpha_1}\ldots \nabla_{\alpha_j} R_{\mu\nu}$ with $j\leq m+p+q$.
\end{itemize}
From now on, it is assumed that at every occurrence of $D_{A_1} \ldots D_{A_m} \partial_v^p \partial_r^q \beta_B|_{r=0}$ it is expressed with the above quantities.

Consider finally a factor of the form $D_{A_1} \ldots D_{A_k} \partial_v^{p} \partial_r^{q} \mu_{CD}$. We wish to argue that this factor can be written as a sum of monomials whose factors are of the form specified in the statement of the lemma. To this end, we consider the expression for the Ricci component $R_{AB}$ in terms of GNCs (given in more detail in Appendix \ref{app:GNC})
\bea\label{RAB_GNC}
R_{AB}&=&- 2 \partial_v \bar{K}_{AB}- K \bar{K}_{AB} + 4 K_{(A}{}^{C} \bar{K}_{B)C} -  K_{AB} \bar{K}  \nonumber \\
& &+R[\mu]_{AB}-D_{(B}\beta_{A)}- \frac12 \beta_{A}\beta_{B}+\ldots
\eea
where the ellipsis now stands for terms that vanish at $r=0$. This equation allows us to express $\partial_v {\bar K}_{AB}|_{r=0}$ in terms of $R_{AB}$, $R[\mu]_{AB}$, $K_{AB}$, ${\bar K}_{AB}$, $\beta_A$ and $D_A \beta_B$. Next, we act with $\partial_r$ on equation \eqref{RAB_GNC} and use
\be\label{dr_RAB}
\partial_r R_{AB}=\nabla_r R_{AB}+2\Gamma_{r(A}^\mu R_{B)\mu}=-2\partial_v \partial_r {\bar K}_{AB}-\beta_{(A}\partial_r \beta_{B)}-\partial_r D_{(A}\beta_{B)}-2\alpha {\bar K}_{AB}+\ldots
\ee
Here the ellipsis stands for terms that are of the required form and terms involving $\partial_v {\bar K}_{AB}$ that can be written in the required form as described in the previous paragraph. As for the term involving $\alpha$, we have shown it previously that it can be expressed (at $r=0$) in terms of $R_{rv}$, $\mu_{AB}$, $K_{AB}$, ${\bar K}_{AB}$, $\beta_A$, $D_A \beta_B$ and $\partial_v {\bar K}_{AB}$. Combining this with our result on $\partial_v {\bar K}_{AB}$ yields a formula for $\alpha$ in terms of $R_{AB}$, $R_{rv}$, $\mu_{AB}$, $R[\mu]_{AB}$, $K_{AB}$, ${\bar K}_{AB}$, $\beta_A$ and $D_A \beta_B$. Regarding the terms involving $\partial_r \beta_A$ and $\partial_r D_A\beta_B$ in \eqref{dr_RAB}, we have a prescription to write these (at $r=0$) in terms of $\beta_A$, $D_A\beta_B$, $\mu_{AB}$, ${\bar K}_{AB}$, $D_{A_1}{\bar K}_{AB}$, $D_{A_1}D_{A_2}{\bar K}_{AB}$ and covariant components involving the Ricci tensor. Therefore, equation \eqref{dr_RAB} provides an expression for $\partial_v \partial_r{\bar K}_{AB}|_{r=0}$ as a sum of monomials that are products of factors of the following quantities: $\mu_{AB}$, $\beta_A$, $D_A \beta_B$, $K_{AB}$, ${\bar K}_{AB}$, $D_{A_1}{\bar K}_{AB}$, $D_{A_1}D_{A_2}{\bar K}_{AB}$, $\partial_r {\bar K}_{AB}$, $R[\mu]_{AB}$ and covariant quantities of the form $R_{\mu\nu}$, $\nabla_\alpha R_{\mu\nu}$. 

Proceeding inductively, one can fix $\partial_v \partial_r^q \bar{K}_{AB}|_{r=0}$ (by taking multiple derivatives of \eqref{RAB_GNC} w.r.t. $r$ and using previous results) in terms of the quantities
\begin{itemize}
    \item $\mu_{AB}$, $R[\mu]_{AB}$,
    \item $K_{AB}$, $\partial_r^{q'} \bar{K}_{AB}$, $D_{A_1}\partial_r^{q'} \bar{K}_{AB}$, $D_{A_1}D_{A_2}\partial_r^{q'} \bar{K}_{AB}$ with $q'\leq q$
    \item factors of $\beta_A$ and $D_A\beta_B$
    \item covariant components of the form $\nabla_{\alpha_1}\ldots \nabla_{\alpha_j}R_{\mu\nu}$ with $j\leq q$
\end{itemize}

Similarly, an inductive argument establishes that taking multiple $v$-derivatives of our expressions for $\partial_v \partial_r^q {\bar K}_{AB}|_{r=0}$ allows us to write $\partial_v^p \partial_r^q \mu_{AB}|_{r=0}$ using only the factors specified below:
\begin{itemize}
    \item $\mu_{AB}$, $R[\mu]_{AB}$,
    \item $\partial_v^{p'}K_{AB}$, $D_{A_1}\partial_v^{p'} {K}_{AB}$, $D_{A_1}D_{A_2}\partial_r^{p'} {K}_{AB}$ with $p'\leq p-1$
    \item $\partial_r^{q'} \bar{K}_{AB}$, $D_{A_1}\partial_r^{q'} \bar{K}_{AB}$, $D_{A_1}D_{A_2}\partial_r^{q'} \bar{K}_{AB}$ with $q'\leq q-1$
    \item factors of $\beta_A$ and $D_A\beta_B$
    \item covariant components of the form $\nabla_{\alpha_1}\ldots \nabla_{\alpha_j}R_{\mu\nu}$ with $j\leq p+q-2$
\end{itemize}
Taking $D_A$ derivatives of this result then lets us write $D_{(A_1} \dots D_{A_k)} \partial_v^p \partial^q_r \mu_{AB}$ in terms of the factors
\begin{itemize}
    \item $\mu_{AB}$, $D_{A_1}\ldots D_{A_j}R[\mu]_{AB}$ with $0\leq j\leq k$,
    \item $D_{A_1}\ldots D_{A_j}\partial_v^{p'}K_{AB}$ with $0\leq j \leq k+2$ and $p'\leq p-1$ 
    \item $D_{A_1}\ldots D_{A_j}\partial_v^{q'}\bar{K}_{AB}$ with $0\leq j \leq k+2$ and $p'\leq p-1$ 
    \item $D_{A_1}\ldots D_{A_j}\beta_A$ with $0\leq j \leq k+1$ 
    \item covariant components of the form $\nabla_{\alpha_1}\ldots \nabla_{\alpha_j}R_{\mu\nu}$ with $j\leq p+q+m-2$
\end{itemize}

Combining this with our previous results on primitive factors of $\alpha$ and $\beta$, we can determine $D_{(A_1} \dots D_{A_k)} \partial_v^p \partial^q_r \psi|_{r=0}$ with $\psi\in \{\mu_{AB}, \alpha, \beta_A \}$ as a sum of monomials with each monomial being a product of factors
\begin{itemize}
    \item[(i)] $\mu_{AB}$, $D_{A_1}\ldots D_{A_j}R[\mu]_{AB}$
    \item[(ii)] $D_{A_1}\ldots D_{A_j}\partial_v^{p'}K_{AB}$ and $D_{A_1}\ldots D_{A_j}\partial_v^{q'}\bar{K}_{AB}$
    \item[(iii)] $D_{A_1}\ldots D_{A_j}\beta_A$ 
    \item[(iv)] covariant components of the form $\nabla_{\alpha_1}\ldots \nabla_{\alpha_j}R_{\mu\nu}$
\end{itemize}

To obtain the same dependencies as in the statement of the lemma, as well as uniqueness, we first note that (on $\cN$) the covariant quantities $\nabla_{\alpha_1}\ldots \nabla_{\alpha_j} R_{vv}$ with $\alpha_j \in \{v,A\}$ and $\nabla_{\alpha_1}\ldots \nabla_{\alpha_j} R_{rr}$ with $\alpha_j \in \{r,A\}$ can be determined by all the other factors (i)-(iv) listed above. To see this, we consider the identity (see Appendix \ref{app:GNC})
\be
R_{rr}=-\partial_r {\bar K}-\bar{K}_{AB}\bar{K}^{AB}.
\ee
Hence, $R_{rr}$ is clearly expressible in terms of $\bar{K}_{AB}$ and $\partial_r\bar{K}_{AB}$. Taking derivatives of this result w.r.t. $r$ and $x^A$, one can inductively express $\nabla_{\alpha_1}\ldots \nabla_{\alpha_j} R_{rr}$ with $\alpha_j \in \{r,A\}$ in terms of factors of the form $D_{A_1}\ldots D_{A_j}\partial_v^{p'}K_{AB}$ and other factors listed in (i)-(iv). A very similar argument establishes that $\nabla_{\alpha_1}\ldots \nabla_{\alpha_j} R_{vv}$ with $\alpha_j \in \{v,A\}$ is also redundant in the list of factors above. 

Furthermore, we may express all the quantities listed in (i)-(iv) with totally symmetrized derivatives such as $D_{(A_1} \dots D_{A_j)}$. The reason for this is that an anti-symmetrization of a quantity $D_{A_1} \dots D_{A_j}\Psi$ over a subset of the indices $A_1, \ldots A_j$ is expressible via the Riemann tensor $R[\mu]_{ABCD}$ and fewer than $j$ derivatives of the quantity $\Psi$ (using the Bianchi identities). Similarly, the quantities $\nabla_{\alpha_1}\ldots \nabla_{\alpha_j}R_{\mu\nu}$ can be expressed with the totally symmetrized derivatives $\nabla_{(\alpha_1}\ldots \nabla_{\alpha_m)}R_{\mu\nu}$ and (covariant derivatives of) the Riemann tensor associated with $\nabla$. To eliminate the dependency on $R_{\mu\nu\rho\sigma}$, one can use the formulae of Appendix \ref{app:GNC} and the procedure described above to write the components of the Riemann tensor (at $r=0$) in terms of $\mu_{AB}$, $\beta_A$, $D_A\beta_B$, $K_{AB}$, $\bar{K}_{AB}$, $\partial_v K_{AB}$, $\partial_r \bar{K}_{AB}$, $D_C K_{AB}$, $D_C \bar{K}_{AB}$, $R[\mu]_{AB}$ and Ricci components $R_{\mu\nu}$. Then one can argue that any derivative of the Riemann tensor can be expressed in terms of the quantities listed in the statement of the lemma by using induction on the number of derivatives and by making use of identities of the form
\be
\nabla_{\alpha_1}\nabla_{(\alpha_2}\ldots \nabla_{\alpha_j)}R_{\mu\nu}=\nabla_{(\alpha_1}\nabla_{\alpha_2}\ldots \nabla_{\alpha_j)}R_{\mu\nu}+\ldots
\ee
where the ellipsis on the RHS stands for terms involving fewer than $j$ derivatives of the curvature tensor.

Finally, the Bianchi identities give us a freedom when expressing a primitive factor $D_{(A_1} \dots D_{A_j)} \partial_v^p \partial^q_r \psi$ in terms of the factors in the statement of the lemma. It can be seen that this is all the remaining freedom that one has using induction, because the leading derivative terms of the Bianchi identities cancel and these are the only possible linear dependencies among the leading derivative terms used in our induction argument. Q.E.D.

\begin{lemma}
\label{lem:0b}
Any of the expressions $D_{(A_1} \cdots D_{A_k)}  \partial_v^{p} \partial_r^{q} \psi$ with 
$\psi \in \{\mu_{AB}, \alpha, \beta_A\}$ can be expressed (on $\mathcal N$) as a sum of monomials where each monomial is a product of factors of the following form (and possible factors of $\mu^{AB}$)
\begin{itemize}
\item $D_{(A_1} \cdots D_{A_j} K_{A)B}^{}$, $D_{(A_1} \cdots D_{A_j} \bar K_{A)B}^{}$,
\item $D_{(A_1} \cdots D_{A_j} \beta_{B)}$,
\item A GNC component of $\nabla_{(\alpha_1} \cdots \nabla_{\alpha_j)} R_{\mu\nu\sigma\rho}$. By the Bianchi-identities, 
not all these components are independent. An independent 
set for $j \ge 1$ can be obtained by choosing e.g. $\nabla_{(\alpha_1} \cdots \nabla_{\alpha_j} R^\mu{}_{\alpha_{j+1} \alpha_{j+2})\nu}$.
\end{itemize}
\end{lemma}
\noindent
{\em Proof:} By lemma \ref{lem:0a}, we only need to eliminate $D_{(A_1} \cdots D_{A_j)} \partial_{v}^N K_{AB}^{}$, $D_{(A_1} \cdots D_{A_k)} \partial_{r}^{\bar N} \bar K_{AB}^{}$
in favor of $D_{(A_1} \cdots D_{A_j} K_{A)B}^{}$, $D_{(A_1} \cdots D_{A_j} \bar K_{A)B}^{}$, 
we need to eliminate $D_{(A_1} \cdots D_{A_j)} \beta_B$ in favor of $D_{(A_1} \cdots D_{A_j} \beta_{B)}$, and we need 
to eliminate $D_{(A_1} \cdots D_{A_j)}  R_{ABCD}[\mu]$, modulo GNC components of $\nabla_{(\alpha_1} \cdots \nabla_{\alpha_j)} R_{\mu\nu\sigma\rho}$.

First we consider the evolution equation for the expansion and shear in the $r$-direction, which is 
\be
\label{Ray1}
R_{rArB} = \bar K_A{}^C \bar K_{BC}-\partial_r \bar K_{AB}, 
\ee
which lets us eliminate $\partial_r \bar K_{AB}$ in terms of $\bar K_{AB}$, plus a GNC component of the Riemann tensor. Next, we consider, on $\mathcal N$, the Riemman 
component
\be
\label{twist}
R_{rvAB} = 2 K_{[A}{}^C \bar K_{B]C} + D_{[A} \beta_{B]},
\ee
which lets us write $D_A \beta_B$ in terms of $D_{(A} \beta_{B)}$ plus terms containing $K_{AB}$,$\bar K_{AB}$ or a GNC component of the Riemann tensor. The 
evolution equation for the expansion and shear in the $v$-direction is on $\mathcal N$ given by 
\be
\label{Ray2}
R_{vAvB} = K_A{}^C K_{BC}-\partial_v K_{AB}, 
\ee
which lets us eliminate $\partial_vK_{AB}$ in terns of $K_{AB}$, plus a GNC component of the Riemann tensor. Then we consider the Gauss-Codacci equation on $\mathcal N$, 
\be
\label{GaCo}
R_{ABCD} =R[\mu]_{ABCD}+2 K_{B[C} \bar{K}_{D]A} + 2K_{A[D} \bar{K}_{C]B}
\ee
which lets us eliminate $R[\mu]_{ABCD}$  in favor of terms containing $K_{AB}$,$\bar K_{AB}$ or a GNC component of the Riemann tensor.
Finally, we have 
\be\label{DK_asym}
\begin{split}
R_{ABrC}&=  -\left(D_{A}+\frac12 \beta_A\right)\bar{K}_{BC} + \left(D_{B}+\frac12 \beta_B\right)\bar{K}_{AC}\\
R_{ABvC}&=  -\left(D_{A}-\frac12\beta_A\right){K}_{BC} + \left(D_{B}-\frac12 \beta_B\right){K}_{AC}.
\end{split}
\ee
This lets us eliminate $D_{[A}{K}_{B]C}, D_{[A}\bar {K}_{B]C}$ in favor of terms containing $K_{AB}$,$\bar K_{AB}, \beta_A$ 
or a GNC component of the Riemann tensor

We now continue this process by taking suitable derivatives of the above equations, using the non-zero Christoffel symbols (on $\mathcal N$) given in Appendix \ref{app:GNC}.

First we show inductively that $\partial_r^{\bar N} \bar K_{AB}$ can be expressed with only $\bar{K}_{AB}$ and covariant components of the form $\nabla_{(\alpha_1} \cdots \nabla_{\alpha_j)} R_{\mu\nu\sigma\rho}$ with $j\leq \bar{N}-1$. We have already shown above that this is true for $\bar N=1$. Now assume that $\partial_r^{\bar q} \bar K_{AB}$ with $q\leq \bar{N}-1$ can be expressed using $\bar{K}_{AB}$ and GNC components $\nabla_{(\alpha_1} \cdots \nabla_{\alpha_j)} R_{\mu\nu\sigma\rho}$ with $j\leq \bar{q}-1$. Now act ${\bar N}$ times with $\partial_r$ on \eqref{Ray1}, noting that \eqref{Ray1} holds off $\mathcal N$. We can convert any partial derivatives of Riemann tensor components to components of covariant derivatives of the Riemann tensor. This lets us eliminate $\partial_r^{\bar N} \bar K_{AB}$ in terms of $\partial_r^q \bar K_{AB}$ with $0\leq q\leq \bar{N}-1$, $\partial_r^{q'} \Gamma^A_{rB}$ with $0\leq q' \leq \bar{N}-1$, plus GNC components of the Riemann tensor. Note that other Christoffel symbols cannot appear in this expression since $\Gamma^\mu_{rr}$ and $\Gamma^v_{r\mu}$ vanish (on and off $\mathcal N$). Furthermore, $\Gamma^A_{rB}=\bar{K}^A{}_B$ on and off $\mathcal N$. By the induction hypothesis we can eliminate the terms $\partial_r^q \bar K_{AB}$ in favor of $\bar K_{AB}$ and covariant derivatives of the Riemann tensor, closing the induction loop.

A very similar inductive argument lets us write $\partial_v^N K_{AB}$ in terms of $K_{AB}$ and GNC components $\nabla_{(\alpha_1} \cdots \nabla_{\alpha_j)} R_{\mu\nu\sigma\rho}$ with $j\leq {N}-1$. This argument is based on acting with $\partial_v$ multiple times on \eqref{Ray2} and noting that $\Gamma^r_{v\mu}=\Gamma^\mu_{vv}=0$ and $\Gamma^A_{vB}=K^A{}_B$ on $\mathcal N$.

Next, we express the factors $D_{(A_1} \cdots D_{A_j)}  K_{AB}^{}$, $D_{(A_1} \cdots D_{A_j)} \bar K_{AB}^{}$, $D_{(A_1} \cdots D_{A_j)} \beta_B$ and $D_{(A_1} \cdots D_{A_{j-1})}  R_{ABCD}[\mu]$ in terms of the quantities listed in the statement of the lemma by using induction on $j$. These expressions have been obtained for $j=1$ above. Now suppose we have the corresponding expressions for $D_{(A_1} \cdots D_{A_{j-1})}  K_{AB}^{}$, $D_{(A_1} \cdots D_{A_{j-1})} \bar K_{AB}^{}$, $D_{(A_1} \cdots D_{A_{j-1})} \beta_B$ and $D_{(A_1} \cdots D_{A_{j-2})}  R_{ABCD}[\mu]$. We next use an identity of the form
\bea\label{K_symd}
D_{(A_1}\ldots D_{A_j)} K_{BC} &=&  -\frac{2j}{j+1}\left(D_{(A_1}\ldots D_{A_{j-1})} D_{[A_j} K_{B]C} +\ldots + D_{(A_2}\ldots D_{A_{j})} D_{[A_1} K_{B]C}\right) \nonumber \\
& & +D_{(A_1}\ldots D_{A_j} K_{B)C} +\ldots
\eea
where the ellipsis in the second line stands for a sum of monomials containing factors of the form $D_{(A_1}\ldots D_{A_m)}R[\mu]_{ABCD}$ with $m\leq j-2$ and a factor $D_{(A_1}\ldots D_{A_k)} K_{BC}$ or a factor $D_{(A_1}\ldots D_{A_{k-1})}D_{[A_k} K_{B]C}$ with $k\leq j-2$. Then we consider  $\nabla_{A_1} \cdots \nabla_{A_{j-1}}R_{ABvC}$ and write it in terms of primitive factors. This lets us eliminate $D_{(A_1} \cdots D_{A_{j-1})}D_{[A}K_{B]C}$ in favor of $D_{(A_1} \cdots D_{A_k)}K_{BC}$, $D_{(A_1} \cdots D_{A_l)}\beta_B$, $D_{(A_1} \cdots D_{A_{m-1})}R[\mu]_{ABCD}$ with $k,l,m\leq j-1$ and covariant derivatives of the Riemann tensor. Making use of the induction hypothesis yields an expression for $D_{(A_1} \cdots D_{A_j)}K_{BC}$ in terms of the desired quantities. A similar argument establishes the corresponding result for $D_{(A_1}\ldots D_{A_j)} \bar{K}_{BC}$. 

Writing $\nabla_{A_1} \cdots \nabla_{A_j-1}R_{ABCD}$ in terms of primitive factors yields an expression relating $D_{(A_1} \cdots D_{A_{j-1})}  R_{ABCD}[\mu]$ to $D_{(A_1} \cdots D_{A_k)}\bar{K}_{BC}$, $D_{(A_1} \cdots D_{A_l)}\bar{K}_{BC}$, $D_{(A_1} \cdots D_{A_{m-1})}  R_{ABCD}[\mu]$ with $k,l,m\leq j-1$ and covariant components. By employing the induction hypothesis, we get an expression for $D_{(A_1} \cdots D_{A_{j-1})}  R_{ABCD}[\mu]$ in terms of the quantities listed in the statement of the lemma. 

Regarding $D_{(A_1} \cdots D_{A_j)} \beta_B$, we first use an identity similar to \eqref{K_symd}:
\bea\label{beta_symd}
D_{(A_1}\ldots D_{A_j)} \beta_{B} &=&  -\frac{2j}{j+1}\left(D_{(A_1}\ldots D_{A_{j-1})} D_{[A_j} \beta_{B]} +\ldots + D_{(A_2}\ldots D_{A_{j})} D_{[A_1} \beta_{B]}\right) \nonumber \\
& & +D_{(A_1}\ldots D_{A_j} \beta_{B)} +\ldots
\eea
where again the ellipsis in the second line stands for a sum of monomials containing factors of $D_{(A_1}\ldots D_{A_m)}R[\mu]_{ABCD}$ with $m\leq j-2$ and a factor $D_{(A_1}\ldots D_{A_k)} \beta_{B}$ or a factor $D_{(A_1}\ldots D_{A_{k-1})}D_{[A_k} \beta_{B]}$ with $k\leq j-2$. The terms in the first line of the RHS of this equation can be eliminated by writing $\nabla_{(A_1} \cdots \nabla_{A_{l-1})}R_{rvAB}$ in terms of primitive factors. The terms in the ellipsis can be dealt with by invoking the induction assumption. This yields an expression for $D_{(A_1} \cdots D_{A_j)} \beta_B$ in terms of the desired quantities, closing the induction loop.

Finally, consider the case of $D_{(A_1} \cdots D_{A_j)}\partial_r^{\bar N} \bar{K}_{AB}$ with $\bar{N}\geq 1$. Let us take derivatives w.r.t. $x^A$ on our previous expression relating $\partial_r^{\bar N}\bar{K}_{AB}$ to $\bar K_{AB}$ and covariant components involving the Riemann tensor. This gives expressions relating $D_{(A_1} \cdots D_{A_j)}\partial_r^{\bar N} \bar{K}_{AB}$ to $D_{(A_1} \cdots D_{A_k)}\bar{K}_{AB}$, $D_{(A_1} \cdots D_{A_l)}\beta_{B}$, $D_{(A_1} \cdots D_{A_k)}R[\mu]_{ABCD}$ and covariant components. Using previous substitutions yields the desired expression for $D_{(A_1} \cdots D_{A_j)}\partial_r^{\bar N} \bar{K}_{AB}$. A similar argument establishes that $D_{(A_1} \cdots D_{A_j)}\partial_r^{N} {K}_{AB}$ is also expressible in the desired form. Q.E.D.

\subsection{Covariance of GNC tensors and BRST-formalism}

We now consider how quantities on our cut $C$ transform under a non-trivial change of GNCs, i.e., a change with non-constant $a(x^A)>0$. It follows from the results of section \ref{sec:GNC} that under such a change, on $C$ we have\footnote{The first two transformations hold on all of $\cN$, not just on $C$.}
\bea
\label{eq:homog1}
&{}&\mu_{AB} \to \, \mu_{AB}^{} \qquad \partial^N_v K_{AB}  \to  \, a^{+(N+1)} \partial_v^N K_{AB}  \qquad
\partial_r^{\bar N} \bar K_{AB} \to \, a^{-(\bar N+1)} \partial_r^{\bar N} \bar K_{AB}^{}   \nonumber \\
 &{}& \nabla_{(\alpha_1} \cdots \nabla_{\alpha_j)} R_{\mu\nu\sigma\rho}
\to \, a^b 
\nabla_{(\alpha_1} \cdots \nabla_{\alpha_j)} R_{\mu\nu\sigma\rho} \qquad 
\beta_A \to \,  \beta_A + 2\partial_A \log a 
\eea
where $b$ is the boost weight of the GNC component of the covariant tensor 
$\nabla_{(\alpha_1} \cdots \nabla_{\alpha_j)} R_{\mu\nu\sigma\rho}$, see eq. \eqref{bw}. A useful way to think about the replacements \eqref{eq:homog1} on the cut $C$ is that $\bbeta = \beta_A \ud x^A$ is a
gauge potential and that $K_{AB}, \bar K_{AB}$ and the GNC components of 
$\nabla_{(\alpha_1} \cdots \nabla_{\alpha_j)} R_{\mu\nu\sigma\rho}$ are charged under the gauge group ${\mathbb R}_+$, 
with charges $+1,-1$ and $b$ respectively. The function $a$ corresponds to a finite gauge transformation 
associated with the gauge group ${\mathbb R}_+$. It is therefore useful to define the gauge covariant derivative
\be
\cD_A := D_A - \tfrac{1}{2} b\beta_A,
\ee
with $b$ the boost weight ($=$ charge) of the quantity that it is acting on. See the start of section \ref{sec:GNC} for the geometrical interpretation of $\cD$ as a connection on the normal bundle of $C$.

Our eventual aim is to write an expression for an entropy as an integral of a boost-weight zero $n-2$ form over the cut $C$. For this entropy to be invariant under a change of GNCs, the $n-2$ form should be gauge-invariant up to addition of an exact form. The following lemma characterizes the most general structure of such a form: 

\begin{lemma}
\label{lem:0c}
Let $\bw_m$ be a local $m$-form on $C$ of boost weight 0 built from contractions of 
$D_{(A_1} \cdots D_{A_k)}  \partial_v^{p} \partial_r^{q} \psi$ with 
$\psi \in \{\mu_{AB}, \beta_A, \alpha\}$, or $\mu^{AB}$, or $\epsilon_{A_1\dots A_{n-2}}[\mu]$,  or $D_{(A_1} \cdots D_{A_k)} R_{ABCD}[\mu]$, such that $\bw_m \to \bw_m + \ud \bw_{m-1}$ on 
$C$ under a change of GNCs. Then 
\be
\bw_m = \bbeta \wedge \bI_{m-1} + \bI_{m} -\ud \bb_{m-1}
\ee
where $\bbeta = \beta_A \ud x^A$; $\bI_m$ and $\bI_{m-1}$ are gauge-invariant local forms such that $\bI_{m-1}$ is identically closed ($\ud \bI_{m-1}=0$), and $\bb_{m-1}$ is a local form on $C$. The forms $\bI_m$, $\bI_{m-1}$ can be expressed as boost-weight zero contractions of the following quantities: 
\begin{itemize}
\item 
$\mu_{AB}$, $\mu^{AB}$, $\epsilon_{A_1\dots A_{n-2}}[\mu]$,
\item
$\cD_{(A_1} \cdots \cD_{A_j} K_{A)B}^{}$, $\cD_{(A_1} \cdots \cD_{A_j} \bar K_{A)B}^{}$, and 
\item 
GNC components of the covariant tensors $\nabla_{(\alpha_1} \cdots \nabla_{\alpha_j)} R[g]_{\mu\nu\sigma\rho}$ (By the Bianchi identities, not all these Riemann components are independent, see lemma \ref{lem:0b}.) 
\end{itemize}

\end{lemma}

To prove lemma \ref{lem:0c} and other similar lemmas which we shall need later on, we now reformulate 
the gauge transformations \eqref{eq:homog1} using the BRST method (see e.g. \cite{Henneaux}). First, we write $a = e^{t\Lambda/2}$, and then obtain the transformation, on $\cN$, of any quantity under an infinitesimal gauge transformation by differentiating its transformation law w.r.t. $t$ and evaluating at $t=0$. This gives -- in general very complicated -- formulas
for the infinitesimal change of any monomial $D_{(A_1} \cdots D_{A_k)}  \partial_v^{p} \partial_r^{q} \psi$
under an infinitesimal change of GNCs on $\cN$. 

On $C$, these formulas simplify if 
we pass from this basis of primitive monomials to one of the bases provided by lemmas \ref{lem:0a} or \ref{lem:0b}. Following the usual BRST approach, we now declare $\Lambda$ to be an anti-commuting field (of boost weight $0$) and the infinitesimal version of the transformations \eqref{eq:homog1} becomes a ``BRST transformation'', which we denote as $\gamma$:
\be
\label{eq:homog2}
\begin{split}
\gamma \mu_{AB} =& \, 0\\
\gamma \partial_v^N K_{AB}  = & +\tfrac{1}{2}(N+1) \Lambda \partial^N_v K_{AB} \\
\gamma \partial_r^{\bar N} \bar K_{AB}  = & -\tfrac{1}{2}(\bar N+1) \Lambda \partial^{\bar N}_r \bar K_{AB} \\
\gamma \nabla_{(\alpha_1} \cdots \nabla_{\alpha_j)} R_{\mu\nu\sigma\rho}
=& \, \tfrac{1}{2} b \Lambda \nabla_{(\alpha_1} \cdots \nabla_{\alpha_j)} R_{\mu\nu\sigma\rho} \\
\gamma \beta_A =&\,  \partial_A \Lambda\\
\gamma \Lambda =& \, 0
\end{split}
\ee
where the last transformation is imposed as usual to ensure that $\gamma^2=0$ and where $b$ is 
the boost weight of the corresponding GNC component. The action of $\gamma$ is extended to a product via the graded Leibniz rule. 
The degree in $\Lambda$ of a monomial in the fields and their derivatives is referred to as the ghost number and $\Lambda$ as the ghost field. Consistent with the last two equations equation in 
\eqref{eq:homog2}, we declare that the boost weight of $\Lambda$
is zero, so that $\gamma$ does not change the boost weight of any quantity. We furthermore follow the custom to declare the action of the exterior differential $\ud$ on a ghost number $g$ monomial to be $(-1)^g$ times the usual definition, and with this convention we have $\ud \gamma = -\gamma \ud$, for example. 

For {\it constant} $\Lambda$, the action of $\gamma$ on a quantity of boost weight $b$ is to multiply that quantity by $b\Lambda/2$. However, in general $\Lambda$ is not constant. The transformation law assumed in lemma \ref{lem:0c} may be stated as saying that $\gamma \bw_m = \ud \bw_{m-1}$
and $\bw_m$ has ghost number 0 whereas $\bw_{m-1}$ has ghost number $1$. A form such that $\gamma \bw_m = 0$
is called BRST-closed ($\gamma$-closed) and a form of the type $\gamma \bw_m$ is called BRST-exact.

\begin{lemma} 
\label{lem:0d}
Let $\bw_m$ be a local $m$ form on $C$ of boost weight 0 and ghost number $g>0$ such that $\gamma \bw_m = \ud \bw_{m-1}$. Then $\bw_m$ is a sum of the following:
\begin{itemize}
\item A local $\gamma$-exact local form of boost weight 0. 
\item A local $\ud$-exact local form of boost weight 0.
\item  $\Lambda$ times a boost weight 0 contraction of 
$\cD_{(A_1} \cdots \cD_{A_j} K_{A)B}^{}$, $\cD_{(A_1} \cdots \cD_{A_j} \bar K_{A)B}^{}$, 
of GNC component of covariant tensors
$\nabla_{(\alpha_1} \cdots \nabla_{\alpha_j)} R[g]_{\mu\nu\sigma\rho}$
[Not all of these are independent by the Bianchi identities, see lemma \ref{lem:0b}],
of $\mu^{AB}$ or of $\epsilon[\mu]_{A_1\dots A_{n-2}}$. This case only occurs for $g=1$.
\end{itemize}
\end{lemma}

\noindent
{\em Proof of lemmas \ref{lem:0c} and \ref{lem:0d}:} The proofs of lemmas \ref{lem:0c} and \ref{lem:0d} are quite similar and interrelated so we will give them together. We shall start with the special 
case that $\gamma \bw_m = 0$ where $\bw_m$ is a local form of ghost number $g=0$. 
In $\bw_m$ we make the replacements as in lemma \ref{lem:0b}. Then we further replace 
$D_{(A_1} \cdots D_{A_j} K_{A)B}^{}$, $D_{(A_1} \cdots D_{A_j} \bar K_{A)B}^{}$ with the gauge covariantized forms
$\cD_{(A_1} \cdots \cD_{A_j} K_{A)B}^{}$, $\cD_{(A_1} \cdots \cD_{A_j} \bar K_{A)B}^{}$
at the expense of additional terms of the form $D_{(A_1} \cdots D_{A_j} \beta_{B)}$. Now we impose 
$\gamma \bw_m = 0$. Since 
the factors $D_{(A_1} \cdots D_{A_j} \beta_{B)}$ are the only factors which transform
with derivatives of $\Lambda$ and since the terms with undifferentiated $\Lambda$ 
drop out using that $\bw_m$ has boost weight 0, it follows 
\be
\label{eq:gabw}
0=\gamma \bw_m = \sum_{j\ge 1} \frac{\partial \bw_m}{\partial D_{(A_1} \cdots D_{A_{j-1}} \beta_{A_j)}} D_{(A_1} \cdots D_{A_j)}  \Lambda.
\ee
At any point in $C$ the terms $D_{(A_1} \cdots D_{A_j)} \Lambda$ can be chosen as linearly independent from each other so 
we learn that $\partial \bw_m/\partial D_{(A_1} \cdots D_{A_{j-1}} \beta_{A_j)}=0$ for all $j$. Thus there is no dependence on $\beta_A$
in the chosen basis of monomials and we have demonstrated lemma \ref{lem:0c} in our special case $\gamma \bw_m = 0$. 

We remain in the same special case but now we go to ghost number $g>0$. In the basis of monomials given by $\cD_{(A_1} \cdots \cD_{A_j} K_{A)B}^{}$, $\cD_{(A_1} \cdots \cD_{A_j} \bar K_{A)B}^{}$, by 
GNC components of  $\nabla_{(\alpha_1} \cdots \nabla_{\alpha_j)} R_{\mu\nu\sigma\rho}$, 
by $D_{(A_1} \cdots D_{A_{j-1}} \beta_{A_j)}$ and by $D_{(A_1} \cdots D_{A_j)} \Lambda$, we realize 
that the BRST-transformations \eqref{eq:homog2} have $D_{(A_1} \cdots D_{A_{j-1}} \beta_{A_j)}$ and $D_{(A_1} \cdots D_{A_j)} \Lambda$
as ``contractible pairs'' in the terminology of \cite{Henneaux}, meaning that 
$\gamma D_{(A_1} \cdots D_{A_{j-1}} \beta_{A_j)} = D_{(A_1} \cdots D_{A_j)} \Lambda$, 
$\gamma D_{(A_1} \cdots D_{A_j)} \Lambda = 0$ and the rest of the BRST transformations are independent 
of these variables. By a standard result \cite{Henneaux}, Appendix 2.B, we know that $\bw_m$ is up to a $\gamma$-exact 
local form equal to a local form that has no explicit dependence on $\beta_A$ in the 
chosen basis of monomials and only depends on $\Lambda$ in undifferentiated form. The proof of this works as follows:
First define 
\be
\label{Ndef}
N = \sum_{j\ge 1} \left(
D_{(A_1} \cdots D_{A_{j-1}} \beta_{A_j)} \frac{\partial}{\partial D_{(A_1} \cdots D_{A_{j-1}} \beta_{A_j)}} + 
D_{(A_1} \cdots D_{A_j)} \Lambda \frac{\partial}{\partial(D_{(A_1} \cdots D_{A_j)} \Lambda)}
\right)
\ee
to be the number operator for the contractible pair, and let 
\be
\label{rhodef}
\rho = \sum_{j\ge 1} 
D_{(A_1} \cdots D_{A_{j-1}} \beta_{A_j)} \frac{\partial}{\partial D_{(A_1} \cdots D_{A_{j})} \Lambda}. 
\ee
One has $[N,\gamma]=0$ and $N = \gamma \rho + \rho \gamma$. We now decompose $\bw_m = \sum_{k \ge 0} \bw_m^k$ where $\bw_m^k$ are eigenvectors of $N$: $N \bw_m^k = k \bw_m^k$. Since $N$ and $\gamma$ commute, $\gamma \bw_m=0$ implies $\gamma \bw_m^k=0$. For $k>0$ we have $\bw_m^k = (1/k) N \bw_m^k = (1/k) (\gamma \rho + \rho \gamma)\bw_m^k = \gamma[(1/k) \rho \bw_m^k]$. Therefore $\bw_m = \bw_m^0$ modulo $\gamma$ exact. By construction, $\bw_m^0$ only depends on $\Lambda$ but not on $\beta_A$. Therefore, by lemma \ref{lem:0b},
\be
\bw_m = \Lambda^g \bI_m + \text{$\gamma$-exact,}
\ee
where $\gamma$-exact means an $m$-form in the image of the local $m$-forms and where $\bI_m$ has boost weight zero and is 
gauge invariant, i.e. it is a local $m$-form constructed from $\cD_{(A_1} \cdots \cD_{A_j} K_{A)B}^{}$, $\cD_{(A_1} \cdots \cD_{A_j} \bar K_{A)B}^{}$ and 
GNC components of  $\nabla_{(\alpha_1} \cdots \nabla_{\alpha_j)} R_{\mu\nu\sigma\rho}$.  
Since $\Lambda^g = 0$ for $g>1$, it follows that when $g>1$, $\bw_m$ is $\gamma$-exact. 
When $g=1$, we have $\bw_m = \Lambda \bI_m$ plus $\gamma$-exact. Thus, 
we have demonstrated lemma \ref{lem:0d} in our special case $\gamma \bw_m = 0$. 

\medskip

We now turn to the general case of lemmas \ref{lem:0c} and \ref{lem:0d} where we only assume that $\gamma \bw_m = \ud \bw_{m-1}$ for some local $m-1$
form $\bw_{m-1}$. This case will be analyzed using the standard technique of descent equations. To construct the descent equations, 
we need to appeal to the algebraic Poincare lemma \cite{Wald}:

\begin{lemma}
\label{lem:poincare}
{\bf (Algebraic Poincare Lemma)}
Let $\bxi([\Psi,\Phi], x)$ be an $m$ form on a manifold $C$ that is locally constructed out of smooth fields $\Psi$ and ``background'' fields $\Phi$ such that $\ud \bxi([\Psi,\Phi], x)=0$
for all $\Psi$ in a neighborhood of a special configuration $\Psi_0$. Assume that all $\Psi$ in that neighborhood can be joined to 
$\Psi_0$ by a differentiable path $\Psi_\lambda, \lambda \in [0,1]$ of smooth configurations such that $\Psi_{\lambda}|_x$
depends on finitely many derivatives $\partial^k \Psi|_x$ for any $x \in C$ and such that $\Psi_{\lambda = 1} = \Psi$. 
Then $\bxi([\Psi,\Phi], x) - \bxi([\Psi_0,\Phi],x) = \ud \boldeta([\Psi,\Psi_0,\Phi],x)$
for some local $m-1$ form $\boldeta([\Psi,\Psi_0,\Phi],x)$.
\end{lemma}

\noindent
{\em Proof:} \cite{Wald} We sketch the proof here since we later want to see how $\boldeta$ preserves certain structures of $\bxi$ below. We will suppress the dependence on the background fields $\Phi$ in this proof. We fix an auxiliary 
covariant derivative $D_A$ on $C$ (coordinate indices $A,B,\dots$). Consider the linearization of $\bxi$,
\be
\delta \bxi([\Psi],x)_{B_1 \dots B_m}=  \sum_{j=0}^k \frac{\partial \bxi_{B_1 \dots B_m}}{\partial (D_{(A_1} \dots
D_{A_j)} \Psi)} ([\Psi],x) D_{(A_1} \dots
D_{A_j)} \delta \Psi(x) ,
\ee
where $k$ is the maximum number of derivatives of $\Psi$ that $\bxi$ depends on. Then we let
\be
\bgamma_k([\Psi,\delta \Psi],x)_{C_1 \dots C_{m-1}} :=  
\frac{mk}{(m+1)(k+1)}
 \frac{\partial \bxi_{C_1 \dots C_{m-1}B}}{\partial (D_{(A_1} \dots
D_{A_{k-1}} D_{B)} \Psi)} ([\Psi],x) D_{(A_1} \dots
D_{A_{k-1})} \delta \Psi(x).
\ee
The arguments given in \cite{Wald} show that $\delta \bxi([\Psi],x) - \ud \bgamma_k([\Psi,\delta \Psi],x)$ is a closed local functional depending on up to ($k-1$)
derivatives of $\delta \Psi$ and so verifies the assumptions of the theorem because $\delta \Psi$ is an arbitrary variation. 
The argument is then iterated to determine $\bgamma_{k-1}([\Psi,\delta \Psi],x)$ and so 
on and then set $\bgamma([\Psi,\delta \Psi],x)= \sum_{j \le k}\bgamma_j([\Psi,\delta\Psi],x)$. 
Finally, we set (using a dot to denote a derivative w.r.t. $\lambda$)
\be
\boldeta([\Psi_0, \Psi],x) = \int_0^1 \ud \lambda \bgamma([\Psi_\lambda,\dot \Psi_\lambda],x),
\ee
which is local and satisfies the desired equation.
Q.E.D.

\medskip
\noindent
{\bf Remark:} In the applications below, the space of $\Psi$ will consist of certain monomials as in lemmas \ref{lem:0a},\ref{lem:0b}, and will have the structure of vector space. Thus, 
the interpolating family can be chosen to be simply $\lambda \Psi=\Psi_\lambda$ and $\Psi_0=0$. Furthermore, the forms $\bxi$ in question will be at least quadratic 
in positive boost weight monomials. It is clear from the above proof that also $\boldeta$ will then be at least quadratic in positive boost weight quantities. 

\medskip
\noindent
We now return to the proof of lemmas \ref{lem:0c} and \ref{lem:0d} in the general case. Taking $\gamma$ of $\gamma \bw_m = \ud \bw_{m-1}$, using $\gamma^2 = \ud \gamma + \gamma \ud = 0$, we learn that 
$\ud(\gamma \bw_{m-1})=0$. $\gamma \bw_{m-1}$ depends at least linearly on $\Lambda$, so we can apply the algebraic 
Poincare lemma for $\Psi = \Lambda, \Psi_0 = 0$ (viewing the other fields as background fields), to learn that $\gamma \bw_{m-1} = \ud \bw_{m-2}$, where $\bw_{m-2}$ is again local. 
Iterating this process we get the descent equations
\be
\begin{split}
& \gamma \bw_m = \ud \bw_{m-1}\\
& \gamma \bw_{m-1} = \ud \bw_{m-2}\\
& \dots\\
& \gamma \bw_{k} = 0.
\end{split}
\ee
where the ghost number increases by $1$ at each step. The last equation does not involve a boundary term. Such an equation must exist for some $k$ -- in the worst case, i.e. the longest descent, 
we have $k=0$. The important point is now that (as $\bw_{k}$ has positive ghost number) we can apply our previous results to $\gamma \bw_{k} = 0$. We learn 
$\bw_k = \gamma \bb_k + \Lambda \bI_k$, where the last term $\Lambda \bI_k$ is present only if $\bw_k$ has ghost number 1. Thus, if $\bw_m$ had had 
ghost number 0 (lemma \ref{lem:0c}), the last term is present when $k=m-1$, whereas if $\bw_m$ had had 
ghost number $g>0$ (lemma \ref{lem:0d}), the last term is present only when $g=1$ and $k=m$.  So long as the last term $\Lambda \bI_k$ is not 
present, we can consider $\tilde \bw_{k+1}:=\bw_{k+1} + \ud \bb_k, \tilde \bw_j:= \bw_j$ for $j>k+1$, and then the new ladder 
$\{ \tilde \bw_m, \dots, \tilde \bw_{k+1} \}$ will satisfy by construction the descent
\be
\begin{split}
& \gamma \tilde \bw_m = \ud \tilde \bw_{m-1}\\
& \gamma \tilde \bw_{m-1} = \ud \tilde \bw_{m-2}\\
& \dots\\
& \gamma \tilde \bw_{k+1} = 0,
\end{split}
\ee
i.e. we have shortened the descent. 
We continue with this until we get a nontrivial term $\Lambda \bI_k$. 

\medskip
\noindent
{\em Lemma \ref{lem:0d}:} Since $g \ge 1$, the ghost number of $\tilde \bw_j$ is greater than $1$ for $j<m$ so we can shorten the descent until we reach $\gamma \tilde \bw_m =0$. Our previous result gives $\tilde \bw_{m} = \gamma \bb_{m}+\Lambda^g \bI_m$ and hence $\bw_{m} = \gamma \bb_{m} - \ud \bb_{m-1}+\Lambda^g \bI_m$, establishing the result (as $\Lambda^g=0$ for $g>1$). 

\medskip
\noindent
{\em Lemma \ref{lem:0c}:} In this case we have $g=0$ and we can shorten the descent until we reach $\gamma \tilde \bw_{m-1} =0$, which implies $\tilde \bw_{m-1} = \gamma \bb_{m-1} + \Lambda \bI_{m-1}$, so 
 $ \bw_{m-1} = \gamma \bb_{m-1} + \Lambda \bI_{m-1} - \ud \bb_{m-2}$. This gives 
\be
\begin{split}
&\gamma \bw_m = \ud(\gamma \bb_{m-1} + \Lambda \bI_{m-1}) \quad \Longrightarrow\\
&\gamma(\bw_m + \ud \bb_{m-1}) = \ud( \Lambda \bI_{m-1}) \quad \Longrightarrow \\
&\gamma(\bw_m + \ud \bb_{m-1} - \bbeta \wedge \bI_{m-1}) = -\Lambda \ud \bI_{m-1}.
\end{split}
\ee
The left side depends only on $\Lambda$ in differentiated form by the definition of $\gamma$ and because all forms have boost weight 0, 
whereas the right side depends on $\Lambda$ only in undifferentiated form. Therefore, since the equation must hold for all $\Lambda$, and since 
$D_{(A_1} \dots D_{A_j)} \Lambda$ are linearly independent at each point, both sides must vanish. Vanishing of the LHS implies by our previous results (for ghost number $0$) that $\bw_m + \ud \bb_{m-1} - \bbeta \wedge \bI_{m-1} = \bI_{m}$ and vanishing of the RHS gives $\ud \bI_{m-1} = 0$ which completes the proof. Q.E.D.

\medskip

Although it is not essential for our proof of gauge invariance of the IWW entropy, it is interesting to note that we can be more explicit about the form of $\bI_{m-1}$ in Lemma \ref{lem:0c}. In the following Lemma note that $\ud \bbeta$ (the curvature of the connection on the normal bundle of $C$) can be expressed in terms of the quantities listed in Lemma \ref{lem:0c} via equation \eqref{twist}. 

\begin{lemma}
\label{lem:char_classes} The gauge-invariant forms $\bI_{m-1}$, $\bI_m$ of Lemma \ref{lem:0c} can be defined so that $\bI_{m-1}$ is written in terms of $\ud\bbeta$ and characteristic classes built from the curvature of $\mu_{AB}$ as follows:
\be\label{top}
\bI_{m-1}= \, \sum_j \left(\bigwedge^{j} \ud \bbeta \right) \wedge p_{m-1-2j}(
\underbrace{\bR[\mu],\dots,\bR[\mu]}_{(m-1-2j)/2}
),
\ee
where $\bR[\mu]_A{}^B = R[\mu]_{CDA}{}^B \ud x^C \wedge \ud x^D$ and $p_i$ is an invariant symmetric polynomial of the Lie-algebra $\mathfrak{so}(n-2)$. Such terms can only appear when $m$ is odd.
\end{lemma}
\noindent {\em Proof.} 
Given our metric corresponding to the $\psi=(\alpha,\beta_A,\mu_{AB})$ in its GNC system, let $\psi_0$ be such that  $\alpha_0=(\beta_0)_A=0$ and $(\mu_0)_{AB}(v,r,x^A)=\mu_{AB}(0,0,x^A)$. We can construct a 1-parameter family $\psi_\lambda$ with $\psi_1=\psi$ by setting $\alpha_\lambda = \lambda \alpha$, $(\beta_\lambda)_A = \lambda \beta_A$ and $(\mu_\lambda)_{AB} = (1-\lambda) (\mu_0)_{AB}+\lambda \mu_{AB}$. 
From the algebraic Poincar\'e lemma we have $\bI_{m-1}[\psi]=\bI_{m-1}[\psi_0]+\ud\bJ_{m-2}[\psi]$ where $\bJ_{m-2}$ is a local form such that $\ud\bJ_{m-2}$ is gauge invariant. The latter condition gives $\gamma \ud \bJ_{m-2}=0$ hence $\ud\gamma \bJ_{m-2}=0$ but $\gamma \bJ_{m-2}$ has ghost number $1$ so we can use the algebraic Poincar\'e lemma interpolating $\Lambda$ to $0$ (treating other fields as background fields) and deduce that $\gamma \bJ_{m-2}$ is exact. We can now apply Lemma \ref{lem:0c} to deduce $\bJ_{m-2} = \bbeta \wedge \bI_{m-3} + \bI_{m-2} - \ud\bb_{m-3}$ where $\bI_{m-3}$ is gauge-invariant and closed, and $\bI_{m-2}$ is gauge-invariant. This gives
\be
 \bI_{m-1}[\psi] = \bI_{m-1}[\psi_0]+\ud \bbeta \wedge \bI_{m-3}[\psi]+\ud\bI_{m-2}[\psi]
\ee
We now repeat the process starting from $\bI_{m-3}$ to obtain
\be
 \bI_{m-3}[\psi] = \bI_{m-3}[\psi_0]+\ud \bbeta \wedge \bI_{m-5}[\psi]+\ud\bI_{m-4}[\psi]
\ee
and hence
\be
 \bI_{m-1}[\psi] = \bI_{m-1}[\psi_0]+\ud \bbeta \wedge \bI_{m-3}[\psi_0]+\ud \bbeta \wedge \ud \bbeta \wedge \bI_{m-5}[\psi]+\ud\bI'_{m-2}[\psi]
\ee
where $\bI'_{m-2} = \ud\bbeta \wedge \bI_{m-4} + \bI_{m-2}$ is gauge-invariant. Proceeding inductively we reach
\be
 \bI_{m-1}[\psi] = \sum_j \left( \bigwedge^j \ud \bbeta \right) \wedge \bI_{m-1-2j}[\psi_0]+ \ud\bI''_{m-2}[\psi]
\ee
On the RHS, $\bI_k[\psi_0]$ is an identically closed $k$-form constructed locally and covariantly only from $(\mu_0)_{AB}\equiv \mu_{AB}|_C$: these are the characteristic classes built from 
$\bR[\mu]_A{}^B = R[\mu]_{CDA}{}^B \ud x^C \wedge \ud x^D$ [see for example eq. 6.32 of \cite{Kreuzeretal}, replacing spacetime by $(C, \mu_{AB})$]. The result now follows upon substituting the above equation into our formula for $\bw_m$ in Lemma \ref{lem:0c} and setting $\bI'_m = \bI_m + \ud \bbeta \wedge \ud \bI''_{m-2}$ (which is gauge invariant) and $\bb'_{m-1} = \bb_{m-1}+ \bbeta \wedge \ud \bI''_{m-2}$. Q.E.D.

\section{Entropy current associated with a null surface}

\label{sec:entropy_current}

\subsection{Covariant phase space formalism}

We shall use the covariant phase space formalism of Iyer and Wald \cite{Iyer:1994ys} and so we summarize the main results of that formalism that we shall need. Consider a Lagrangian $n$-form $\bL$ depending locally and covariantly on some fields $\Psi^i = (g_{\mu\nu}, \Phi_I, A_{J\mu})$ including a Lorentzian metric $g$. When the fields include 
$1$-form fields $\bA_J$, we also require that $\bL$ is gauge invariant.
Local and covariant means by definition that in any coordinate system and at any point $x$, $\bL|_x$ is expressible 
in terms of a finite number of derivatives $\partial_{\alpha_1} \dots \partial_{\alpha_d} \Psi|_x$, and that, for every 
diffeomorphism $f$, $f^* \bL[\Psi] = \bL[f^* \Psi]$. By the Thomas replacement lemma \cite{Iyer:1994ys}, $\bL$ can be written as $L \bepsilon$, with $\bepsilon$ the positively oriented 
volume form defined from $g_{\mu\nu}$, where $L$ can be expressed entirely in terms of 
$g_{\mu\nu}, g^{\mu\nu}, \nabla_{\alpha_1} \cdots \nabla_{\alpha_k} R_{\mu\nu\alpha\beta}$, $\epsilon_{\alpha_1 \dots \alpha_n}$, $\nabla_{\alpha_1} \cdots \nabla_{\alpha_k} \Phi_I$, and $ \nabla_{\alpha_1} \cdots \nabla_{\alpha_k} F_{\mu_1 \dots \mu_{p+1} J}$ when $p$-form fields with curvatures $\bF_J$ are present. 
However, for most parts of this paper we will not consider matter fields and treat (local and covariant) Lagrangians depending only on the metric. Note that 
anti-symmetrized combinations of covariant derivatives can be rewritten in terms of curvature tensors, so without loss of generality we will usually consider 
symmetrized derivatives as the variable fields in our Lagrangians and other functionals. There are further dependencies among the derivatives of the curvature components 
arising from the Bianchi identities which are usually understood to be eliminated by choosing a suitably linearly independent set, see below.

In such a case, the only equation of motion is the Einstein equation $E^{\mu\nu} = 0$ where
\be
 E^{\mu\nu} \bepsilon \equiv \bE_g^{\mu\nu} = \sum_{j \ge 0} (-1)^j \partial_{\alpha_1} \dots \partial_{\alpha_j} \frac{\partial \bL}{\partial(\partial_{\alpha_1} \dots \partial_{\alpha_j} g_{\mu\nu})}.
\ee 
$E^{\mu\nu}$ specifies the classical dynamics so any classical physical quantity should be specified only in terms of this dynamical content plus potentially some global quantities of topological nature. The symplectic
$(n-1)$ potential is defined by $\btheta[\Psi; \delta \Psi]$
\be
\label{delL}
\delta \bL[\Psi] = \bE_{\Psi^i}[\Psi] \cdot \delta \Psi^i + \ud \btheta[\Psi; \delta \Psi].
\ee
We use the general notation 
\be\label{eq:notation}
\delta \Psi = \frac{\ud}{\ud \lambda} \Psi(\lambda), \quad \delta^2 \Psi = \frac{1}{2} \frac{\ud^2}{\ud \lambda^2} \Psi(\lambda)
\ee
for a 1-parameter family of dynamical fields. More generally, we write $\delta_ 1 \Psi, \delta_2 \Psi, \delta_1 \delta_2 \Psi$ for families depending on multiple parameters.
The definition of $\btheta[\Psi; \delta \Psi]$ leaves some ambiguities because we may add a $\ud$-exact covariant term or topological term to the Lagrangian $\bL \to \bL + \ud \bmu + \bL^{\rm top}$
without changing the equations of motion, $E_g^{\mu\nu}$. Furthermore, we may add a local and covariant $\ud$-exact term to $\btheta$ itself, $\btheta \to \btheta + \ud \bY$. Altogether, 
these ambiguities result in the freedom to modify $\btheta \to \btheta + \delta \bmu + \btheta^{\rm top} + \ud \bY$ without changing the local dynamical content of the theory. 

The symplectic $(n-1)$-form is, for a purely gravitational theory,
\be
\bomega[g; \delta_1 g, \delta_2 g] = \delta_1 \btheta[g; \delta_2 g] - \delta_2 \btheta[g; \delta_1 g].
\ee
The Noether current is 
\be
\bJ_X[g] = \btheta[g; \cL_X g] - X \cdot \bL[g],
\ee
again for a purely gravitational theory, and the above ambiguities of $\btheta$ and $\bL$ 
propagate into ambiguities of $\bJ_X$ and of $\bomega$. One can establish \cite{Iyer:1994ys} the existence of a non-unique
Noether-charge $(n-2)$-form $\bQ_X$ satisfying 
\be
\ud \bQ_X[g]= \bJ_X[g] + \bC_X[g],
\ee
where $\bC_X$, which we call the constraint, does not depend on 
derivatives of $X$, and $\bC_X=0$ when the equations of motion are fulfilled. For a purely gravitational theory as considered here, 
the constraint is \cite{Bhattacharyya:2021jhr}
\be
\label{CX}
(\bC_X)_{\alpha_1 \dots \alpha_{n-1}} = 2 E^{\mu\nu} X_\nu \epsilon_{\mu\alpha_1 \dots \alpha_{n-2}}, 
\ee
To see this, note that equation \eqref{delL} and the (off-shell) Bianchi identity $\nabla_\mu E^{\mu\nu}=0$ give
\be
\begin{split}
\ud \bJ_X[g] =& \, \ud \btheta[g, \cL_X g] - \ud(X \cdot \bL[g]) \\
=& \, \cL_X \bL[g] - \bE^{\mu\nu}_g[g] \cL_X g_{\mu\nu} - (\ud (X \cdot \bL[g]) + X \cdot \ud \bL[g]) \\
=& \, -2 \bepsilon E^{\mu\nu} \nabla_\mu X_\nu \\
=& \, -2 \bepsilon \nabla_\mu(E^{\mu\nu} X_\nu) \\
=& \, -\ud \bC_X[g],
\end{split}
\ee 
using Cartan's magic formula for the Lie derivative. Hence $\ud(\bJ_X[g] + \bC_X[g] )=0$. Since this is true for any $X$ and any $g_{\mu\nu}$, the algebraic Poincare lemma \ref{lem:poincare} yields the existence of a 
corresponding local $\bQ_X$. Note also that the expression \eqref{CX} is unaffected by the above ambiguity to change 
$\bL \to \bL + \ud \bmu + \bL_{\rm top}$ and the corresponding ambiguity $\btheta \to \btheta + \delta \bmu + \btheta^{\rm top} + \ud \bY$, 
even though $\bQ_X$ itself is affected by this ambiguity. 

A general result from \cite{Iyer:1994ys} is that, up to the ambiguities affecting $\btheta$ and $\bL$ just described, $\bQ_X$ may be written as
\be
\label{eq:Qdef}
Q_{\mu_1 \dots \mu_{n-2}}[g]= -E^{\alpha\beta\mu\nu}_R[g] (\nabla_\mu X_\nu) \epsilon_{\alpha\beta \mu_1 \dots \mu_{n-2}} 
+ W^{\alpha\beta\nu} [g] X_\nu \epsilon_{\alpha\beta \mu_1 \dots \mu_{n-2}}, 
\ee
where $E^{\alpha\beta\mu\nu}_R$ is 
\be
E^{\alpha\beta\mu\nu}_R \bepsilon := \sum_{k\ge 0}(-1)^k 
\nabla_{(\alpha_1} \cdots \nabla_{\alpha_k)}  \frac{\partial \bL}{\partial (\nabla_{(\alpha_1} \cdots \nabla_{\alpha_k)} R_{\mu\nu\alpha\beta})}.
\ee
The ambiguities affecting $\btheta$ and $\bL$ result in the change $\bQ_X[g] \to \bQ_X[g] + X \cdot \bmu[g] + \bY[g; \cL_X g] + \ud \bZ_X[g] + \bQ_X^{\rm top}[g]$, 
which represents the total ambiguity in defining the Noether charge. 

Another useful general identity which holds for general 1-parameter families of fields which are not necessarily solutions is obtained starting from
\be
\begin{split}
\delta \bC_X[g] &= \ud \delta \bQ_X[g] - \delta \bJ_X[g] \\
&= \ud \delta \bQ_X[g]- \delta ( \btheta[g; \cL_X g] - X \cdot \bL[g]) \\
&= \ud \delta \bQ_X[g] - \delta  \btheta[g; \cL_X g] 
+ X \cdot (\bE_g[g] \, \delta g + \ud \btheta[g; \delta g])\\
&= \ud ( \delta \bQ_X[g] - X \cdot \btheta[g; \delta g]) - \bomega[g; \delta g, \cL_X g] 
- X \cdot ( \bE_g[g] \, \delta g),
\end{split}
\ee
using Cartan's magic formula $\cL_X \btheta = \ud(X \cdot \btheta)+X \cdot \ud \btheta$ and 
the definition of $\bomega$ in the last step.
Now take a $C^1$-codimension 1 surface $\cS$ in spacetime, take $X$ tangent to $\cS$, and pull back 
the above equation for forms to $\cS$. Since the pull back of $X \cdot (\bE_g \, \delta g)$ vanishes, we get:
\be
\label{eq:ADM}
\ud (\delta \bQ_X[g] - X \cdot \btheta[g; \delta g]) - \delta \bC_X[g] = \bomega[g; \delta g, \cL_X g]
\quad \text{on $\cS$.}
\ee
It is important to note that this is an off-shell identity: neither $g$ nor $\delta g$ have to satisfy any equations. 

\subsection{Covariant phase space derivation of IWW entropy current}

\label{sec:IWW}

The aim of this section is to demonstrate the relation between the entropy determined by Wall's procedure and the Iyer-Wald entropy. We will apply the results of the previous section to the case where $\cS$ is a smooth null hypersurface $\cN$ ruled by affinely parameterized null geodesics. We 
choose an arbitrary cut $C$ and introduce GNCs as described in section \ref{sec:GNCintro}. We now define the vector field
\be
\label{Kdef}
K = v \partial_v - r \partial_r
\ee
generating a boost in the coordinates $r,v$. We take $X=K$ and consider the constraint $\bC_K$, which we write 
\be
\label{C_K_def}
(\bC_K)_{\mu_1 \dots \mu_{n-1}} = 2K^\nu E_{\nu}{}^\alpha \epsilon_{\alpha \mu_1 \dots \mu_{n-1}}, 
\ee
The pull-back of $\bC_K$ to $\cN$ is $2v E_{vv} \sqrt{\mu} \ud v \wedge \ud^{n-2}x$ in GNCs. To analyze the structure of $E_{vv}$ it is useful 
to introduce an equivalence relation on functionals on $\cN$:

\begin{definition}
\label{def:2} Two functionals $A,B$ each of which is a sum of primitive monomials (definition \ref{def:primitive}) with definite boost weight $p$ are said to be equivalent $A \sim B$ if $A-B$ contains only monomials having at least two primitive factors with positive boost weight.
\end{definition}

In the following, functionals of boost weight $p$ will be denoted with symbols such as $A^{(p)}, X^{(p)}, \dots$. Consider a functional $A^{(2)}$ of boost weight $+2$ (later we will take this to be $E_{vv}$). One can use a ``partial integration'' argument to show that 
there exist $X^{(-k)}, W^{(1)A}, k \ge 0$ such that (i) $X^{(-k)}$ is a sum of primitive monomials for which  
{\em each} primitive factor has non-positive boost weight, such that (ii)  $W^{(1)A}$ is a sum of primitive monomials for which precisely one primitive factor 
has positive boost weight, and such that (iii)
\be
\label{eq:ABdec}
\begin{split}
A^{(2)} \sim& \, \partial_v \Bigg\{ 
\frac{1}{\sqrt{\mu}} \sum_{j,p,\psi}
X^{(0)A_1\dots A_j}_\psi \partial_v (\sqrt{\mu} D_{(A_1} \cdots D_{A_j)}  \partial_v^{p} \partial_r^{p} \psi)   \\
& \, + \frac{1}{\sqrt{\mu}} \sum_{j}
 X^{(0)A_1\dots A_jABCD}_R \partial_v (\sqrt{\mu} D_{(A_1} \cdots D_{A_j)}  R[\mu]_{ABCD})  \\
& \, +\frac{1}{\sqrt{\mu}} \partial_v \sum_{j,p-q > 0,\psi} \sqrt{\mu} X^{(-p+q)A_1 \dots A_j}_{q,\psi}  
D_{(A_1} \cdots D_{A_j)}  \partial_v^{p} \partial_r^{q} \psi \, + D_A W^{(1)A}
\Bigg\} .
\end{split}
\ee
Note that (i) implies that each primitive monomial in an $X^{(0)}$ term must be a product of primitive factors with zero boost weight. A derivation of the above result was sketched in \cite{Wall:2015raa} and explained in more detail in \cite{Bhattacharyya:2021jhr}.

The key trick is now to define a 1-parameter family of metrics interpolating between a stationary ``background'' metric (similarly as in \cite{Iyer:1994ys}) and some other metric. Let $g_{\mu\nu}(x^\alpha)$ be the given metric expressed in GNCs and read off $\psi \in \{ \alpha, \beta_A, \mu_{AB}\}$. The background fields are defined by
\be
\label{background_def}
\bar{\psi} := \sum_{p\ge 0} \frac{1}{(p!)^2} \chi(c_p rv) (rv)^p (\partial_r \partial_v)^p \psi |_{r=v=0} 
\ee
where $\chi$ is smooth and of compact support and identically equal to $1$ in a neighborhood of the origin
and $\{ c_p \}$ is a sequence of positive numbers increasing sufficiently fast in order that the series converges\footnote{This is just a convenient 
trick to avoid having to consider Taylor series with remainder, which would be fine too but notationally cumbersome.}. From $\bar{\psi}$ we define a corresponding background metric $\bar{g}_{\mu\nu}$ in GNCs. This background metric has the properties:
\begin{itemize}
\item
$\cL_K \bar{g} = 0$. In particular, the background metric $\bar{g}_{\mu\nu}$ is such that $\cN$ is a Killing horizon with bifurcation surface $C$. Note that this background metric depends on the original choice of GNCs.
\item
In the background, any primitive factor with positive boost weight vanishes on $\cN$ and any primitive factor with negative boost weight vanishes on $C$. This implies that when $X^{(k)}$ has positive boost weight $k$, then $X^{(k)}[\bar{g}] = 0$ on $\cN$, 
and when $X^{(-k)}$ has negative boost weight $-k$, then $X^{(-k)}[\bar{g}] = 0$ on $C$. 
\item
If $X^{(0)}$ is a sum of primitive monomials where each monomial involves only primitive factors of boost weight $0$ then $X^{(0)}[\bar{g}] = X^{(0)}[g]$ on $C$.
\end{itemize}

Next for $\lambda \in [0,1]$ (the interpolation parameter), and for an arbitrary smooth $\delta \psi$ we set
\be
\psi(\lambda) := \bar{\psi} + \lambda  \delta \psi
\ee
and from $\psi(\lambda)$, we define a corresponding 1-parameter family of metrics $g_{\mu\nu}(\lambda,x^\alpha)$ in GNCs.\footnote{We assume that $\delta \mu_{AB}$ is chosen such that $\mu_{AB}(\lambda)$ is Riemannian for $\lambda \in [0,1]$ which ensures that $g_{\mu\nu}(\lambda)$ is Lorentzian near $\cN$.} This 1-parameter family has the properties
\begin{itemize}
\item
$g|_{\lambda=0}=\bar{g}$
\item If we choose $\delta \psi = \psi - \bar \psi$ then $g|_{\lambda = 1}$ is the original metric $g$. Note that we do not make this choice at the beginning of the following argument.
\end{itemize}

We now take $A^{(2)}=E_{vv}$ in \eqref{eq:ABdec} and consider the first variation of this equation within our 1-parameter family. Using the properties just described, we get on $\cN$ (schematically denoting by $Y^{(k)}$ the monomials $D_{(A_1} \cdots D_{A_j)}  \partial_v^{p} \partial_r^{q} \psi$
with $k=p-q$ or the monomials $D_{(A_1} \cdots D_{A_j)}  R[\mu]_{ABCD}$ with $k=0$)
\be
\label{eq:Cvv}
\frac{\partial}{\partial \lambda} E_{vv}|_{\lambda=0} = \partial_v^2 \left\{   X^{(0)} \frac{\partial}{\partial \lambda} Y^{(0)} + 
\sum_{k > 0} X^{(-k)} \frac{\partial}{\partial \lambda} Y^{(k)} \right\}_{\lambda=0}+ \partial_v D_A \frac{\partial}{\partial \lambda} W^{(1)A} |_{\lambda=0}
\ee
because all other terms have an unvaried monomial with positive boost weight and such a monomial vanishes in the background $\bar{g}$. 
The terms with $k>0$ in curly brackets can be rewritten as $\frac{\partial}{\partial \lambda} \sum_{k > 0} X^{(-k)} Y^{(k)}|_{\lambda=0}$ 
because $Y^{(k)}[\bar{g}]=0$ for $k>0$. 

We will now show that the terms with $k=0$ are related 
to the Noether charge using the restriction of eq. (\ref{eq:ADM}) (with $X=K$) to $\cN$\footnote{
When considering the restriction of forms to $\cN$ it is useful to work with the Hodge duals of the forms; for example we write $(\bQ_X)_{\mu_1 \dots \mu_{n-2}} = \frac{1}{2} \epsilon_{\mu_1 \dots \mu_{n-2} \alpha\beta}  (\star \bQ_X)^{\alpha\beta}$.
}
and evaluated on the background $\bar{g}$:
\be
\partial_v\left( \frac{1}{\sqrt{\mu}} \frac{\partial}{\partial \lambda} ( \star Q^{vr}_K \sqrt{\mu} ) - v \, (\star \theta)^r\left[g; \frac{\partial}{\partial \lambda} g\right]  \right)_{\lambda=0}
+ D_A  \frac{\partial}{\partial \lambda}  (\star Q^{Ar}_K)|_{\lambda=0}   \\
= 2v \, \frac{\partial}{\partial \lambda}  E_{vv}|_{\lambda=0}  ,
\ee
using that 
\be
\bomega\left[g; \frac{\partial}{\partial \lambda} g, \cL_K g \right]_{\lambda=0}=0,
\ee
as $\cL_K g|_{\lambda=0}=0$ by construction. We now assume that $\delta \psi$ has compact support so $\frac{\partial}{\partial \lambda} g|_{\lambda=0}$ also has compact support. If we integrate the above equation over the $v>0$ portion of $\cN$, use the structure (\ref{eq:Cvv}),  and perform integrations by part in $v$, we obtain
\be
\label{eq:Qvr}
\int_C \Big( \frac{\partial}{\partial \lambda}  ( \star Q^{vr}_K \sqrt{\mu})  + 2\sqrt{\mu}  X^{(0)} \frac{\partial}{\partial \lambda} Y^{(0)}  \Big)_{\lambda=0}  = 0.
\ee
In deriving this equation it has been used that $X^{(-k)}(g|_{\lambda=0}) = 0$ for a negative boost weight $-k$ quantity on $C$, we have used that all terms with an explicit dependence on $v$ vanish at $C$, and that divergence terms (i.e. $D_A  \tfrac{\partial}{\partial \lambda} (\star Q^{Ar}_K)$ and $D_A \tfrac{\partial}{\partial \lambda} W^{(1)A}$) integrate to zero. 

On $\cN$, we can write 
\be
\label{eq:ERvrvr}
( \star Q_K)^{vr} =  q^{(0)}   + \dots
\ee
where $q^{(0)}$ is a sum of primitive monomials for which {\em all} primitive factors have boost weight $0$\footnote{By 
\eqref{eq:Qdef}, we may determine $q^{(0)}$ by considering all contributions to $E^{rvrv}_R$ made up exclusively of boost weight 0 primitive factors \cite{Iyer:1994ys}.} and the ellipsis represents a sum of primitive monomials, each containing at least one primitive factors of strictly positive boost weight and at least one primitive factor of strictly negative boost weight (since the total boost weight is $0$).  The Iyer-Wald entropy is defined as the integral over a horizon cross-section of $2\pi \sqrt{\mu} q^{(0)}$ \cite{Iyer:1994ys}. The terms in the ellipsis of \eqref{eq:ERvrvr} are $O(\lambda^2)$ on $C$ and therefore their $\lambda$-derivative vanishes on $C$ for $\lambda=0$, so $q^{(0)}$ is the only surviving term on $C$ after we take a variation off our stationary background $g|_{\lambda=0}$. Therefore we have derived 
\be
\int_C \Big(  \frac{1}{2} \frac{\partial}{\partial \lambda} (q^{(0)} \sqrt{\mu}) +\sqrt{\mu}  X^{(0)} \frac{\partial}{\partial \lambda} Y^{(0)}  \Big)_{\lambda=0} = 0.
\ee
By construction, the integrand here has the general form
\be
\biggl( \frac{1}{2\sqrt{\mu}} \frac{\partial}{\partial \lambda}  (q^{(0)} \sqrt{\mu}) 
  +  X^{(0)} \frac{\partial}{\partial \lambda} Y^{(0)} \biggr)_{\lambda=0} = \sum_{p,q} A^{(0)B_1 \dots B_q}_p[\bar{\psi}] 
  \bar D_{(B_1} \cdots \bar D_{B_q)} (\partial_v \partial_r)^p \frac{\partial}{\partial \lambda} \psi |_{\lambda=0},
\ee
where $\bar D_A$ is the covariant derivative of $\bar \mu_{AB} = \mu_{AB}|_{\lambda=0}$.
 The integral over $C$ of this quantity vanishes for completely 
arbitrary $\delta \psi = \frac{\partial}{\partial \lambda} \psi |_{\lambda=0}$, and in particular the restrictions $(\partial_v \partial_r)^p \delta \psi |_C$ can be chosen 
independently. Therefore we can write (still on $C$)
\be
\label{eq:sXYW}
\bigg( \frac{1}{2\sqrt{\mu}} \frac{\partial}{\partial \lambda}  (q^{(0)} \sqrt{\mu}) 
  + X^{(0)} \frac{\partial}{\partial \lambda} Y^{(0)} \bigg)_{\lambda=0}
= \bar{D}_A \bar W^{(0)A}
\ee
with
\be 
\bar W^{(0)C} = \sum_{j,p,q} (-1)^j \bar D_{(B_{j+1}} \cdots \bar D_{B_q} A^{(0)CB_1 \dots B_q}_p[\bar \psi] \bar D_{B_1} \cdots \bar D_{B_j)} (\partial_v \partial_r)^p \frac{\partial}{\partial \lambda} \psi |_{\lambda=0}, 
\ee
We will now argue that \eqref{eq:sXYW} holds not just on $C$ but on any constant $v$ cross section $C(v)$ of the horizon.  To this end we note that evaluating the integrals over $C(v)$ instead of $C$ is equivalent to evaluating them over the variation $T_{v}^* \delta \psi$ instead of $\delta \psi$, where $T_v$ is a translation by $v$. This is because, on $\cN$, all boost weight zero quantities  are invariant under $T_v$ in the background $\bar \psi$.
But (\ref{eq:sXYW}) is an identity that holds for arbitrary variations $\delta \psi$ of compact support and in particular also for $T_{v}^* \delta \psi$. Hence we conclude that this equation holds on $C(v)$. Then we can take a derivative with respect to $v$ and note that this equation becomes
\be
\label{deltaQ}
\frac{\partial}{\partial \lambda}  \Big[\frac{1}{2\sqrt{\mu}} \partial_v (q^{(0)} \sqrt{\mu})  + X^{(0)} \partial_v Y^{(0)} \Big]_{\lambda=0} = \frac{\partial}{\partial \lambda}  ( D_A V^{(1)A})_{\lambda=0}
\ee
where
\be 
\label{V1}
V^{(1)C} = \sum_{j,p,q} (-1)^j D_{(B_{j+1}} \cdots D_{B_p} A^{(0)CB_1 \dots B_q}_p[\psi] D_{B_1} \cdots D_{B_j)} (\partial_v \partial_r)^p \partial_v  \psi.
\ee
Thus, in a GNC system, the first variation of $E_{vv}$ around a background $\bar \psi$ of the form described above can be written on $\cN$ as 
\be
\label{eq:Cvv1}
\frac{\partial}{\partial \lambda}  E_{vv}|_{\lambda=0}  = \frac{\partial}{\partial \lambda}  \Big( \partial_v \Big\{ \frac{1}{\sqrt{\mu}} \partial_v (\sqrt{\mu}s^v) + D_A s^A \Big\} \Big)_{\lambda=0}
\ee
where 
\be
\label{IWWentropy}
s^v := -\frac{1}{2} q^{(0)} + \sum_{j\ge 0,p-q > 0,\psi} X^{(-p+q)A_1 \dots A_j}_{q,\psi}  
D_{(A_1} \cdots D_{A_j)}  \partial_v^{p} \partial_r^{q} \psi
\ee
with $q^{(0)}$ as in \eqref{eq:ERvrvr},  $X^{(k)}_{q,\psi}$ as in \eqref{eq:ABdec} for $A^{(2)} = E_{vv}$, and where 
\be
s^A := W^{(1)A}-V^{(1)A}, 
\ee
with $V^{(1)}$ as in \eqref{V1} and $W^{(1)}$ as in \eqref{eq:ABdec} for $A^{(2)} = E_{vv}$. In deriving this result we assumed a perturbation of compact support. If we have instead a perturbation of non-compact support then, for any compact subset of $\cN$, we can define a perturbation of compact support that agrees with the original perturbation on this subset and therefore the above result holds for the original perturbation on this subset. Since the subset is arbitrary, it follows that this result holds on all of $\cN$ even for perturbations of non-compact support. 

If $\cN$ is a black hole event horizon then, when substituted into \eqref{Ssv}, the first term of \eqref{IWWentropy} gives the Iyer-Wald (IW) entropy \cite{Iyer:1994ys}. The other terms in \eqref{IWWentropy} are those predicted by the argument of Wall \cite{Wall:2015raa}. We will therefore refer to \eqref{IWWentropy} as the Iyer-Wald-Wall (IWW) entropy density. The need for the term in \eqref{eq:Cvv1} involving $s^A$ was pointed out in \cite{Bhattacharyya:2021jhr}. 

We will now investigate $E_{vv}$ at the fully non-linear level. Consider a metric $g_{\mu\nu}(x^\alpha)$ (not necessarily a solution) in GNCs with 
corresponding $\psi \in \{ \alpha, \beta_A, \mu_{AB} \}$, define $\bar \psi$ as the series in \eqref{background_def}, choose $\delta \psi$ to be $\psi - \bar \psi$ 
and define $g_{\mu\nu}(\lambda, x^\alpha)$ to be the 1-parameter family of metrics in GNCs corresponding to $\psi(\lambda)$
as in \eqref{background_def}. By construction, this family interpolates between the stationary metric $g|_{\lambda=0}$ and $g|_{\lambda=1}$ which is equal to 
the given metric $g$. Now let
\be
\label{eq:Cvvdecomp}
F[\psi] \equiv E_{vv} - \partial_v \Big\{ \frac{1}{\sqrt{\mu}} \partial_v (\sqrt{\mu}s^v) + D_A s^A \Big\} .
\ee
Using \eqref{eq:ABdec} and the expressions for $s^v$ and $s^A$ gives (recall definition \ref{def:2})
\be
\label{Fexp2}
 F \sim \partial_v \Big[\frac{1}{2\sqrt{\mu}} \partial_v (q^{(0)} \sqrt{\mu})  + X^{(0)} \partial_v Y^{(0)} - D_A V^{(1)A}\Big] 
\ee
We expand $F$ to all orders in $\lambda$ around the background $\lambda=0$. The zeroth order term must vanish on $\cN$ since all positive boost weight quantities vanish on $\cN$ in the background. The linear term must vanish on $\cN$ by \eqref{deltaQ} or \eqref{eq:Cvv1}. Hence $F$ contains only terms of quadratic or higher order in $\lambda$. 

If we consider a typical monomial in $F$ then it is a product of factors with different boost weights. If we evaluate this on $C$ then the zero boost weight factors coincide with those of $\bar\psi$. For a non-zero boost weight factor we have $D_{A_1} \ldots D_{A_j} \partial_v^p \partial_r^q \psi(\lambda)|_C = \lambda \bar{D}_{A_1} \ldots \bar{D}_{A_j} \partial_v^p \partial_r^q \delta \psi|_C$ if $p \ne q$. Thus the power of $\lambda$ in any given monomial equals the number of factors in that monomial that have non-zero boost weight. Since non-zero boost weight arises from $v$ or $r$ derivatives, the number of such factors will be bounded if the total number of derivatives in each monomial of $E_{vv}$ is bounded above (as is the case if we consider an EFT in which we include terms in the Lagrangian only up to some fixed total number of derivatives). In such a case, the expansion in $\lambda$ terminates on $C$. 

We will now show that $F \sim 0$. Note that the expression in square brackets in \eqref{Fexp2} involves no terms of boost weight higher than $1$. Therefore expanding out the $v$-derivative gives 
\be
F \sim \sum_{n,q,\psi} A_{np\psi}^{(0) A_1 \ldots A_n}[\psi] {D}_{(A_1} \ldots {D}_{A_n)} \partial_v^{q+2} \partial_r^q \psi 
\ee
where each $A^{(0)}$ is a sum of monomials within which each factor has zero boost weight. Quantities that are ignored in the $\sim$ relation involve monomials with at least two positive boost weight factors, and such terms are of quadratic order, or higher, in $\lambda$. Hence 
\be
F|_C= \lambda \sum_{n,q,\psi} A_{np\psi}^{(0) A_1 \ldots A_n}[\bar \psi] ({\bar D}_{(A_1} \ldots {\bar D}_{A_n)} \partial_v^{q+2} \partial_r^q \delta \psi)_C + O(\lambda^2)
\ee
However, as we have explained above, the terms in $F$ linear in $\lambda$ must cancel for {\it any} $\psi$. In particular, if we fix $\bar \psi$ and then construct $\psi$ by specifying $\delta \psi$ then the linear terms above must cancel. Since this $\delta \psi$ is arbitrary, this implies $A_{np\psi}^{(0) A_1 \ldots A_n}[\bar \psi] =0$ on $C$. Our choice of $\bar \psi$ here was arbitrary so this must hold for any $\bar\psi$. Now given $\psi$, let $\psi_v=T_v^* \psi$ where $T_{v_0}$ is the diffeomorphism corresponding to translation in $v$ by $v_0$. We then have 
\be
0=A_{np\psi}^{(0) A_1 \ldots A_n}[\overline{\psi_v}] |_C= A_{np\psi}^{(0) A_1 \ldots A_n}[\psi_v] |_C = A_{np\psi}^{(0) A_1 \ldots A_n}[\psi] |_{C(v)}
\ee
where $C(v)$ is a constant $v$ cut of $\cN$. But $v$ is arbitrary here so it follows that $A_{np\psi}^{(0) A_1 \ldots A_n}[\psi]$ vanishes on all of $\cN$. This implies $F \sim 0$ as claimed.

We summarize the discussion so far in the following lemma:

\begin{lemma} 
\label{lem:1}
Consider a null surface $\cN$ ruled by affinely parameterized null geodesics and a generally covariant 
Lagrangian $\bL$. Define a foliation $\{C(v)\}$ adapted to GNCs. Then we can define $s^v, s^A$ and $F$ satisfying (\ref{eq:Cvvdecomp}), where: 
\begin{itemize}
\item 
Each $s^v, s^A, F$ is a sum of primitive monomials.
\item 
Each primitive monomial in $s^v, s^A, F$ has boost weight $0,1,2,$ respectively.
\item 
Each primitive monomial in $F$ contains at least two primitive factors with strictly positive boost weight. 
\item 
$s^v$ contains those primitive monomials in $E^{rvrv}_R$
which are a product of primitive factors with zero boost weight (Iyer-Wald dynamical entropy), and all other primitive monomials in $s^v$ contain precisely one primitive factor with positive boost weight. $s^{A}$ consists of a sum of primitive monomials 
of boost weight 1, each of which has precisely one primitive factor of boost weight 1. 
\end{itemize}
\end{lemma}
This justifies equation \eqref{s_def}. The main novelty of our approach compared to earlier work \cite{Wall:2015raa,Bhattacharya:2019qal,Bhattacharyya:2021jhr} has been to demonstrate the connection to the Iyer-Wald entropy. Note that we could add extra terms to $s^v$ and $s^A$ that are of quadratic (or higher) order in primitive factors of positive boost weight without affecting the above results. This will be important below.

\subsection{Transformation of IWW entropy under a change of GNCs} 

\label{sec:gauge_inv}

The above procedure for determining $s^v$ and $s^A$ worked with the GNC components of the metric. Even though GNCs are defined geometrically starting from a cross section $C$ of $\cN$, their construction requires apart from the chosen cut $C$ a choice of affine parameter for each null generator of $\cN$. Under the remaining reparameterization freedom, $\alpha, \beta_A, \mu_{AB}$
and their derivatives in general transform in a very complicated way (e.g. equation \eqref{beta_trans}), so it is not clear that for a given cut $C$, $s^v$ is fully covariant, i.e. 
a functional of $g_{\mu\nu}$ that is independent of arbitrary choices apart from the cut $C$. Thus, we must investigate if, and how, $s^v$ transforms under a change of GNCs.

Consider a change of coordinates $x^\mu = x^\mu(x^{\prime \alpha})$ in a neighborhood of $\mathcal N$ preserving the Gaussian Null Form, see 
\eqref{affine1}. Thus, on $\cN$, we have $(v, x^A) = (a(x^{\prime C}) v', x^{\prime A})$, and off of $\cN$, the -- in general very complicated -- form of the coordinate
transformation is described in sec. \ref{sec:GNC}. The GNC components of $g$ in the primed coordinates are 
denoted by $\psi' \in \{\mu'_{A'B'}, \beta'_{A'}, \alpha'\}$. We want to understand the 
behavior of $s^v$ under such a transformation. If the multiplication factor $a$ in \eqref{affine1} is constant, then this 
behavior is trivially determined by the boost weight of each quantity, so we get $s^v[\psi] = s^v[\psi']$ ($a$ constant), because $s^v$
has boost weight 0. But for non-constant $a$, the transformation is not at all evident.  In this subsection, we shall show (Proposition \ref{prop:1}) that $s^v$ can be adjusted by terms 
quadratic in positive boost weight quantities, and a total divergence (which does not affect the total entropy), such that the modified $s^v$ is manifestly invariant under a change of GNCs. 

To investigate this, we first define $\bl = g_{\mu\nu} l^{\nu} \ud x^{\mu}$, and define, uniquely, an $(n-1)$ 
form $\bepsilon_l$ on $\cN$ by demanding that $\bl \wedge \bepsilon_l = \bepsilon$ and $n\cdot \bepsilon_l=0$. In our GNCs this gives 
$\bepsilon_l = \sqrt{\mu} \ud v \wedge \ud x^1 \ldots \wedge \ud x^{n-2}$. The constraint $(n-1)$ form $\bC_l$ is defined as in \eqref{C_K_def},
 giving $\bC_l = 2E_{vv} \bepsilon_l$. Similarly we define $\bF_l = F\bepsilon_l$. We now define the entropy current as a $(n-2)$ form $\bS$ on $\mathcal N$ by
\be
\begin{split}
\bS &:= (s^v \partial_v + s^A \partial_A) \cdot \bepsilon_l \\
&= s^v \sqrt{\mu} \ud x^1 \wedge \cdots \wedge \ud x^{n-2} + \sum_{A=1}^{n-2} (-1)^A s^A \sqrt{\mu} \ud v \wedge \ud x^1 \wedge \cdots \widehat{\ud x^A} \wedge \cdots  \ud x^{n-2}.
\end{split}
\ee
In forms notation, \eqref{eq:Cvvdecomp} can be rewritten as 
\be
\cL_l \ud \bS - \frac{1}{2}\bC_l = \bF_l \quad \text{on $\mathcal N$.}
\ee
In our second set of GNCs we have $l^{\prime \mu} = al^\mu$ on $\cN$ and so the primed version of the above equation gives
\be
\cL_{al} \ud \bS' - \frac{1}{2}\bC_{al} = \bF'_{al} \quad \text{on $\mathcal N$.}
\ee
In this formula, $\bS'$ is the same functional as $\bS$ but evaluated on the transformed $\psi' \in \{\mu'_{AB}, \beta'_A, \alpha'\}$ and in the 
transformed coordinates. $\bS'$ can be transformed back to the original coordinates in the usual way using the Jacobian of the coordinate transformation. 

Since $\bC_l$ is defined by contracting $l$ with a local, covariant tensor, we have $\bC_{al} = a\bC_l$. We also have $\cL_{al} \ud \bS' = \ud(al \cdot \ud \bS') = a \cL_l \ud \bS' + \ud a \wedge l\cdot \ud \bS'=a \cL_l \ud \bS'$ where the final equality uses $l\cdot \ud \bS' \propto \ud x^{\prime 1} \wedge \ldots \wedge \ud x^{\prime (n-2)}$ and $\ud a= (\partial_{A'}a) \ud x^{A'}$. Combining these results gives
\be
\label{LldS}
\cL_l \ud (\bS -  \bS') = \bF_l^{}- \frac{1}{a} \bF_{l'}'
\quad \text{on $\cN$}
\ee
We shall now show that this equation implies that the IWW entropy is gauge invariant. In our first set of GNCs, let $\tilde{\psi}$ arise from a ``background'' metric (not necessarily a solution) for which all positive boost weight primitive factors vanish on $\cN$. This need not be a background constructed as in the previous subsection (e.g. negative boost weight quantities need not vanish on $C$). The most interesting case is when $\tilde{\psi}$ corresponds to a stationary black hole solution with event horizon $\cN$. Now we perturb this background: let $\psi(\lambda)$ be a 1-parameter family such that $\psi(\lambda=0)=\tilde{\psi}$. Let the corresponding metric be $g_{\alpha\beta}(\lambda,x^\mu)$. 

Consider transforming this metric to the ``primed'' set of GNCs. These are defined by a function $a>0$ on the cut $C$. On $\cN$ the coordinate transformation does not depend on $\lambda$. However, since the definition of GNCs depends on the metric, the change of coordinates does depend on $\lambda$ away from $\cN$: $x^\mu = x^\mu(\lambda,x^{\prime \alpha})$. Under this coordinate transformation, we obtain $g=g_{\alpha\beta}'(\lambda,x^{\prime \mu}) \ud x^{\prime \alpha} \ud x^{\prime \beta}$,
corresponding to $\psi'(\lambda, x^{\prime \mu}) \in \{\mu_{A'B'}'(\lambda, x^{\prime \mu}), 
\beta_{A'}'(\lambda, x^{\prime \mu}), \alpha'(\lambda, x^{\prime \mu})\}$.
The expression for derivatives $D'_{(A_1} \cdots D'_{A_j)} \partial_{v'}^p \partial_{r'}^q \psi'(\lambda, x^{\prime \mu})$
in terms of $D_{(A_1} \cdots D_{A_j)} \partial_{v}^p \partial_{r}^q \psi(\lambda, x^{\mu})$ will in general be very complicated on $\cN$ (with explicit factors of $v$ but not $r$), 
but it follows from Lemma \ref{lem:bw_def} that a positive boost weight primitive factor always transforms to an expression only involving positive boost 
weight primitive factors times appropriate powers of $v$ (which has boost weight $-1$). Hence $\psi'(\lambda=0)$ has the property that positive boost weight primitive factors vanish on $\cN$. 
As a consequence, because $\bF_l^{}$ is at least quadratic in positive boost weight primitive factors, 
we obtain from \eqref{LldS}
\be
\cL_l \ud \frac{\partial}{\partial \lambda} (\bS - \bS') |_{\lambda = 0} =  0. 
\ee
Now assume that the perturbation $\partial \psi(\lambda, x^\mu)/\partial \lambda|_{\lambda=0}$ has compact support (we relax this below). Integrating up the above equation gives
\be
\ud \frac{\partial}{\partial \lambda} (\bS - \bS') |_{\lambda = 0} =  0.
\ee
We now use the algebraic Poincar\'e Lemma (lemma \ref{lem:poincare}) taking $\Psi_\lambda=\lambda (\partial_\lambda \psi|_{\lambda=0})$ and $\Phi=(\tilde{\psi},a)$. This gives
\be
\label{part1}
\frac{\partial}{\partial \lambda} (\bS - \bS') |_{\lambda = 0} = \ud \bb_{n-3}, 
\ee
where $\bb_{n-3}$ is a local form depending on $\partial_\lambda \psi|_{\lambda=0}$, $\tilde{\psi}$ and $a$. Next we investigate $(\bS - \bS') |_{\lambda = 0}$.
Because all positive boost quantities built from $\psi(\lambda=0, x^\mu)$ vanish on $\mathcal N$, we have $s^A|_{\lambda=0} = 0$ and
 $s^v |_{\lambda=0}= (1/2)q^{(0)}[\psi|_{\lambda=0}]$ which is proportional to $E^{rvrv}_R[\psi|_{\lambda=0}]$ (as $q^{(0)}$ only contains zero boost weight primitive factors),
Then, because $E^{\mu\nu\sigma\rho}_R$ is a covariant tensor we easily see 
\be
\label{part2}
(\bS - \bS') |_{\lambda = 0} = 0 \qquad {\text{on $\cN$}}
\ee
Combining \eqref{part1} and \eqref{part2} we find 
\be
\label{ds_change}
\bS - \bS' = \lambda \ud \bb_{n-3} + O(\lambda^2) \qquad {\text{on $\cN$}}
\ee
Integrating this equation over $C$ gives 
\be
S_{\rm IWW}[C]-S_{\rm IWW}'[C]=O(\lambda^2)
\ee
where the LHS refers to the IWW entropy defined w.r.t. the two different sets of GNCs. Hence we have shown that {\it the IWW entropy is gauge invariant to linear order} in perturbations around our background metric. Note that this is an off-shell result: neither the background nor the perturbation need satisfy any equations of motion. In deriving this result we assumed that the perturbation has compact support. For a non-compactly supported perturbation we can pick a perturbation of compact support that agrees with the given perturbation in a neighbourhood of $C$ and then apply the above result. 

We shall now extend this to a fully nonlinear result, eliminating any reference to the background metric. We assume that we have some given metric $g_{\alpha\beta}$ and a set of GNCs, with corresponding quantities $\psi$. We now {\it define} a background metric $\tilde{\psi}$ such that, on $C$, primitive factors of zero {\it or negative} boost weight agree with those of our original metric. We do this by setting
\be
\label{expansion1}
 \tilde{\psi} := \sum_{q \ge p} \frac{1}{p!q!} \chi(c_p v) \chi(c_q r)  v^p r^q \partial_v^p  \partial_r^q  \psi |_{r=v=0}
\ee
where $\chi$ is of compact support, with $\chi$ identically equal to $1$ in a neighbourhood of $0$, and where $c_p$ goes to infinity suitably rapidly so as to make the sums convergent. The corresponding metric $\tilde{g}_{\alpha\beta}$ is our background metric. We now define the one parameter family ($\lambda \in [0,1]$)
\be
\psi(\lambda) := \tilde{\psi} + \lambda \delta \psi, 
\ee 
where $\delta \psi$ is of compact support. We let $g_{\alpha\beta}(\lambda,x^\mu)$ denote the corresponding family of metrics in GNCs. We choose $\delta \psi$ such that $\delta \psi = \psi-\tilde{\psi}$ in a neighbourhood of $C$ so, on $C$,  all derivatives of $g_{\alpha\beta}(\lambda=1,x^\mu)$ coincide with those of the original metric. Since we will in the end be interested in the behavior of $s^v$
on the chosen $C$, we may therefore replace $g_{\alpha\beta}(x^\mu)$ with $g_{\alpha\beta}(\lambda=1,x^\mu)$. 

We now apply \eqref{ds_change} to $g_{\alpha\beta}(\lambda,x^\mu)$. By construction, the $O(\lambda^2)$ terms must on $C$ be at least quadratic in positive boost weight quantities. This is because each factor of $\lambda$ must be accompanied by $\delta \psi$, and on $C$ the only non-vanishing quantities linear in $\delta \psi$ are those with positive boost weight (e.g. $D^k \partial_v^p \partial_r^q \delta \psi|_C=0$ if $q \ge p$). Furthermore, in such quantities we can replace $\delta \psi$ with $\psi$ (e.g. $D^k \partial_v^p \partial_r^q \delta \psi|_C = D^k \partial_v^p \partial_r^q \psi|_C$ for $p >q$). Similarly, on $C$ we can write terms involving $\tilde{\psi}$ in terms of $\psi$ using the definition of $\tilde{\psi}$ (e.g. $D^k \partial_v^p \partial_r^q \tilde{\psi}|_C = D^k \partial_v^p \partial_r^q \psi|_C$ for $q \ge p$). Thus we can write everything in \eqref{ds_change} locally in terms of $\psi$ (and $a$): the dependence on the background has been eliminated.  

Now setting $\lambda=1$, so that \eqref{ds_change} is being evaluated for our original metric, we learn that there exists a local form $\bb_{n-3}[\psi,a]$ on $C$ such that $\bS - \bS' - \ud \bb_{n-3}$ is at least quadratic in positive boost weight quantities. We now dualize $\bb_{n-3}$ by setting 
$(\bb_{n-3})_{A_1 \dots A_{n-3}} = \epsilon[\mu]_{CA_1 \dots A_{n-3}} B^{vC}, (\bb_{n-3})_{vA_1 \dots A_{n-4}} = \tfrac{1}{2}\epsilon[\mu]_{CDA_1 \dots A_{n-4}} B^{CD}$. We then have, on $C$, 
\be
\label{trafo}
s^{v'} - s^{v} =  D_A B^{vA}+\ldots  \qquad 
\frac{1}{a} s^{A'} - s^A = \frac{1}{\sqrt{\mu}} \partial_v (\sqrt{\mu} B^{vA}) + D_C B^{AC} +\ldots
\ee
where the ellipses represent terms that, when expressed w.r.t. the first set of GNCs, are at least quadratic in positive boost weight primitive factors. These terms depend on $a$ in an extremely complicated way, but we will now show by a general argument using lemmas \ref{lem:0a}-\ref{lem:0d} that 
they can always be reabsorbed in a redefinition of $s^v$, thereby rendering this quantity invariant (up to total derivative) under a change of GNCs. 
For this purpose, it is useful to pass to an infinitesimal version of the transformation \eqref{trafo}, which in 
view of \eqref{cocycle} is fully equivalent to transformation law \eqref{trafo} under finite GNC coordinate transformations. 
Consider an $a$ of the form $e^{t\Lambda/2}$ and take a derivative of the above equation with respect to $t$ at $t=0$. Calling the corresponding 
infinitesimal transformation $\gamma$ (given in the basis of monomials of lemma \ref{lem:0b} concretely by \eqref{eq:homog2}), we see that, on $C$,
\be
\gamma s^v[\psi] = D_A X^A[\psi,\Lambda] + Y[\psi,\Lambda], 
\ee
where $X^A, Y$ are local and $Y$ is linear in $\Lambda$ and at least quadratic in positive boost weight factors. Taking $\gamma$ of this equation, we find a consistency condition on $Y$, which is, on $C$, 
\be
\gamma Y[\psi,\Lambda] = -D_A( \gamma X^A[\psi,\Lambda] ). 
\ee
We can use lemma \ref{lem:0d} to characterize the solutions to this cohomological problem, identifying $Y \sqrt{\mu} \ud^{n-2} x =: \bw_{n-2}$, 
$-\gamma (\star X_A \ud x^A) =: \bw_{n-3}$. Lemma \ref{lem:0d} made no assumption that $\bw_{n-2}$ be quadratic in positive boost weight factors, but the method for constructing the solutions to the cohomological problem given in the proof is constructive, and we can see from the proof that the solution provided by the lemma can be chosen to be at least quadratic in positive boost weight: Firstly, the forms in the descent equations constructed in the proof do not leave the space of local forms at least quadratic in positive boost weight factors and with overall boost weight zero. 
Indeed, the BRST operator $\gamma$ preserves this space, and applying the algebraic Poincare lemma to construct the subsequent rungs in the descent equations, we do not leave this space either, as we can see from its proof. Secondly, the arguments used to analyze the bottom rung of the descent equations which relies on the method of contractible pairs
can be carried out in that space, too, because the BRST operator $\gamma$, the exterior differential $\ud$ on $C$, and the operators $N,\rho$ [see \eqref{Ndef},\eqref{rhodef}] all commute with the boost weight number operator.\footnote{
That is, the operator
$
N_b =\sum_{N, \bar N, \psi} (N-\bar N) \partial_v^N \partial_r^{\bar N}
D_{(A_1} \cdots D_{A_j)} \psi \frac{\partial}{\partial(\partial_v^N \partial_r^{\bar N}
D_{(A_1} \cdots D_{A_j)} \psi)}.
$
}

Thus, lemma \ref{lem:0d} gives, on $C$, 
\be
Y[\psi,\Lambda] = \gamma Z[\psi] + D_A U^A[\psi,\Lambda] + \Lambda V[\psi], 
\ee
for local forms $Z,U^A,V$ at least quadratic in positive boost weight and $V$ invariant under changes of GNCs, and all quantities have overall boost weight zero. Therefore, on $C$,
\be
\gamma s^v[\psi] = \gamma Z[\psi] + D_A (X^A[\psi,\Lambda] + U^A[\psi,\Lambda]) + \Lambda V[\psi], 
\ee
which must hold for all $\Lambda, \psi$. The terms containing undifferentiated $\Lambda$ in the BRST transformation $\gamma$ \eqref{eq:homog2} must 
cancel in the above expression because $Z,s^v$ have boost weight 0, so the explicit term $\Lambda V[\psi]$ with $\Lambda$ in undifferentiated form 
must be cancelled by the total divergence term. Thus, $V$ has to be a total divergence of a local quantity of boost weight 0 which is at least quadratic in positive boost weight
monomials, $V = D_A W^A$. Therefore 
\be
\gamma s^v[\psi] = \gamma (Z[\psi] - \beta_A W^A[\psi]) + D_A (X^A[\psi,\Lambda] + U^A[\psi,\Lambda] + \Lambda W^A[\psi]). 
\ee
Now we redefine $s^v[\psi] \to s^v[\psi] - Z[\psi] + \beta_A W^A[\psi]$, which still satisfies \eqref{eq:Cvvdecomp}
for a new $F[\psi]$ that is at least quadratic in positive boost weight, and which satisfies additionally $\gamma s^v[\psi] = D_A (X^A[\psi,\Lambda] + U^A[\psi,\Lambda] + \Lambda W^A[\psi])$.
This already shows that our modified IWW entropy $\int_C s^v \sqrt{\mu} \ud^{n-2}x$ is gauge invariant. 

We can say more about the structure of the redefined $s^v$
using Lemmas \ref{lem:0c} and \ref{lem:char_classes} for $m=n-2$, setting $\bw_{n-2} = s^v \sqrt{\mu} \ud^{n-2}x$. The non-covariant terms $\bbeta \wedge \bI_{n-3}$
in Lemma \ref{lem:0c} involving the characteristic classes $\bI_{n-3}$ as described in Lemma
\ref{lem:char_classes} (which can only appear if $n$ is odd)
consist exclusively of monomials in the primitive basis such that each primitive factor has zero boost weight. Such terms are contained in the IW-part of the entropy -- i.e. the term $q^{(0)}$ in \eqref{IWWentropy} --
which in turn is given by that part the manifestly covariant expression $E_R^{rvrv}$ consisting only of zero boost weight terms. Thus, non-covariant terms $\bbeta \wedge \bI_{n-3}$ must in fact be absent in $s^v$ and we have shown:

\begin{proposition}
\label{prop:1}
(Invariance of modified IWW-entropy density) $s^v$ can be modified by terms at least quadratic in positive boost weight [so it still satisfies the properties claimed in lemma \ref{lem:1} and \eqref{eq:Cvvdecomp}] in such a way that 
it is a sum of: 
\begin{itemize}
\item
A local functional of boost weight 0 that is a contraction of the following factors: $\cD_{(A_1} \cdots \cD_{A_j} K_{A)B}^{}$, $\cD_{(A_1} \cdots \cD_{A_j} \bar K_{A)B}^{}$, 
GNC components of $\nabla_{(\alpha_1} \cdots \nabla_{\alpha_j)} R[g]_{\mu\nu\sigma\rho}$ [by the Bianchi-identities, not all these components are independent, see lemma \ref{lem:0b}],
$\mu^{AB}$, or $\epsilon[\mu]_{A_1\dots A_{n-2}}$. 
\item
A total divergence $D_A \xi^A$, where $\xi^A$ is a local functional of boost weight 0.
\end{itemize}
In particular, on $C$, $s^v$ is invariant up to at total divergence under a change of the affine parameter in GNCs. 
\end{proposition}

\noindent
{\bf Remark:} Proposition \ref{prop:1} uses that $\bL$ is locally and covariantly constructed out of the metric $g_{\mu\nu}$, meaning that it is a functional of $g_{\mu\nu}$, 
$\epsilon_{\mu_1 \dots \mu_n}$ and covariant derivatives of the Riemann tensor (in a purely gravitation theory). Hence Proposition 1 does not cover the case of gravitational 
Lagrangians that are covariant only up to $\ud$-exact terms like Chern-Simons terms in odd dimensions $n$, 
such as $\tr(\bGamma \wedge \ud \bGamma + \frac{2}{3} \bGamma \wedge  \bGamma \wedge  \bGamma)$ in $n=3$, 
where $\bGamma$ is the spin-connection associated with a 3-bein of $g_{\mu\nu}$. Such terms will result in terms of the type \eqref{top} in $s^v$. For example 
from a gravitational Chern-Simons term in $\bL$ in $n=5$, we can see from the formulas for Christoffel symbols in GNCs in Appendix \ref{app:GNC} that we 
would get a Chern-Simons term of the form $\bbeta \wedge \ud \bbeta$ in $s^v \sqrt{\mu} \ud^3 x$, consistent with Lemmas \ref{lem:0c} and \ref{lem:char_classes}.

On the other hand, fully covariant topological 
terms in $\bL$ are covered by our analysis. For example, in any even dimension an Euler density in $\bL$, i.e. 
$\bolde_n[g]:= \bR[g]^{\mu_1\mu_2} \wedge \dots \wedge \bR[g]^{\mu_{n-1}\mu_n} \epsilon_{\mu_1 \dots \mu_n}$ will result in an Euler density term $\bolde_{n-2}[\mu]$ in $s^v \sqrt{\mu} \ud^{n-2} x$, 
whereas a Chern class in $\bL$, i.e. $\bc_n[g]:= \bR[g]^{\mu_1}{}_{\mu_2} \wedge \dots \wedge \bR[g]^{\mu_{n-1}}{}_{\mu_1}$ will result in a term of the form 
$\bc_{n-4}[\mu] \wedge \ud \bbeta$ in $s^v \sqrt{\mu} \ud^{n-2} x$. (Recall that $\ud\bbeta$ can be eliminated in favour of Riemann components using \eqref{twist}.) All of these terms are built from boost weight zero quantities and are already present in the IW-part of $s^v$.

\section{Improved entropy current} 

\label{sec:improved_entropy}

\subsection{Vacuum gravity}

\label{sec:vacuum_gravity}

For simplicity we shall consider vacuum gravity in this subsection, i.e., no matter fields. (In the next subsection we shall consider gravity coupled to a scalar field.) If the spacetime dimensionality is odd then we assume parity symmetry, again for simplicity. Relaxing this assumption should be straightforward. We assume the Lagrangian takes the form\footnote{
In this section we no longer use $n$ to denote spacetime dimension. 
} 
\be
\label{eq:expansion}
\bL = \left[ -2\Lambda + R + \sum_{n \ge 2} \ell^n L_n \right] \bepsilon
\ee
with a single UV length scale $\ell$. $L_n$ is a local covariant scalar, depending only on the metric and its derivatives, and of dimension $n+2$, i.e., it involves $n+2$ derivatives of the metric. We define the dimension of $\ell$ to be $-1$ and the dimension of $\Lambda$ to be $+2$. Each derivative $\partial_\mu$ carries one index, and $g_{\mu\nu}$ has two indices, so non-zero $L_n$ is possible only for even $n$.\footnote{
In an even number of dimensions, there may be dependence on the volume form (for a parity violating theory) but this has an even number of indices, which does not affect the argument. In an odd number of dimensions our assumption of parity symmetry excludes dependence on the volume form.
}  
It will be convenient in this section to imagine that the $L_n$ are known for all $n$. It is then not hard to obtain results for which only finitely many of the $L_n$ are known, as we will discuss at the end of this section. 

The Einstein equation has the form
\be
\label{ELEFT}
 R_{\mu\nu}-\frac{1}{2}Rg_{\mu\nu} +\Lambda g_{\mu\nu} = \sum_{n \ge 2} \ell^n H_{n \, \mu\nu}
 \ee
where again only even $n$ appears in the sum and $H_{n \, \mu\nu}$ has dimension $n+2$.

Recall from section \ref{sec:EFT_2nd_law} that we are not interested in all solutions of \eqref{ELEFT}. Instead, we only consider solutions that fall within the regime of validity of EFT. This means that we restrict attention to solutions that, in some set of GNCs near $\cN$, are slowly varying compared to the scale $\ell$, i.e., $\ell/L \ll 1$ where $L$ is the length scale of variation of the fields. In the case of a dynamical black hole, $L$ will be the minimum of the size of the black hole and any length/time scales associated with the dynamics. Writing this out formally we have

\begin{definition}[Validity of EFT condition.] 
\label{def:EFT_valid}
We consider a 1-parameter family of smooth solutions of \eqref{ELEFT} with parameter $L$, such that $\cN$ is a smooth null hypersurface for all members of the family. We assume that there exist GNCs defined near $\cN$ such that, near $\cN$, if $T_n$ is a quantity of dimension $n$ (in the sense of Def. \ref{def:dim}) constructed from $\{\alpha,\beta_A,\mu_{AB}\}$ and their derivatives then there is a dimensionless constant $C_n$ (independent of $L$) such that $|T_n| \le C_n/L^n$. We also require $|\Lambda| L^2 \le 1$. Then EFT is valid for sufficiently small $\ell/L$ (i.e. large enough $L$).
\end{definition}

For example, on the RHS of \eqref{ELEFT}, $|H_{n\, \mu\nu}|$ is bounded above by $C_n/L^{n+2}$ so $\ell^n H_{n\, \mu\nu}={\cal O}(\ell^n/L^{n+2})$. As in section \ref{sec:EFT_2nd_law}, from now on we shall not indicate the $L$-dependence explicitly below so we would write $\ell^n H_{n\mu\nu}= {\cal O}(\ell^n)$. Factors of $L$ can be reinstated by dimensional analysis.

Recall the definition of $F$ in equation (\ref{eq:Cvvdecomp}). We can decompose this into a sum of terms arising from the various terms in \eqref{ELEFT}
\be
\label{exp:F}
F = F_0 + \sum_{n \ge 2} \ell^n F_n
\ee
where $F_n$ has dimension $n+2$ and has the same general structure as described in lemma \ref{lem:1}. As above, only even $n$ can appear in the sum. This is because only even powers of $\ell$ appear in the Einstein equation. 

$F_0$ is the expression coming from the Einstein-Hilbert part of the Lagrangian and reads 
\be
F_0=-K_{AB}K^{AB},
\ee
In the EFT spirit, we might expect $F_0$, which is sign definite, to dominate the higher order terms in $\ell$. This would require in particular that if $K_{AB}$ vanishes, then so should $F_n, n>0$. However, this is not true in general. Nevertheless, we can still make progress with this idea. To do this we need to understand the structure of the terms $F_n$. We start by using the equation of motion \eqref{ELEFT} to obtain the following result:

\begin{lemma}
\label{lem:elim_ric}
On-shell, $F$ can be written as in \eqref{exp:F} where each $F_n$ is a polynomial in $D_{(A_1} \cdots D_{A_j)} \partial_v^N K_{AB}$, $D_{(A_1} \cdots D_{A_j)} \partial_r^{\bar N} \bar K_{AB}$, $D_{(A_1} \cdots D_{A_j)} \beta_B$, $D_{(A_1} \cdots D_{A_j)} R[\mu]_{ABCD}$, $\mu^{AB}$, $\epsilon[\mu]_{A_1 \dots A_{n-2}}$ and $\Lambda$. Each term in this polynomial contains at most $n+2$ derivatives.
\end{lemma}

\noindent {\it Proof}. By lemma \ref{lem:0a}, we can write $F$ in terms of $D_{(A_1} \cdots D_{A_j)} \partial_v^N K_{AB}, D_{(A_1} \cdots D_{A_j)} \partial_r^{\bar N} \bar K_{AB}$, 
$D_{(A_1} \cdots D_{A_j)} \beta_B$, $D_{(A_1} \cdots D_{A_j)} R[\mu]_{ABCD}$, $\mu^{AB}$, $\epsilon[\mu]_{A_1 \dots A_{n-2}}$, and GNC components of covariant derivatives of the Ricci tensor. The idea now is to use the Einstein equation \eqref{ELEFT} to eliminate the terms involving the Ricci tensor. In doing this we replace each occurrence of a Ricci tensor either with a multiple of $\Lambda g_{\mu\nu}$ or with powers of $\ell^2$ times curvature terms of higher dimensions. Then we decompose the GNC components of the latter terms again as in lemma \ref{lem:0a}, eliminate any occurrence of a Ricci tensor again and so forth. At each step, any Ricci terms arise with explicit factors of $\ell^2$ and therefore get pushed to higher order in $\ell$. The end result is that, on-shell, $F$ can be written in terms of the quantities stated in the lemma, up to terms vanishing to infinite order in $\ell$. Finally we can express this on-shell result in the form \eqref{exp:F} where each $F_n$ can be written in terms of the quantities just listed, i.e., explicit occurrence of Ricci tensors and their covariant derivatives have been eliminated. 

Since $F_n$ has dimension $n+2$, each term in $F_n$ contains at most $n+2$ derivatives. Terms with fewer than $n+2$ derivatives must have enough factors of $\Lambda$ to make the dimension up to $n+2$. Q.E.D.

So far we assumed that our EFT is defined to all orders in $n$. If only terms with $n \le N-2$ are known then the Einstein equation will contain  unknown ``errors terms'' of order $\ell^N$ (equation \eqref{Einstein_EFT}). The off-shell expression for $F$ will have a similar form. The above argument still works and any terms of order $\ell^N$ or higher can be absorbed into the error term in $F$. The on-shell expression for $F$ will be a sum  of terms $F_n$ with $n \le N-2$, each of the form just described, and an error term of order $\ell^N$. 

We will now argue that the on-shell structure of $F$ can be further rearranged in a ``nice'' way. The argument involves induction on the order in $\ell$: 

\paragraph{Induction hypothesis.}

For each even $n \ge 0$, there are local tensors $X_{n AB}^{}, \varsigma_n^v$ having boost weights $(1,0)$
and local tensors $y_n^A, O_{n}$, each depending polynomially on the quantities listed in lemma~\ref{lem:elim_ric}, such that on-shell we have
\begin{equation}
  \begin{split}
    F=& -(K_{AB} + X_{nAB})(K^{AB} + X^{AB}_n) - \sum_{j=0}^n \partial_v \left[ \frac{1}{\sqrt{\mu}} \partial_v \left( \sqrt{\mu} \varsigma^v_j \right) \right] - {\sum_{j=2}^n D_A y^A_j} +O_{n+2}^{},
  \end{split}
\end{equation}
such that {$X_{n AB}$ is symmetric, $O_{n+2}, D_A y^A_j$, $\varsigma_n^v$} has the same general structure as $F$ described in lemma~\ref{lem:1}
(in particular they are at least quadratic in positive boost weight quantities) and
\begin{equation}
  \label{eq:EFT}
  O_{n+2}={\cal O}(\ell^{n+2})
\end{equation}
as well as $X_{nAB}={\cal O}(\ell^2)$. Furthermore $\varsigma^v_j$ and $y^A_j$ depend explicitly on $\ell$ only through an overall factor $\ell^j$.

\subparagraph{Induction start ($n=0$).}

In this case, $X_{0AB} = y_0^A = \varsigma^v_{0} = 0$ and $O_2 = \sum_{n \ge 2} \ell^{n} F_n={\cal O}(\ell^2)$ so the induction hypothesis is true for $n=0$.

\subparagraph{Induction step ($n-2 \to n, n \ge 2$).}

Consider $O_n$ written in terms of the quantities described in lemma~\ref{lem:elim_ric}. The induction hypothesis gives $O_n={\cal O}(\ell^n)$ and that $O_n$ is a sum of terms where each term is at least quadratic in terms of positive boost weight, and so must contain at least two factors of the form $D_{(A_1} \cdots D_{A_j)} \partial_v^{p-1} K_{AB}$ ($p \ge 1$) (as other possible factors listed in lemma~\ref{lem:elim_ric} have non-positive boost weight). In terms of $\mu_{AB}$ this means that $O_n$ must take the schematic form
\begin{equation}
  \label{eq:interm1}
  O_n = \sum_{k,k',p,p'} A_{n,k,p,k',p'} D^{k} \partial_v^{p} \mu D^{k'} \partial_v^{p'} \mu + O_{n+2}
\end{equation}
where $p,p'>0$ in all terms, the coefficient $A_{n,k,p,k',p'}$ has boost weight $2-p-p'$ and, aside from an overall factor of $\ell^n$, depends only on $\beta$, $D^k\beta$, $\Lambda$, $\mu$ and derivatives of $\mu$. $O_{n+2}$ satisfies~(\ref{eq:EFT}). Here and in the following, tensor indices $A,B,C,\ldots$ are suppressed.
The total number of derivatives in each term of the above sum is at most $n+2$. Our aim is to show that the above expression for $O_n$ can be rearranged to ensure that the induction hypothesis is true for $n$.

The next sequence of steps is designed to bring each term in the sum in~(\ref{eq:interm1}) into the form
$A_{n,k,p} D^{k} \partial_v^{p} \mu \partial_v \mu$, i.e.\ linear in $\partial_v \mu$ and at least quadratic in positive boost weight quantities,
for a new $A_{n,k,p}$, possibly at the expense of
further terms to be absorbed in $O_{n+2}$, or terms involving $y^A_n, \varsigma^v_n$
as in the induction hypothesis. {This is done by moving, one by one, the derivatives $D^{k'}$ from the factor $D^{k'} \partial_v^{p'} \mu$ over to the other terms as if we were performing a partial integration, but keeping the ``boundary terms''. These boundary terms are then absorbed in $y^A_{n}$ as in the induction hypothesis.} As a result,
\begin{equation}
  O_n =
  \sum_{k,p,p'} A_{n,k,p,p'} D^{k} \partial_v^{p} \mu \partial_v^{p'} \mu + D_A y^A_{n} + O_{n+2},
\end{equation}
with the summation indices subject to $p,p' > 0$.
Next, we would like to bring $p'-1$ of the $v$-derivatives on $\partial_v^{p'} \mu$ in the sum over to the other terms, leaving us finally with
expressions that are linear in $\partial_v \mu$ and quadratic in positive boost weight. This is done by a similar ``partial integration'', moving
all boundary terms into a new quantity $\varsigma^v_n$, as in the induction hypothesis, so in particular quadratic in positive boost weight quantities. \footnote{\label{foot_v_deriv} {Note that we might encounter $\partial_v$ acting on a term $D^j\beta_A$ or $D^j \partial_r^q \mu$ contained in $A_{n,k,p,k',p'}$. We deal with such terms as in the proof of lemma~\ref{lem:elim_ric}, i.e., converting them to the basis of lemma~\ref{lem:0a} and using the Einstein equation to eliminate any Ricci terms by expressing them in terms of $\Lambda$ and quantities of order $\ell^2$: the latter terms can be absorbed into $O_{n+2}$ (as $A_{n,k,p,k',p'}$ already contains a factor $\ell^n$).}}
We can do this by an induction on $p'+p$ on the terms in the sum, lowering at each step this counter by at least one until we reach
terms of the form $A_{n,k,p} D^{k} \partial_v^{p} \mu \partial_v \mu$ or $A_{n,k,p} D^{k} \partial_v \mu \partial_v^p \mu$ (where $p>0$ in each case). The latter term then can be rewritten in the same form as the former by the same ``partial integration'' w.r.t.\ $D$ argument that we just used, which generates further additions to $y^A_n$.
To see how this induction works in more detail, note that the induction hypothesis is true for $p+p' \le 3$ (since then either $p=1$ or $p'=1$) so assume $p+p'>3$. We claim that there exist numbers $a_j$ such that
\begin{align}
  \partial_v \left[ \frac{1}{\sqrt{\mu}} \partial_v \left( \sqrt{\mu}A_{n,k,p,p'} \sum_{j=1}^{p'+p-3} a_j
    D^{k} \partial_v^{j} \mu \partial_v^{p+p'-2-j} \mu \right) \right] = A_{n,k,p,p'} D^{k} \partial_v^{p} \mu \partial_v^{p'} \mu
  + \dots,
  \nonumber\\*
  \label{eq:distr_dv}
\end{align}
where the ellipsis represent any terms of the form $A_{n,\bar k,\bar p,\bar k',\bar p'} D^{\bar k} \partial_v^{\bar p} \mu \partial_v^{\bar p'} \mu$
having $\bar p + \bar p' < p + p'$, $\bar p, \bar p' >0$ (which can be dealt with inductively), or terms having $\bar p + \bar p' = p + p'$, $\bar p, \bar p' >0$, but $\bar p'=1$ or $\bar p=1$. We will justify this claim by showing that in order for the terms represented by the ellipsis to have the desired property, the $a_j$'s must satisfy a linear system of algebraic equations that has a unique solution. Expanding out the l.h.s.\ of~\eqref{eq:distr_dv}, we find
\begin{eqnarray}
  \text{l.h.s.\ of~\eqref{eq:distr_dv}}&=&A_{n,k,p,p'}\left[\vphantom{\sum\limits_{j=3}^{p'+p-3}}a_1 D^k \partial_v \mu \partial_v^{p'+p-1}\mu+(a_2+2a_1)D^k\partial_v^2\mu \partial_v^{p'+p-2}\mu
    \right.\kern-\nulldelimiterspace
    \nonumber\\
    & &\left.\kern-\nulldelimiterspace
    \qquad \qquad+\sum\limits_{j=3}^{p'+p-3}(a_j+2a_{j-1}+a_{j-2})D^k \partial_v^j \mu \partial_v^{p'+p-j}\mu
    \right.\kern-\nulldelimiterspace
    \nonumber\\
    & &\left.\kern-\nulldelimiterspace
    \qquad \qquad+(2a_{p'+p-3}+a_{p'+p-4})D^k \partial_v^{p'+p-2} \mu \partial_v^{2}\mu
    \right.\kern-\nulldelimiterspace
    \nonumber\\
    & &\left.\kern-\nulldelimiterspace\vphantom{\sum\limits_{j=3}^{p'+p-3}}
    \qquad \qquad+a_{p'+p-3}D^k \partial_v^{p'+p-1} \mu \partial_v\mu\right]+\dots .
\end{eqnarray}
{Any} $v$-derivatives of $D^j \beta$ or $D^j \partial_r^q \mu$ (arising from $v$-derivatives of $A_{n,k,p,p'}$) are dealt with as explained in footnote \ref{foot_v_deriv}, and contribute to the terms of the form $A_{n,\bar k,\bar p,\bar k',\bar p'} D^{\bar k} \partial_v^{\bar p} \mu \partial_v^{\bar p'} \mu$ with $\bar p + \bar p' < p + p'$ which are denoted by the ellipsis. (The $v$-derivatives of $\sqrt{\mu}$ also generate such terms, and further such terms arise from commutators of $D$ and $\partial_v$.) Now we require~that
\begin{eqnarray}
  \text{l.h.s.\ of~\eqref{eq:distr_dv}}&=&A_{n,k,p,p'}\Bigl[a_1 D^k \partial_v \mu \partial_v^{p'+p-1}\mu+D^k\partial_v^p\mu \partial_v^{p'}\mu
    \nonumber\\
    & &\qquad \qquad+a_{p'+p-3}D^k \partial_v^{p'+p-1} \mu \partial_v\mu\Bigr]+\dots
  \,. \label{eq:req_lin}
\end{eqnarray}
Then the $p'+p-3$ coefficients $a_j$ must satisfy the following $p'+p-3$ linear equations
\begin{eqnarray}
  a_2+2a_1&=&0,
  \nonumber\\
  a_j+2a_{j-1}+a_{j-2}&=&0, \qquad \qquad j\neq p
  \nonumber\\
  a_p+2a_{p-1}+a_{p-2}&=&1,
  \label{eq:aj_lin_sys}
  \\
  2a_{p'+p-3}+a_{p'+p-4}&=& 0\,.
  \nonumber
\end{eqnarray}
The coefficient matrix of this linear system has the form
\begin{equation}
  M_{p'+p-3}=\left(
  \begin{array}{cccccc}
    2&1&0&0&\dots &0
    \\
    1&2&1&0&\dots &0
    \\
    0&1&2&1&\dots &0
    \\
    \dots &\dots &\dots &\dots &\dots &\dots
    \\
    0& \dots & \dots & 0&1&2
  \end{array}
  \right) .
\end{equation}
It is easy to check that such a matrix has a nonzero determinant. In fact, the determinant of an $N\times N$ matrix of this form is $N+1$ which is found by induction on $N$: expanding the determinant along the first row of the matrix gives the simple recurrence relation $\det M_{N+1}=2\det M_N-\det M_{N-1}$. Therefore, $M_{p'+p-3}$ is invertible and the linear system above has a unique solution for the coefficients $a_j$ so we can indeed rewrite terms of the form $A_{n,k,p,p'} D^{k} \partial_v^{p} \mu \partial_v^{p'} \mu$ as in~\eqref{eq:distr_dv} {(with the first and last terms in square brackets in~\eqref{eq:req_lin} contributing to the ellipsis in~\eqref{eq:distr_dv})}. The terms $a_j A_{n,k,p,p'} D^{k} \partial_v^{j} \mu \partial_v^{p+p'-2-j} \mu$ are absorbed in $\varsigma^v_{n}$. We repeat the procedure until we have obtained
\begin{equation}
  \label{eq:interm3}
  O_n =
  \sum_{k,p} A_{n,k,p} D^{k} \partial_v^{p} \mu \partial_v \mu
  + \partial_v \left[ \frac{1}{\sqrt{\mu}} \partial_v \left( \sqrt{\mu} \varsigma^v_{n} \right) \right] + D_A y^A_{n} + O_{n+2},
\end{equation}
where $p>0$. Since $A_{n,k,p,k',p'}$ depend only on $D^k\beta$ and derivatives of $\mu$, the quantity $\varsigma^v_{n}$ have the same dependency on $\mu$ and $\beta$ as in the induction hypothesis.

The first term in~\eqref{eq:interm3} depends linearly on $K_{AB}$ so, reinstating indices, we can write it as $-2K^{AB} \Delta X_{AB}$ where $\Delta X_{AB}$ is {symmetric in $AB$ and} of order $\ell^n$ because all of the terms above, except $O_{n+2}$, depend on $\ell$ only via an overall factor of $\ell^n$. Recalling our original induction hypothesis we can now ``recomplete the square'' on $K_{AB}$ as follows:
\begin{equation}
        {-(K_{AB} + X_{n-2\,AB})(K^{AB} + X_{n-2}^{AB})} -2 K^{AB}\Delta X_{AB} = -(K_{AB} + X_{nAB})(K^{AB} + X^{AB}_n)+{\cal O}(\ell^{n+2})
\end{equation}
where $X_n \equiv X_{n-2} + \Delta X$ and we used $X_{n-2}={\cal O}(\ell^2)$ (from our induction hypothesis). The error term above can be absorbed into $O_{n+2}$. This closes the induction loop.

The inductive structure of the term $F$ can be combined with lemma~\ref{lem:1} to get the following central result of this section, proposition~\ref{prop:2}. We set, for each given even $n \ge 2$
\begin{equation}
  S^v_n := s^v + \sum_{j=2}^n \varsigma^v_j, \quad S^A_n := s^A, \quad Y^A_n = \sum_{{j=2}}^n {y^A_j}
\end{equation}
and otherwise keep the same notations as in the inductive structure.
\begin{proposition}
  \label{prop:2}
  Consider a generally covariant EFT Lagrangian of the form~\eqref{eq:expansion} and a $1$-parameter family of smooth solutions (with parameter $L$) satisfying our validity of EFT condition (definition~\ref{def:EFT_valid}) with a smooth null hypersurface $\cN$ ruled by affinely parameterized null geodesics. Then in the GNCs of definition~\ref{def:EFT_valid} on $\cN$ we have the (on-shell) equation
  \begin{align}
    \partial_v \left[ \frac{1}{\sqrt{\mu}} \partial_v \left( \sqrt{\mu} S^v_n \right) + D_A S^A_n \right]
    = -(K_{AB} + X_{nAB})(K^{AB} + X^{AB}_n) - D_A Y^A_n +O_{n+2}^{},
    \nonumber\\*
  \end{align}
  such that $S^v_n$, $S^A_n$, $X_{nAB}$, $Y_n^A$, $O_{n+2}$ have the same general structure as follows from lemma~\ref{lem:1} and the induction hypothesis above.
  In particular, $O_{n+2} ={\cal O}(\ell^{n+2})$, {$X_{nAB}={\cal O}(\ell^2)$ is symmetric}, $Y_n^A={\cal O}(\ell^2)$.
\end{proposition}

As discussed in section~\ref{sec:EFT_2nd_law}, in practice the EFT Lagrangian will not be known to all orders, instead only the terms with $n \le N-2$ will be known, for some $N \ge 2$. The equation of motion will take the form~\eqref{Einstein_EFT}. In this case we could repeat all of the above arguments, now with lemma~\ref{lem:elim_ric} only determining $F$ up to unknown terms of order $\ell^N$ and the above induction stopping when we reach $n=N-2$. Equivalently we could simply appeal to the above results applied to the full EFT, including the unknown terms with $n \ge N$, and note that these unknown terms will only affect the final answer at ${\cal O}(\ell^N)$. Either way, we obtain equation~\eqref{gen_Ray} by setting $n=N-2$ in the above proposition (and dropping the $n$ index on $S^v$, $S^A$, $X_{AB}$ and $Y^A$).

We can now define the entropy by equation~\eqref{SSv2} and this will obey the second law to quadratic order, in the sense of EFT, as explained in section~\ref{sec:EFT_2nd_law}.

\subsection{Scalar-tensor effective field theories}
\label{sec:scalar_tensor_all_orders}

Now we briefly discuss how the previous construction (valid for vacuum gravity) is modified when we include the simplest form of matter: a real scalar field. We again exclude the case of parity violating theories in odd spacetime dimension. The Lagrangian for a scalar-tensor EFT takes the form
\begin{equation}
  \label{eq:st_eft_lagr}
  \bL = \left[ R-V(\phi)+X + \sum_{n \ge 2} \ell^n L_n \right] \bepsilon
\end{equation}
where the scalar potential $V(\phi)$
has dimension $+2$ and the scalar kinetic term is given by $X\equiv -\tfrac12 (\partial\phi)^2$. Here, we assumed again that there is a single UV length scale $\ell$ (of dimension $-1$) suppressing the terms $L_n$ which are therefore subleading corrections to Einstein's theory with a minimally coupled scalar. We further assume that $L_n$ are diffeomorphism covariant scalars (of dimension $n+2$) that are locally constructed out of the metric $g$, the scalar field $\phi$ and derivatives of these fields. The fact that $L_n$ is a local covariant scalar implies that $n$ must be even. Moreover, in each monomial in $L_n$ the number of derivatives acting on the metric and the scalar field must be $n+2$ in total.

The gravitational and scalar equations of motion have the form
\begin{eqnarray}
  \label{eq:gr_eom_st_eft}
  R_{\mu\nu}-\frac{1}{2}Rg_{\mu\nu} - \frac{1}{2} \partial_\mu \phi \partial_\nu \phi - \frac{1}{2}(X-V) g_{\mu\nu} &=& \sum_{n \ge 2} \ell^n H_{n \, \mu\nu}
  \\[\jot]
  \label{eq:sc_eom_st_eft}
  \Box\,\phi - V'(\phi)&=& \sum_{n \ge 2} \ell^n I_{n}
\end{eqnarray}
where $n$ runs over even numbers and $H_{n \, \mu\nu}$ and $I_n$ are local covariant tensors of dimension $n+2$ constructed out of the metric and the scalar field.

Validity of the EFT requires that we only consider $1$-parameter families of smooth solutions to~\eqref{eq:gr_eom_st_eft}--\eqref{eq:sc_eom_st_eft} that satisfy the conditions of definition~\ref{def:EFT_valid}, with the additions that any tensor field $T_n$ of dimension $n$ locally constructed out of $\alpha$, $\beta_A$, $\mu_{AB}$, $\phi$ and their derivatives must satisfy a bound $|T_n|\leq C_n/L^n$. Note in particular that this means that the scalar potential $V(\phi)$ must obey $L^2 d^kV/d\phi^k\leq 1$ for any $k\geq 0$.

Now we are in the position to state the corresponding result on the existence of an entropy current for theories of the form~\eqref{eq:st_eft_lagr}.
\begin{proposition}
  \label{prop:st}
  \hskip-0.8355pt
  Consider a diffeomorphism-covariant scalar-tensor EFT with a Lagrangian given by~\eqref{eq:st_eft_lagr} and a $1$-parameter family of smooth solutions (with parameter $L$) of this theory. Suppose that the $1$-parameter family of solutions fall within the regime of validity of the EFT (in the sense of definition~\ref{def:EFT_valid} and the addition provided above) and the solutions possess a smooth null hypersurface $\cN$ ruled by affinely parameterized null geodesics (for any $L$). Then in the GNCs of definition~\ref{def:EFT_valid} on $\cN$ there exist quantities $S^v_n, S^A_n, X_{nAB}, P_n, Y_n^A, O_{n+2}$ with the following properties:
  \begin{itemize}
  \item[(i)]
    they satisfy an on-shell relation of the form
    \begin{eqnarray}
      \partial_v \left[ \frac{1}{\sqrt{\mu}} \partial_v \left( \sqrt{\mu} S^v_n \right) + D_A S^A_n \right]
      &=& -(K_{AB} + X_{nAB})(K^{AB} + X^{AB}_n)+
      \nonumber\\[\jot]
      & &-\frac12(\partial_v\phi+P_n)^2 - D_A Y^A_n +O_{n+2}^{};
      \label{eq:st_prop}
    \end{eqnarray}
  \item[(ii)]
    \sloppy{each of $S^v_n$, $S^A_n$, $X_{nAB}$, $P_n$, $Y_n^A$, $O_{n+2}$ is a sum of monomials such that each monomial is a product of factors of $D_{(A_1} \!\cdots D_{A_j)} \partial_v^N K_{AB}$, $D_{(A_1} \!\cdots D_{A_j)} \partial_r^{\bar N} \bar K_{AB}$, $D_{(A_1} \!\cdots D_{A_j)} \partial_v^N \phi$, $D_{(A_1} \!\cdots D_{A_j)} \partial_r^{\bar N} \phi$, $D_{(A_1} \!\cdots D_{A_j)} \beta_B$, $\mu^{AB}$, $D_{(A_1} \!\cdots D_{A_j)} R[\mu]_{ABCD}$, $\epsilon[\mu]_{A_1 \dots A_{n-2}}$ and $d^kV/d\phi^k$;}
  \item[(iii)]
    $S^v_n$, $S^A_n$, $X_{nAB}$, $P_n$, $Y_n^A$, $O_{n+2}$ have the same general structure as follows from lemma~\ref{lem:1}. In particular, $O_{n+2} ={\cal O}(\ell^{n+2})$, $X_{nAB},P_n, Y_n^A={\cal O}(\ell^2)$.
  \end{itemize}
\end{proposition}

To prove this statement, one can follow the algorithm devised for vacuum gravity with minor modifications. Hence, to avoid repetition, we merely highlight the key differences in this section rather than giving a complete proof.

We start by noting that a primitive factor $D_{(A_1} \cdots D_{A_j)} \partial_v^p\partial_r^{q} \phi$ with $p,q\geq 1$ can be written on $\cN$ in terms of the factors of $\mu^{AB}$, $D_{(A_1} \cdots D_{A_j)} \partial_v^N K_{AB}$, $D_{(A_1} \cdots D_{A_j)} \partial_r^{\bar N} \bar K_{AB}$, $D_{(A_1} \cdots D_{A_j)} \partial_v^N \phi$, $D_{(A_1} \cdots D_{A_j)} \partial_r^{\bar N} \phi$, $D_{(A_1} \cdots D_{A_j)} \beta_B$, $D_{(A_1} \cdots D_{A_j)} R[\mu]_{ABCD}$, and covariant quantities of the form $\nabla_{(\alpha_1}\ldots \nabla_{\alpha_j)}\Box \,\phi$, $\nabla_{(\alpha_1}\ldots \nabla_{\alpha_j)}R_{\mu\nu}$ so that the expression has polynomial dependence on each of these quantities. To prove this statement, one starts with the expression of $\Box \,\phi$ in GNCs:
\begin{equation}
  \Box \,\phi=2\partial_r\partial_v\phi+\mu^{AB}D_A D_B\phi+K\partial_r\phi+{\bar K}\partial_v\phi+\beta^A D_A\phi+\ldots
\end{equation}
where the ellipsis stands for terms vanishing at $r=0$. Taking derivatives of this identity and arguing inductively as in the proof of lemma~\ref{lem:0b} lets us eliminate all mixed $r-v$ derivatives of $\phi$, in favour of the quantities listed above, establishing our claim.

The next step is to apply this result to $F$ (defined in equation~\eqref{eq:Cvvdecomp}) and use the gravitational and scalar equations of motion~\eqref{eq:gr_eom_st_eft}--\eqref{eq:sc_eom_st_eft} to eliminate the dependencies on $\nabla_{(\alpha_1}\ldots \nabla_{\alpha_j)}\Box \,\phi$ and $\nabla_{(\alpha_1}\ldots \nabla_{\alpha_j)}R_{\mu\nu}$. This yields the scalar-tensor version of lemma~\ref{lem:elim_ric}: on-shell, $F$ can be written as
\begin{equation}
  F = F_0 + \sum_{n \ge 2} \ell^n F_n, \qquad {\mathrm{with}} \qquad F_0=-K_{AB}K^{AB}-\frac12 (\partial_v\phi)^2
\end{equation}
where each $F_n$ depends polynomially on $\mu^{AB}$, $D_{(A_1} \cdots D_{A_j)} \partial_v^N K_{AB}$, $D_{(A_1} \cdots D_{A_j)} \partial_r^{\bar N} \bar K_{AB}$, $D_{(A_1} \!\cdots D_{A_j)} \partial_v^N \phi$, $D_{(A_1} \!\cdots D_{A_j)} \partial_r^{\bar N} \phi$, $D_{(A_1} \!\cdots D_{A_j)} \beta_B$, $D_{(A_1} \!\cdots D_{A_j)} R[\mu]_{ABCD}$, $\epsilon[\mu]_{A_1 \dots A_{n-2}}$ and $d^kV/d\phi^k$.

To obtain~\eqref{eq:st_prop}, one can argue inductively in $n$. At $n=0$~\eqref{eq:st_prop} holds with $X_{0AB}$, $P_0$, $Y_0^A$ vanishing, $S^v_0=s^v$, $S^A_0=s^A$ and $O_2 = \sum_{n \ge 2} \ell^{n} F_n={\cal O}(\ell^2)$. Next, we assume that~\eqref{eq:st_prop} is true for any $n'\leq n-2$ and argue that it must hold for $n'=n$. This amounts to showing that $O_n$ can be brought to a form that has the same structure as the r.h.s.\ of~\eqref{eq:st_prop}. To do this, we first write $O_n$ in terms of the quantities listed at the end of the previous paragraph. By assumption, $O_n$ must be ${\cal O}(\ell^n)$ and it must take the form (tensor indices are suppressed in the rest of the discussion)
\begin{equation}
  O_n = \sum_{\substack{k,k',p,p',
      \\
      \psi,\psi'\in\{ \mu,\phi\} }} A_{n,k,p,k',p',\psi,\psi'} D^{k} \partial_v^{p} \psi D^{k'} \partial_v^{p'} \psi' + O_{n+2}
\end{equation}
where $p,p'>0$ in all terms, i.e.\ $O_n$ must contain at least two factors with quadratic boost weight and these factors can only involve $\mu$ and $\phi$. Now one can proceed as in the case of vacuum gravity and perform a sequence of ``integrations by parts'' to produce terms that are linear either in $\partial_v\mu$ or in $\partial_v\phi$. This of course comes at the cost of boundary terms to be absorbed in $Y^A_n$, $S^v_n$ or terms that can be absorbed into $O_{n+2}$. (The latter class of terms may arise when a $v$-derivative is moved from one of the two positive boost weight factors to the coefficients $A_{n,k,p,k',p',\psi,\psi'}$. The reason for this is that these coefficients can depend on $D^m\beta$, $D^m\partial_r^q \mu$ or $D^m\partial_r^q \phi$. When $\partial_v$ is relocated to act on such terms, the resulting objects need to be expressed with the desired set of primitive factors and some covariant components involving $R_{\mu\nu}$ and $\Box\,\phi$. However, the covariant components can be shifted to higher order in $\ell$ by using the gravitational or scalar equations of motion.) Note in particular that to determine the boundary terms to be added to $S^v_n$, one needs to use an identity of the form (cf.\ equation~\eqref{eq:distr_dv})
\begin{equation*}
  \partial_v \left[ \frac{1}{\sqrt{\mu}} \partial_v \left( \sqrt{\mu}A_{n,k,p,p',\psi,\psi'} \sum_{j=1}^{p'+p-3} a_j
    D^{k} \partial_v^{j} \psi \partial_v^{p+p'-2-j} \psi' \right) \right] = A_{n,k,p,p',\psi,\psi'} D^{k} \partial_v^{p} \psi \partial_v^{p'} \psi'
  + \dots
\end{equation*}
where $\psi,\psi'\in \{ \mu,\phi \} $, and the ellipsis stands for terms of the form $A_{n,\bar k,\bar p,\bar k',\bar p',\psi,\psi'} D^{\bar k} \partial_v^{\bar p} \psi \partial_v^{\bar p'} \psi'$
with either (1) $\bar p + \bar p' < p + p'$, $\bar p, \bar p' >0$, (2) $\bar p + \bar p' = p + p'$, $ \bar p >0$, $\bar p'=1$, or (3) $\bar p + \bar p' = p + p'$, $ \bar p' >0$, $\bar p=1$. The coefficients $a_j$ in this identity are to be determined by solving the linear system~\eqref{eq:aj_lin_sys} (for any $\psi,\psi'\in \{ \mu,\phi \} $). In the end, this procedure lets us write
\begin{eqnarray}
  O_n &=&
  \sum_{\substack{k,p>0,
      \\
      \psi\in\{ \mu,\phi\} }} \left(A_{n,k,p,\psi} D^{k} \partial_v^{p} \psi \partial_v \mu+B_{n,k,p,\psi} D^{k} \partial_v^{p} \psi \partial_v \phi\right)
  \nonumber\\[\jot]
  & &+ \partial_v \left[ \frac{1}{\sqrt{\mu}} \partial_v \left( \sqrt{\mu} \varsigma^v_{n} \right)\right] + D_A y^A_{n} + O_{n+2}.
\end{eqnarray}
The quantities $y^A_{n}$, $\varsigma^v_{n}$ can be absorbed into $Y^A_n$, $S^v_n$, and they have the same dependencies on $\mu$, $\phi$ and $\beta$ as required by the proposition. The terms in the first line of the r.h.s.\ of this equation can be dealt with by ``completing the square'' on $K_{AB}$ and $\partial_v\phi$, thereby producing terms with definite signs and an error term that is higher order in $\ell$ and to be absorbed into $O_{n+2}$, concluding the sketch of the proof of proposition~\ref{prop:st}.

\section*{Acknowledgments}

We are very grateful to Iain Davies for suggesting a simplification to the inductive argument of section 4. We are grateful to Bob Wald, Aron Wall and Zihan Yan for helpful discussions. SH thanks Trinity College, Cambridge U. for support as a VFC, and DAMTP, Cambridge U. He is grateful to the Max-Planck Society for supporting the collaboration between MPI-MiS and Leipzig U., grant Proj. Bez. M.FE.A.MATN0003. ADK acknowledges support from the European Union’s H2020 ERC Consolidator Grant “GRavity from Astrophysical to Microscopic Scales” (Grant No. GRAMS815673). This work was supported by the EU Horizon 2020 Research and Innovation Programme under the Marie Sklodowska-Curie Grant Agreement No. 101007855. HSR is supported by STFC grant no. ST/T000694/1.

\begin{appendices}

\section{Curvature components in Gaussian null coordinates}\label{app:GNC}

Below we provide the expressions for the curvature components in Gaussian null coordinates. The metric is given by
\be
g=2\ud v\ud r-r^2\alpha \ud v^2-2r\beta_A \ud x^A \ud v +\mu_{AB}\ud x^A \ud x^B
\ee
The nonzero components of the inverse metric are
\be
g^{rr}=r^2(\alpha+\beta^2), \qquad g^{rv}=1, \qquad g^{rA}=r\beta^A, \qquad g^{AB}=\mu^{AB}
\ee
where $\mu^{AB}$ is the inverse of $\mu_{AB}$, $\beta^A\equiv \mu^{AB}\beta_B$ and $\beta^2\equiv \beta_A \beta^A$.
We introduce the following variables
\be
K_{AB}\equiv \frac12 \partial_v \mu_{AB}, \qquad \bar K_{AB}\equiv \frac12 \partial_r \mu_{AB}
\ee
The nonzero Christoffel symbols are given by
\bea
\Gamma^r_{rv}&=&-\tfrac12 \partial_r(r^2\alpha)-\tfrac12 r\beta^A \partial_r(r\beta_A) \nonumber \\
\Gamma^A_{rv}&=&-\tfrac12 \mu^{AB}\partial_r (r\beta_B) \nonumber \\
\Gamma^r_{rA}&=&-\tfrac12 \partial_r(r\beta_A)+r\beta^B {\bar K}_{AB} \nonumber \\
\Gamma^A_{rB}&=&{\bar K}^A{}_B \nonumber \\
\Gamma^v_{vv}&=&\tfrac12 \partial_r(r^2\alpha) \nonumber \\
\Gamma^r_{vv}&=&-\tfrac12 r^2 \partial_v \alpha+\tfrac12 r^2(\alpha+\beta^2)\partial_r(r^2\alpha)-r^2\beta^A\left(\partial_v \beta_A+\tfrac12 r D_A\alpha \right) \nonumber \\
\Gamma^A_{vv}&=&-\tfrac12 r\beta^A \partial_r(r^2\alpha)+r\mu^{AB}\left(-\partial_v \beta_B+\tfrac12 r D_B \alpha\right) \nonumber \\
\Gamma^v_{vA}&=&\tfrac12 \partial_r(r\beta_A) \nonumber \\
\Gamma^r_{vA}&=&\tfrac12 r^2(\alpha+\beta^2)\partial_r(r\beta_A)-\tfrac12 r^2 D_A\alpha- r^2\beta^B D_{[A}\beta_{B]}+r\beta^BK_{AB} \nonumber \\
\Gamma^A_{vB}&=&\tfrac12 r\beta^A\partial_r(r\beta_B)+ r\mu^{AC} D_{[C}\beta_{B]}+K^A{}_{B} \nonumber \\
\Gamma^v_{AB}&=&-{\bar K}_{AB} \nonumber \\
\Gamma^r_{AB}&=&-r^2\left(\alpha+\beta^2\right){\bar K}_{AB}-rD_{(A}\beta_{B)}-K_{AB} \nonumber \\
\Gamma^C_{AB}&=&\Gamma[\mu]^C_{AB}-r\beta^C{\bar K}_{AB}
\eea
Then the components of the Riemann tensor are as follows
\bea
R_{rvrv}&=&\tfrac14\mu^{AB}\partial_r(r\beta_{A}) \partial_r(r\beta_B) + \tfrac{1}{2} \partial_r^2 (r^2\alpha) \nonumber \\
R_{rvrA}&=&-\tfrac12 \bar{K}_{A}{}^B \partial_r (r\beta_B) + \tfrac12 \partial_r^2 (r\beta_{A}) \nonumber \\
R_{rvvA}&=& \tfrac12 \partial_v\partial_r (r\beta_A)+ \tfrac12 K_{A}{}^B \partial_r (r\beta_B)- \bar{K}_{A}{}^B r\partial_v \beta_B - \tfrac14 r\beta^{B} \partial_r(r\beta_{A}) \partial_r(r\beta_{B})  
 \nonumber \\
& &-  \tfrac{1}{2} r\mu^{BC}\partial_r(r\beta_C) D_{[A}\beta_{B]} -  \tfrac{1}{2} D_{A}\partial_r (r^2\alpha)+ \tfrac{1}{2} r^2\bar{K}_{AB} D^{B}\alpha +  \tfrac{1}{4}r\beta^B \bar{K}_{AB}  \
\partial_r (r^2\alpha)  \nonumber \\
R_{rArB}&=&\bar{K}_{A}{}^{C} \bar{K}_{BC} -  \partial_r \bar{K}_{AB} \nonumber \\
R_{rAvB}&=&-  \partial_v \bar{K}_{AB}+K_{A}{}^{C} \bar{K}_{BC} +r\beta^C \bar{K}_{B[C} \partial_r(r\beta_{A]})  \nonumber \\
& &- r \bar{K}_{B}{}^C D_{[A}\beta_{C]} -\tfrac12 D_{B}\partial_r(r\beta_{A})- \tfrac14 \partial_r(r\beta_{A}) \partial_r(r\beta_{B}) -  \tfrac{1}{2} \bar{K}_{AB} \partial_r (r^2\alpha)  \nonumber \\
R_{vAvB}&=&-  \partial_v K_{AB}+K_{A}{}^{C} K_{BC} - 2 r\beta^C \bar{K}_{A[B} \partial_r(r\beta_{C]}) - 2 r\beta^C \bar{K}_{B[A} \partial_r(r\beta_{C]}) \nonumber \\
& &+ \tfrac14 r^2(\alpha+\beta^2) \partial_r(r\beta_{A}) \partial_r(r\beta_{B})  -  \tfrac12 r^2\partial_r(r\beta_{(A}) D_{B)}\alpha +  r^2\beta^{C} \partial_r (r\beta_{(A}) D_{B)}\beta_{C} \nonumber \\
& &+ r K_{C(A} D^{C}\beta_{B)}-r K_{C(A} D_{B)}\beta^{C}  - r D_{(B}\partial_v\beta_{A)} + \tfrac{1}{2}r^2 D_{B}D_{A}\alpha \nonumber \\
& &+  \tfrac{1}{2} r^3\bar{K}_{AB} \beta^{C} D_{C}\alpha -  \tfrac{1}{2} r^2\beta^{C} \partial_r(r\beta_{(A|}) D_{C}\beta_{|B)} \nonumber \\
& & +\mu^{CD} r^2  D_{[C}\beta_{A]} D_{[D}\beta_{B]} + \tfrac{1}{2} K_{AB} \partial_r (r^2\alpha) + \tfrac{1}{2} r^2(\alpha+\beta^2)\bar{K}_{AB} \partial_r (r^2\alpha)  \nonumber \\
& &-  \tfrac{1}{2} r D_{(A}\beta_{B)} \partial_r (r^2\alpha) -  \tfrac{1}{2} r^2\bar{K}_{AB} \partial_v \alpha \nonumber \\
R_{ABvC}&=&- 2 D_{[A}K_{B]C}+\partial_r (r\beta_{[A}) K_{B]C} +  2K_{D[A} \bar{K}_{B]C} r \beta^{D}  +\partial_r (r\beta_{[A}) \bar{K}_{B]C} r^2(\alpha+\beta^2) \nonumber \\
& & +  r^2\bar{K}_{C[A} D_{B]}\alpha +  \tfrac{1}{2} r \partial_r (r\beta_{[A}) D_{B]}\beta_{C}  - \tfrac{1}{2} r D_{C}\beta_{[A} \partial_r (r\beta_{B]})  \nonumber \\
& &   +  r D_{C}D_{[A}\beta_{B]} +   r^2 \bar{K}_{C[A} \beta^{D} D_{B]}\beta_{D}- r^2\beta_D \bar{K}_{C[A}  D^{D}\beta_{B]}  \nonumber \\
R_{ABrC}&=&2D_{[B}\bar{K}_{A]C}-2r \bar{K}_{D[A} \bar{K}_{B]C} \beta^{D} -
\partial_r (r\beta_{[A})\bar{K}_{B]C} \nonumber \\
R_{ABCD}&=&R[\mu]_{ABCD}+2 K_{B[C} \bar{K}_{D]A} + 2K_{A[D} \bar{K}_{C]B} + 2r^2(\alpha+\beta^2)\bar{K}_{A[D} \bar{K}_{C]B}  \nonumber \\
& & + r\bar{K}_{D[B} D_{A]}\beta_{C} -   r\bar{K}_{C[B} D_{A]}\beta_{D}  +  r\bar{K}_{B[D} D_{C]}\beta_{A} -  r \bar{K}_{A[D} D_{C]}\beta_{B}  \nonumber 
\eea
The components of the Ricci tensor are given by the following expressions
\bea
R_{vv}&=&-  \mu^{AB} \partial_v K_{AB}+K_{AB} K^{AB} -  r\bar{K} \beta^{A} \partial_r(r\beta_{A}) + 2 r \bar{K}^{AB} \beta_{A} \partial_r(r \beta_{B})    \nonumber \\
& &- r^3\bar{K}_{AB} \beta^{A} D^{B}\alpha+ \tfrac{1}{2}r^3 \bar{K} \beta^{A} D_{A}\alpha -\tfrac12 r^2 \partial_r(r\beta_A) D^{A}\alpha + r^2 \beta^{A} \partial_r(r\beta_B) D_{A}\beta^{B} \nonumber \\
& &- r D^{A}\partial_v \beta_A +r\beta^{A} D_{A}\partial_r (r^2\alpha) - r^2 \beta^{A} \partial_r(r\beta_B) D^{B}\beta_{A} + \tfrac{1}{2}r^2 D^A D_{A}\alpha- \tfrac{1}{2} r^2\bar{K} \partial_v \alpha   \nonumber \\
& &-  r^2 D_{[A}\beta_{B]} D^{B}\beta^{A}  + \tfrac{1}{2} K \partial_r (r^2\alpha) + \tfrac{1}{2} r^2(\alpha+\beta^2)\bar{K}  \partial_r (r^2\alpha)- r \beta^{A} \partial_v \partial_r(r\beta_{A})  \nonumber \\
& &- r^2 \bar{K}_{AB} \beta^{A} \beta^{B} \partial_r (r^2\alpha) -  \tfrac{1}{2} r D_{A}\beta^{A} \partial_r (r^2\alpha) + \tfrac{1}{2} r^2(\alpha+\beta^2) \partial_r^2 (r^2\alpha) \nonumber \\
& & + \tfrac12 r^2(\alpha+\beta^2) \partial_r(r\beta_{A}) \partial_r(r\beta_B) \mu^{AB} - \tfrac12 r^2 \beta^{A}\beta^B \partial_r(r\beta_{A})\partial_r(r\beta_{B}) \nonumber \\
R_{rv}&=&-\mu^{AB} \partial_v \bar{K}_{AB}+K^{AB} \bar{K}_{AB} - \tfrac12 r \bar{K} \beta^{A} \partial_r(r\beta_{A})  + r \bar{K}^{AB} \beta_{A}\partial_r(r\beta_B)-  \tfrac{1}{2} \partial_r^2 (r^2\alpha)   \nonumber \\
& &- \tfrac12 \mu^{AB}\partial_r(r\beta_A) \partial_r(r\beta_B)-\tfrac12 D^{A}\partial_r(r\beta_A) -  \tfrac{1}{2} \bar{K} \partial_r (r^2\alpha) - \tfrac12 r\beta^{A} \partial_r^2 (r\beta_A)  \nonumber \\
R_{rr}&=&\bar{K}_{AB} \bar{K}^{AB} -  \mu^{AB} \partial_r \bar{K}_{AB} \nonumber \\
R_{rA}&=&D_{B}\bar{K}_{A}{}^{B}-D_{A}\bar{K} -\tfrac12\partial_r^2 (r\beta_A)-\tfrac12\bar{K} \partial_r(r\beta_A) + \bar{K}_{A}{}^B \partial_r (r\beta_B) \nonumber \\
& &-r\beta^{B}\left(2 \bar{K}_{A}{}^{C} \bar{K}_{BC}  -  \bar{K}_{AB} \bar{K}  -   \partial_r \bar{K}_{AB} \right) \nonumber \\
R_{vA}&=&D_{B}K_{A}{}^{B}-  D_{A}K+\tfrac12 \partial_v\partial_r (r\beta_A) +\tfrac12 K \partial_r (r\beta_A) - r\bar{K}_{A}{}^B \partial_v(\beta_B)\nonumber \\
& &-2r^2\partial_r(r\beta_{[A}) \bar{K}_{B]C} \beta^{B} \beta^{C} +\tfrac12 r\beta^B \partial_r(r\beta_A)\partial_r(r\beta_B)  \nonumber \\
& &- r^2(\alpha+\beta^2)\bar{K}_{A}{}^C  \partial_r(r\beta_C)  -  r^2 \bar{K} \beta^{B} D_{[A}\beta_{B]} + r \partial_r(r\beta_{[A}) D_{B]}\beta^{B} \nonumber \\
& &+ 2r^2 \bar{K}_{C[A} \beta^{B} D^{C}\beta_{B]} - 2r^2\bar{K}_{C[A} \beta^{B} D_{B]}\beta^{C}  -2 r K_{A}{}^{C} \bar{K}_{BC} \beta^{B} + r K_{AB} \bar{K} \beta^{B}  \nonumber \\
& & -  \tfrac{1}{2} D_{A}\partial_r (r^2\alpha) + r D^{B}D_{[A}\beta_{B]}+ r^2\bar{K}_{AB} D^{B}\alpha -  \tfrac{1}{2}r^2 \bar{K} D_{A}\alpha \nonumber \\
& & + r\beta^{B}\left(D_{B}\partial_r(r\beta_A)- \tfrac12 D_{A}\partial_r(r\beta_B)\right)   + r\bar{K}_{AB} \beta^{B} \partial_r (r^2\alpha) \nonumber \\
& &+\tfrac12  r^2(\alpha+\beta^2) \left(\partial_r^2 (r\beta_{A}) +{\bar K}\partial_r(r\beta_A)\right)+ r \beta^{B} \partial_v \bar{K}_{AB} \nonumber \\
R_{AB}&=&R[\mu]_{AB}- 2 \partial_v \bar{K}_{AB}- K \bar{K}_{AB} + 4 K_{(A}{}^{C} \bar{K}_{B)C} -  K_{AB} \bar{K}  \nonumber \\
& &+ r^2(\alpha+\beta^2)\left(2 \bar{K}_{A}{}^{C} \bar{K}_{BC}  -  \bar{K}_{AB} \bar{K}\partial_r \bar{K}_{AB}\right)  - 4r^2 \bar{K}_{C(A} \bar{K}_{B)D} \beta^{C} \beta^{D}  \nonumber \\
& &+ 2 r \partial_r(r\beta_{(A}) \bar{K}_{B)C} \beta^{C}- 2 r \bar{K}_{AB} \beta^{C} \partial_r(r\beta_C)  -  \bar{K}_{AB} \partial_r (r^2\alpha)   \nonumber \\
& & + 2 r\beta^{C} D_{[A}\bar{K}_{C]B}+ 2 r \beta^{C} D_{[B}\bar{K}_{C]A}  +  r\bar{K} D_{(B}\beta_{A)} \nonumber \\
& &-D_{(B}\partial_r(r\beta_{A)})- \tfrac12 \partial_r(r\beta_{A}) \partial_r(r\beta_{B})   - r\bar{K}_{AB} D_{C}\beta^{C} + 2r \bar{K}_{C(A} D^{C}\beta_{B)}  \nonumber 
\eea

\subsection{Expressions on ${\cal N}$}

The nonzero Christoffel symbols on ${\cal N}$ are given by
\bea
\Gamma^r_{rA}=-\Gamma^v_{vA}=\mu_{AB}\Gamma^B_{rv}&=&-\tfrac12 \beta_A \nonumber \\
\mu_{AC}\Gamma^C_{vB}=-\Gamma^r_{AB}&=&K_{AB} \nonumber \\
\mu_{AC}\Gamma^C_{rB}=-\Gamma^v_{AB}&=&{\bar K}_{AB} \nonumber \\
\Gamma^C_{AB}&=&\Gamma[\mu]^C_{AB}
\eea
The Riemann components on ${\cal N}$ simplify as follows
\bea
R_{rvrv}&=&\tfrac14\beta^2 + \alpha \nonumber \\
R_{rvrA}&=&-\tfrac12 \bar{K}_{A}{}^B \beta_B + \partial_r \beta_{A} \nonumber \\
R_{rvvA}&=& \tfrac12 \partial_v\beta_A+ \tfrac12 K_{A}{}^B \beta_B \nonumber \\
R_{rArB}&=&\bar{K}_{A}{}^{C} \bar{K}_{BC} -  \partial_r \bar{K}_{AB} \nonumber \\
R_{rAvB}&=&-  \partial_v \bar{K}_{AB}+K_{A}{}^{C} \bar{K}_{BC}  -\tfrac12 D_{B}\beta_{A}- \tfrac14 \beta_A\beta_B  \nonumber \\
R_{vAvB}&=&-  \partial_v K_{AB}+K_{A}{}^{C} K_{BC} \nonumber \\
R_{ABvC}&=&- 2 D_{[A}K_{B]C}+\beta_{[A} K_{B]C}  \nonumber \\
R_{ABrC}&=&2D_{[B}\bar{K}_{A]C} -\beta_{[A}\bar{K}_{B]C} \nonumber \\
R_{ABCD}&=&R[\mu]_{ABCD}+2 K_{B[C} \bar{K}_{D]A} + 2K_{A[D} \bar{K}_{C]B}  \nonumber 
\eea
The components of the Ricci tensor on $\cal N$ are given by
\bea
R_{vv}&=&-  \mu^{AB} \partial_v K_{AB}+K_{AB} K^{AB}  \nonumber \\
R_{rv}&=&-\mu^{AB} \partial_v \bar{K}_{AB}+K^{AB} \bar{K}_{AB} - \alpha  - \tfrac12 \mu^{AB}\beta_A \beta_B-\tfrac12 D^{A}\beta_A  \nonumber \\
R_{rr}&=&\bar{K}_{AB} \bar{K}^{AB} -  \mu^{AB} \partial_r \bar{K}_{AB} \nonumber \\
R_{rA}&=&-\partial_r \beta_A+\tfrac12 \beta^B \bar{K}_{AB}+\left(D_{B}+\tfrac12 \beta_B\right)\bar{K}_{A}{}^{B}-\left(D_{A} +\tfrac12 \beta_A\right)\bar{K}  \nonumber \\
R_{vA}&=&\tfrac12 \partial_v\beta_A+ \tfrac12{K}_{A}{}^B \beta_B+\left(D_{B}-\tfrac12 \beta_B\right)K_{A}{}^{B}-  \left(D_{A}-\tfrac12 \beta_A\right)K  \nonumber \\
R_{AB}&=&R[\mu]_{AB}- 2 \partial_v \bar{K}_{AB}- K \bar{K}_{AB} + 4 K_{(A}{}^{C} \bar{K}_{B)C} -  K_{AB} \bar{K}  \nonumber \\
& &-D_{(B}\beta_{A)}- \tfrac12 \beta_{A}\beta_{B}  \nonumber 
\eea

\end{appendices}

\end{document}